\documentclass[12pt,preprint,sort]{aastex62}
\usepackage{natbib}
\usepackage{hyperref}
\usepackage{subfigure}
\usepackage{xspace}

\usepackage{graphicx}

\usepackage{color}

\newcommand{\Fermi}{{\em Fermi~}}
\newcommand{\Swift}{{\em Swift~}}

\shorttitle{Joint \Fermi/\Swift Observations of GRBs}
\shortauthors{The \Fermi LAT Collaboration}

\begin{document}

\title{Investigating the Nature of Late-Time High-Energy GRB Emission Through Joint \Fermi/\Swift Observations}

\author{M.~Ajello}
\affiliation{Department of Physics and Astronomy, Clemson University, Kinard Lab of Physics, Clemson, SC 29634-0978, USA}
\author{L.~Baldini}
\affiliation{Universit\`a di Pisa and Istituto Nazionale di Fisica Nucleare, Sezione di Pisa I-56127 Pisa, Italy}
\author{G.~Barbiellini}
\affiliation{Istituto Nazionale di Fisica Nucleare, Sezione di Trieste, I-34127 Trieste, Italy}
\affiliation{Dipartimento di Fisica, Universit\`a di Trieste, I-34127 Trieste, Italy}
\author{D.~Bastieri}
\affiliation{Istituto Nazionale di Fisica Nucleare, Sezione di Padova, I-35131 Padova, Italy}
\affiliation{Dipartimento di Fisica e Astronomia ``G. Galilei'', Universit\`a di Padova, I-35131 Padova, Italy}
\author{R.~Bellazzini}
\affiliation{Istituto Nazionale di Fisica Nucleare, Sezione di Pisa, I-56127 Pisa, Italy}
\author{E.~Bissaldi}
\affiliation{Dipartimento di Fisica ``M. Merlin" dell'Universit\`a e del Politecnico di Bari, I-70126 Bari, Italy}
\affiliation{Istituto Nazionale di Fisica Nucleare, Sezione di Bari, I-70126 Bari, Italy}
\author{R.~D.~Blandford}
\affiliation{W. W. Hansen Experimental Physics Laboratory, Kavli Institute for Particle Astrophysics and Cosmology, Department of Physics and SLAC National Accelerator Laboratory, Stanford University, Stanford, CA 94305, USA}
\author{R.~Bonino}
\affiliation{Istituto Nazionale di Fisica Nucleare, Sezione di Torino, I-10125 Torino, Italy}
\affiliation{Dipartimento di Fisica, Universit\`a degli Studi di Torino, I-10125 Torino, Italy}
\author{E.~Bottacini}
\affiliation{W. W. Hansen Experimental Physics Laboratory, Kavli Institute for Particle Astrophysics and Cosmology, Department of Physics and SLAC National Accelerator Laboratory, Stanford University, Stanford, CA 94305, USA}
\affiliation{Department of Physics and Astronomy, University of Padova, Vicolo Osservatorio 3, I-35122 Padova, Italy}
\author{J.~Bregeon}
\affiliation{Laboratoire Univers et Particules de Montpellier, Universit\'e Montpellier, CNRS/IN2P3, F-34095 Montpellier, France}
\author{P.~Bruel}
\affiliation{Laboratoire Leprince-Ringuet, \'Ecole polytechnique, CNRS/IN2P3, F-91128 Palaiseau, France}
\author{R.~Buehler}
\affiliation{Deutsches Elektronen Synchrotron DESY, D-15738 Zeuthen, Germany}
\author{R.~A.~Cameron}
\affiliation{W. W. Hansen Experimental Physics Laboratory, Kavli Institute for Particle Astrophysics and Cosmology, Department of Physics and SLAC National Accelerator Laboratory, Stanford University, Stanford, CA 94305, USA}
\author{R.~Caputo}
\affiliation{Center for Research and Exploration in Space Science and Technology (CRESST) and NASA Goddard Space Flight Center, Greenbelt, MD 20771, USA}
\author{P.~A.~Caraveo}
\affiliation{INAF-Istituto di Astrofisica Spaziale e Fisica Cosmica Milano, via E. Bassini 15, I-20133 Milano, Italy}
\author{G.~Chiaro}
\affiliation{INAF-Istituto di Astrofisica Spaziale e Fisica Cosmica Milano, via E. Bassini 15, I-20133 Milano, Italy}
\author{S.~Ciprini}
\affiliation{Space Science Data Center - Agenzia Spaziale Italiana, Via del Politecnico, snc, I-00133, Roma, Italy}
\affiliation{Istituto Nazionale di Fisica Nucleare, Sezione di Perugia, I-06123 Perugia, Italy}
\author{J.~Cohen-Tanugi}
\affiliation{Laboratoire Univers et Particules de Montpellier, Universit\'e Montpellier, CNRS/IN2P3, F-34095 Montpellier, France}
\author{D.~Costantin}
\affiliation{Dipartimento di Fisica e Astronomia ``G. Galilei'', Universit\`a di Padova, I-35131 Padova, Italy}
\author{F.~D'Ammando}
\affiliation{INAF Istituto di Radioastronomia, I-40129 Bologna, Italy}
\affiliation{Dipartimento di Astronomia, Universit\`a di Bologna, I-40127 Bologna, Italy}
\author{F.~de~Palma}
\affiliation{Istituto Nazionale di Fisica Nucleare, Sezione di Torino, I-10125 Torino, Italy}
\author{N.~Di~Lalla}
\affiliation{Universit\`a di Pisa and Istituto Nazionale di Fisica Nucleare, Sezione di Pisa I-56127 Pisa, Italy}
\author{M.~Di~Mauro}
\affiliation{W. W. Hansen Experimental Physics Laboratory, Kavli Institute for Particle Astrophysics and Cosmology, Department of Physics and SLAC National Accelerator Laboratory, Stanford University, Stanford, CA 94305, USA}
\author{L.~Di~Venere}
\affiliation{Dipartimento di Fisica ``M. Merlin" dell'Universit\`a e del Politecnico di Bari, I-70126 Bari, Italy}
\affiliation{Istituto Nazionale di Fisica Nucleare, Sezione di Bari, I-70126 Bari, Italy}
\author{A.~Dom\'inguez}
\affiliation{Grupo de Altas Energ\'ias, Universidad Complutense de Madrid, E-28040 Madrid, Spain}
\author{C.~Favuzzi}
\affiliation{Dipartimento di Fisica ``M. Merlin" dell'Universit\`a e del Politecnico di Bari, I-70126 Bari, Italy}
\affiliation{Istituto Nazionale di Fisica Nucleare, Sezione di Bari, I-70126 Bari, Italy}
\author{A.~Franckowiak}
\affiliation{Deutsches Elektronen Synchrotron DESY, D-15738 Zeuthen, Germany}
\author{Y.~Fukazawa}
\affiliation{Department of Physical Sciences, Hiroshima University, Higashi-Hiroshima, Hiroshima 739-8526, Japan}
\author{S.~Funk}
\affiliation{Friedrich-Alexander-Universit\"at Erlangen-N\"urnberg, Erlangen Centre for Astroparticle Physics, Erwin-Rommel-Str. 1, 91058 Erlangen, Germany}
\author{P.~Fusco}
\affiliation{Dipartimento di Fisica ``M. Merlin" dell'Universit\`a e del Politecnico di Bari, I-70126 Bari, Italy}
\affiliation{Istituto Nazionale di Fisica Nucleare, Sezione di Bari, I-70126 Bari, Italy}
\author{F.~Gargano}
\affiliation{Istituto Nazionale di Fisica Nucleare, Sezione di Bari, I-70126 Bari, Italy}
\author{D.~Gasparrini}
\affiliation{Space Science Data Center - Agenzia Spaziale Italiana, Via del Politecnico, snc, I-00133, Roma, Italy}
\affiliation{Istituto Nazionale di Fisica Nucleare, Sezione di Perugia, I-06123 Perugia, Italy}
\author{N.~Giglietto}
\affiliation{Dipartimento di Fisica ``M. Merlin" dell'Universit\`a e del Politecnico di Bari, I-70126 Bari, Italy}
\affiliation{Istituto Nazionale di Fisica Nucleare, Sezione di Bari, I-70126 Bari, Italy}
\author{F.~Giordano}
\affiliation{Dipartimento di Fisica ``M. Merlin" dell'Universit\`a e del Politecnico di Bari, I-70126 Bari, Italy}
\affiliation{Istituto Nazionale di Fisica Nucleare, Sezione di Bari, I-70126 Bari, Italy}
\author{M.~Giroletti}
\affiliation{INAF Istituto di Radioastronomia, I-40129 Bologna, Italy}
\author{D.~Green}
\affiliation{Department of Astronomy, University of Maryland, College Park, MD 20742, USA}
\affiliation{NASA Goddard Space Flight Center, Greenbelt, MD 20771, USA}
\author{I.~A.~Grenier}
\affiliation{AIM, CEA, CNRS, Universit\'e Paris-Saclay, Universit\'e Paris Diderot, Sorbonne Paris Cit\'e, F-91191 Gif-sur-Yvette, France}
\author{S.~Guiriec}
\affiliation{The George Washington University, Department of Physics, 725 21st St, NW, Washington, DC 20052, USA}
\affiliation{NASA Goddard Space Flight Center, Greenbelt, MD 20771, USA}
\author{C.~Holt}
\affiliation{Department of Physics and Center for Space Sciences and Technology, University of Maryland Baltimore County, Baltimore, MD 21250, USA}
\author{D.~Horan}
\affiliation{Laboratoire Leprince-Ringuet, \'Ecole polytechnique, CNRS/IN2P3, F-91128 Palaiseau, France}
\author{G.~J\'ohannesson}
\affiliation{Science Institute, University of Iceland, IS-107 Reykjavik, Iceland}
\affiliation{Nordita, Royal Institute of Technology and Stockholm University, Roslagstullsbacken 23, SE-106 91 Stockholm, Sweden}
\author{D.~Kocevski}
\email{kocevski@slac.stanford.edu}
\affiliation{NASA Goddard Space Flight Center, Greenbelt, MD 20771, USA}
\author{M.~Kuss}
\affiliation{Istituto Nazionale di Fisica Nucleare, Sezione di Pisa, I-56127 Pisa, Italy}
\author{G.~La~Mura}
\affiliation{Laborat\'orio de Instrumenta\c{c}\~{a}o e F\'isica Experimental de Part\'iculas, Av. Prof. Gama Pinto, n.2, Complexo Interdisciplinar (3is), 1649-003 Lisboa, Portugal}
\author{S.~Larsson}
\affiliation{Department of Physics, KTH Royal Institute of Technology, AlbaNova, SE-106 91 Stockholm, Sweden}
\affiliation{The Oskar Klein Centre for Cosmoparticle Physics, AlbaNova, SE-106 91 Stockholm, Sweden}
\author{J.~Li}
\affiliation{Deutsches Elektronen Synchrotron DESY, D-15738 Zeuthen, Germany}
\author{F.~Longo}
\affiliation{Istituto Nazionale di Fisica Nucleare, Sezione di Trieste, I-34127 Trieste, Italy}
\affiliation{Dipartimento di Fisica, Universit\`a di Trieste, I-34127 Trieste, Italy}
\author{F.~Loparco}
\affiliation{Dipartimento di Fisica ``M. Merlin" dell'Universit\`a e del Politecnico di Bari, I-70126 Bari, Italy}
\affiliation{Istituto Nazionale di Fisica Nucleare, Sezione di Bari, I-70126 Bari, Italy}
\author{P.~Lubrano}
\affiliation{Istituto Nazionale di Fisica Nucleare, Sezione di Perugia, I-06123 Perugia, Italy}
\author{J.~D.~Magill}
\affiliation{Department of Astronomy, University of Maryland, College Park, MD 20742, USA}
\author{S.~Maldera}
\affiliation{Istituto Nazionale di Fisica Nucleare, Sezione di Torino, I-10125 Torino, Italy}
\author{A.~Manfreda}
\affiliation{Universit\`a di Pisa and Istituto Nazionale di Fisica Nucleare, Sezione di Pisa I-56127 Pisa, Italy}
\author{M.~N.~Mazziotta}
\affiliation{Istituto Nazionale di Fisica Nucleare, Sezione di Bari, I-70126 Bari, Italy}
\author{P.~F.~Michelson}
\affiliation{W. W. Hansen Experimental Physics Laboratory, Kavli Institute for Particle Astrophysics and Cosmology, Department of Physics and SLAC National Accelerator Laboratory, Stanford University, Stanford, CA 94305, USA}
\author{T.~Mizuno}
\affiliation{Hiroshima Astrophysical Science Center, Hiroshima University, Higashi-Hiroshima, Hiroshima 739-8526, Japan}
\author{M.~E.~Monzani}
\affiliation{W. W. Hansen Experimental Physics Laboratory, Kavli Institute for Particle Astrophysics and Cosmology, Department of Physics and SLAC National Accelerator Laboratory, Stanford University, Stanford, CA 94305, USA}
\author{A.~Morselli}
\affiliation{Istituto Nazionale di Fisica Nucleare, Sezione di Roma ``Tor Vergata", I-00133 Roma, Italy}
\author{M.~Negro}
\affiliation{Istituto Nazionale di Fisica Nucleare, Sezione di Torino, I-10125 Torino, Italy}
\affiliation{Dipartimento di Fisica, Universit\`a degli Studi di Torino, I-10125 Torino, Italy}
\author{E.~Nuss}
\affiliation{Laboratoire Univers et Particules de Montpellier, Universit\'e Montpellier, CNRS/IN2P3, F-34095 Montpellier, France}
\author{N.~Omodei}
\affiliation{W. W. Hansen Experimental Physics Laboratory, Kavli Institute for Particle Astrophysics and Cosmology, Department of Physics and SLAC National Accelerator Laboratory, Stanford University, Stanford, CA 94305, USA}
\author{M.~Orienti}
\affiliation{INAF Istituto di Radioastronomia, I-40129 Bologna, Italy}
\author{E.~Orlando}
\affiliation{W. W. Hansen Experimental Physics Laboratory, Kavli Institute for Particle Astrophysics and Cosmology, Department of Physics and SLAC National Accelerator Laboratory, Stanford University, Stanford, CA 94305, USA}
\author{V.~S.~Paliya}
\affiliation{Department of Physics and Astronomy, Clemson University, Kinard Lab of Physics, Clemson, SC 29634-0978, USA}
\author{J.~S.~Perkins}
\affiliation{NASA Goddard Space Flight Center, Greenbelt, MD 20771, USA}
\author{M.~Persic}
\affiliation{Istituto Nazionale di Fisica Nucleare, Sezione di Trieste, I-34127 Trieste, Italy}
\affiliation{Osservatorio Astronomico di Trieste, Istituto Nazionale di Astrofisica, I-34143 Trieste, Italy}
\author{M.~Pesce-Rollins}
\affiliation{Istituto Nazionale di Fisica Nucleare, Sezione di Pisa, I-56127 Pisa, Italy}
\author{F.~Piron}
\affiliation{Laboratoire Univers et Particules de Montpellier, Universit\'e Montpellier, CNRS/IN2P3, F-34095 Montpellier, France}
\author{G.~Principe}
\affiliation{Friedrich-Alexander-Universit\"at Erlangen-N\"urnberg, Erlangen Centre for Astroparticle Physics, Erwin-Rommel-Str. 1, 91058 Erlangen, Germany}
\author{J.~L.~Racusin}
\email{judith.racusin@nasa.gov}
\affiliation{NASA Goddard Space Flight Center, Greenbelt, MD 20771, USA}
\author{S.~Rain\`o}
\affiliation{Dipartimento di Fisica ``M. Merlin" dell'Universit\`a e del Politecnico di Bari, I-70126 Bari, Italy}
\affiliation{Istituto Nazionale di Fisica Nucleare, Sezione di Bari, I-70126 Bari, Italy}
\author{R.~Rando}
\affiliation{Istituto Nazionale di Fisica Nucleare, Sezione di Padova, I-35131 Padova, Italy}
\affiliation{Dipartimento di Fisica e Astronomia ``G. Galilei'', Universit\`a di Padova, I-35131 Padova, Italy}
\author{M.~Razzano}
\affiliation{Istituto Nazionale di Fisica Nucleare, Sezione di Pisa, I-56127 Pisa, Italy}
\affiliation{Funded by contract FIRB-2012-RBFR12PM1F from the Italian Ministry of Education, University and Research (MIUR)}
\author{S.~Razzaque}
\affiliation{Department of Physics, University of Johannesburg, PO Box 524, Auckland Park 2006, South Africa}
\author{A.~Reimer}
\affiliation{Institut f\"ur Astro- und Teilchenphysik and Institut f\"ur Theoretische Physik, Leopold-Franzens-Universit\"at Innsbruck, A-6020 Innsbruck, Austria}
\affiliation{W. W. Hansen Experimental Physics Laboratory, Kavli Institute for Particle Astrophysics and Cosmology, Department of Physics and SLAC National Accelerator Laboratory, Stanford University, Stanford, CA 94305, USA}
\author{O.~Reimer}
\affiliation{Institut f\"ur Astro- und Teilchenphysik and Institut f\"ur Theoretische Physik, Leopold-Franzens-Universit\"at Innsbruck, A-6020 Innsbruck, Austria}
\affiliation{W. W. Hansen Experimental Physics Laboratory, Kavli Institute for Particle Astrophysics and Cosmology, Department of Physics and SLAC National Accelerator Laboratory, Stanford University, Stanford, CA 94305, USA}
\author{C.~Sgr\`o}
\affiliation{Istituto Nazionale di Fisica Nucleare, Sezione di Pisa, I-56127 Pisa, Italy}
\author{E.~J.~Siskind}
\affiliation{NYCB Real-Time Computing Inc., Lattingtown, NY 11560-1025, USA}
\author{G.~Spandre}
\affiliation{Istituto Nazionale di Fisica Nucleare, Sezione di Pisa, I-56127 Pisa, Italy}
\author{P.~Spinelli}
\affiliation{Dipartimento di Fisica ``M. Merlin" dell'Universit\`a e del Politecnico di Bari, I-70126 Bari, Italy}
\affiliation{Istituto Nazionale di Fisica Nucleare, Sezione di Bari, I-70126 Bari, Italy}
\author{D.~Tak}
\affiliation{Department of Astronomy, University of Maryland, College Park, MD 20742, USA}
\affiliation{NASA Goddard Space Flight Center, Greenbelt, MD 20771, USA}
\author{J.~B.~Thayer}
\affiliation{W. W. Hansen Experimental Physics Laboratory, Kavli Institute for Particle Astrophysics and Cosmology, Department of Physics and SLAC National Accelerator Laboratory, Stanford University, Stanford, CA 94305, USA}
\author{D.~F.~Torres}
\affiliation{Institute of Space Sciences (CSICIEEC), Campus UAB, Carrer de Magrans s/n, E-08193 Barcelona, Spain}
\affiliation{Instituci\'o Catalana de Recerca i Estudis Avan\c{c}ats (ICREA), E-08010 Barcelona, Spain}
\author{G.~Tosti}
\affiliation{Istituto Nazionale di Fisica Nucleare, Sezione di Perugia, I-06123 Perugia, Italy}
\affiliation{Dipartimento di Fisica, Universit\`a degli Studi di Perugia, I-06123 Perugia, Italy}
\author{J.~Valverde}
\affiliation{Laboratoire Leprince-Ringuet, \'Ecole polytechnique, CNRS/IN2P3, F-91128 Palaiseau, France}
\author{M.~Vogel}
\affiliation{California State University, Los Angeles, Department of Physics and Astronomy, Los Angeles, CA 90032, USA}
\author{K.~Wood}
\affiliation{Praxis Inc., Alexandria, VA 22303, resident at Naval Research Laboratory, Washington, DC 20375, USA}

\begin{abstract}
We use joint observations by the \emph{Neil Gehrels} \Swift X-ray Telescope (XRT) and the \Fermi Large Area Telescope (LAT) of gamma-ray burst (GRB) afterglows to investigate the nature of the long-lived high-energy emission observed by \Fermi LAT.  Joint broadband spectral modeling of XRT and LAT data reveal that LAT nondetections of bright X-ray afterglows are consistent with a cooling break in the inferred electron synchrotron spectrum below the LAT and/or XRT energy ranges.  Such a break is sufficient to suppress the high-energy emission so as to be below the LAT detection threshold.  By contrast, LAT-detected bursts are best fit by a synchrotron spectrum with a cooling break that lies either between or above the XRT and LAT energy ranges.  We speculate that the primary difference between GRBs with LAT afterglow detections and the non-detected population may be in the type of circumstellar environment in which these bursts occur, with late-time LAT detections preferentially selecting GRBs that occur in low wind-like circumburst density profiles.  Furthermore, we find no evidence of high-energy emission in the LAT-detected population significantly in excess of the flux expected from the electron synchrotron spectrum fit to the observed X-ray emission. The lack of excess emission at high energies could be due to a shocked external medium in which the energy density in the magnetic field is stronger than or comparable to that of the relativistic electrons behind the shock, precluding the production of a dominant synchrotron self-Compton (SSC) component in the LAT energy range.  Alternatively, the peak of the SSC emission could be beyond the 0.1--100 GeV energy range considered for this analysis.  
\end{abstract}


\section{Introduction}

Joint observations by NASA's \emph{Neil Gehrels} \Swift and \Fermi missions have led to a unique opportunity to study the broadband properties of gamma-ray bursts (GRBs) over an unprecedentedly broad energy range.  The two missions have the combined capability of probing the emission from GRBs over 11 decades in energy, ranging from optical ($\sim$2 eV) to high-energy gamma rays ($>300$ GeV).  After more than 7 yrs of simultaneous operations,  \Swift and \Fermi have detected thousands of GRBs, with over 100 of these bursts detected at energies greater than 30 MeV by the \Fermi Large Area Telescope (LAT) \citep{Vianello2015}\footnote{https://fermi.gsfc.nasa.gov/ssc/observations/types/grbs/lat\_grbs/}.

The properties of the high-energy emission observed by the LAT can differ considerably from the emission detected at keV and MeV energies by other instruments.  While some bursts show evidence for emission in coincidence with activity at keV and MeV energies as observed by the \Swift Burst Alert Telescope (BAT) and \Fermi Gamma-ray Burst Monitor (GBM) \citep{Ackermann2010}, others also exhibit high-energy emission that is temporally extended, lasting longer than the emission observed at lower energies \citep{GRB110731A, GRB130427A_LAT}.  There also appears in some cases to be a delay in the onset of the LAT-detected emission with respect to the emission observed at lower energies \citep{Abdo2009a, Abdo2009b, Ackermann2013}. The delayed onset and long-lived component of the LAT-detected emission suggest that GRB afterglows commonly observed in X-ray, optical, and radio wavelengths may also produce significant gamma-ray emission \citep{KumarBarniolDuran2009, Razzaque2010a, Ghisellini2010, DePasquale2010}.  In this interpretation, the coincident emission detected by the LAT is thought to be an extension of the prompt emission spectrum commonly attributed to shocks internal to the relativistic outflow \citep{Ackermann2010, Maxham2011, Zhang2011, Yassine2017}, while the late-time emission is due to the high-energy extension of the electron synchrotron spectrum produced by the external forward shock associated with the GRB blast wave moving into the circumstellar environment.


Broadband fits to the simultaneous multiwavelength observations of GRB 110731A \citep{GRB110731A} and GRB 130427A \citep{GRB130427A_LAT} show similar late-time spectral and temporal behavior, supporting such an external shock interpretation.  Likewise, a stacking analysis of the LAT data of {\it Swift}-localized bursts that were not detected above 40 MeV has shown evidence for subthreshold emission on timescales that far exceed the typical duration of the prompt emission at keV energies \citep{Beniamini2011, LATStackingAnalysis}.  Furthermore, the strength of this high-energy subthreshold emission correlates directly with the X-ray brightness of the burst's afterglow emission, as measured by the \Swift X-ray Telescope (XRT).  

Despite the growing evidence for an external shock origin of the long-lived high-energy emission observed by the LAT, the fact remains that only $\sim8\%$ of the bursts detected at keV energies within the LAT field of view (FoV) have been detected above 40 MeV \citep{Ackermann2013}.  Therefore, although the signature of the afterglow emission at X-ray wavelengths is largely ubiquitous in GRBs observed by the XRT, the high-energy component is observed in only a small subset of these bursts. This has led to speculation that LAT-detected bursts may represent a unique population of GRBs, either probing a particular type of environment \citep{Racusin2011, Beloborodov2014}, the result of a unique set of afterglow conditions \citep{Ghisellini2010}, or the result of progenitors that produce a rare class of hyperenergetic GRBs \citep{Cenko2011}.

In this paper we attempt to address the conditions that are required to produce the late time high-energy emission detected by the LAT through the use of broadband data collected by both \Swift and \emph{Fermi}. By examining joint XRT and LAT observations of 386 GRBs from 2008 August 4 to 2014 March 23, we can model the broadband spectra of the afterglow emission associated with LAT-detected and non-detected GRBs.  This allows us to determine whether the relative sensitivities of the XRT and LAT are sufficient to account for the majority of LAT nondetections, or whether the LAT-detected bursts differ significantly in their afterglow properties from the general GRB population.  A subset of these bursts is also subjected to detailed broadband spectral fitting of the simultaneous XRT and LAT data.  From these spectral fits, we can determine whether the XRT and LAT data are consistent with being drawn from the same power-law segment (PLS) of an electron synchrotron spectrum, or if a break or suppression of the high-energy emission is required to explain the LAT non-detection. This analysis also allows us to place constraints on the existence of spectral components at high energies that are in excess of that predicted by the electron synchrotron model, such as external inverse Compton (EIC) \citep{Fan2006, He2012, Beloborodov2014A} and synchrotron self-Compton (SSC) \citep{Dermer2000, Zhang2001, Sari2001, Wang2013} contributions. 


The paper is structured as follows: in \S\ref{sec:InstrumentOverview}, we review the characteristics of the \Fermi LAT and \Swift XRT instruments.  In \S\ref{sec:SampleDefinition}, we define the GRB samples considered in this work and outline the analysis performed in \S\ref{sec:Analysis}.  We present the results in \S\ref{sec:Results}, and discuss the implications of our results in \S\ref{sec:Discussion}. Unless specified otherwise, all temporal and spectral indices are defined as $F_\nu\propto E^{-\beta}t^{-\alpha}$, where $\beta=\Gamma-1$, with $\Gamma$ is the photon index.

\section{Instrument Overview } \label{sec:InstrumentOverview}

\subsection{\Swift BAT and \Swift XRT} \label{sec:SwiftInstrumentOverview}

The \emph{Neil Gehrels} \Swift observatory consists of the BAT \citep{Barthelmy05}, the XRT \citep{Burrows05}, and the UltraViolet Optical Telescope (UVOT) \citep{Roming05}. The BAT is a wide-field, coded-mask gamma-ray telescope, covering a FoV of 1.4 sr and an imaging energy range of 15--150 keV. The instrument's coded mask allows for positional accuracy of 1--4 arcminutes within seconds of the burst trigger. The XRT is a grazing-incidence focusing X-ray telescope covering an energy range from 0.3 to 10 keV and providing a typical localization accuracy of $\sim$ 1--3 arcseconds. 

\Swift operates autonomously in response to BAT triggers on new GRBs, automatically slewing to point the XRT at a new source with 1--2 minutes.  Data are promptly downloaded, and localizations are made available from the narrow-field instruments within minutes (if detected).  \Swift then continues to follow up GRBs as they are viewable outside of observing constraints and the observatory is not in the South Atlantic Anomaly (SAA), for at least several hours after each burst, sometimes continuing for days, weeks, or even months if the burst is bright and of particular interest for follow-up.

\subsection{\Fermi LAT} \label{sec:FermiInstrumentOverview}

The \emph{Fermi Gamma-ray Space Telescope} consists of two scientific instruments, the GBM and the LAT.  The LAT is a pair-conversion telescope comprising a $4\times4$ array of silicon strip trackers and cesium iodide (CsI) calorimeters covered by a segmented anti-coincidence detector to reject charged-particle background events. The LAT detects gamma rays in the energy range from 20\,MeV to more than 300\,GeV with a FOV of $\sim 2.4$ steradians, observing the entire sky every two orbits ($\sim$3 hours) while in normal survey mode.  The deadtime per event of the LAT is nominally 26\,$\mu$s, the shortness of which is crucial for observations of high-intensity transient events such as GRBs.  The LAT triggers on many more background events than celestial gamma rays; therefore onboard background rejection is supplemented on the ground using event class selections that are designed to facilitate study of the broad range of sources of interest \citep{Atwood:09}.  

In normal \Fermi operations, the GBM triggers on new GRBs approximately every 1--2 days.  The LAT survey mode rocking profile is occasionally interrupted (approximately once per month) by GBM initiating an autonomous repoint request (ARR) due to high-peak flux or fluence, which has proven to be an effective proxy for bright LAT bursts.  The ARR causes \Fermi to reorient itself such that the GBM localization is placed at the center of the LAT FoV, where it remains for the next 2.5 hours, except when the GRB position is occulted the Earth.  Roughly $\sim12$ GRBs per year simultaneously trigger both the GBM and BAT, but due to extended high-energy $\gamma$-ray emission observed by the LAT in some bursts, a GRB does not necessarily need to be in the LAT FOV at the trigger time to be detected.  In normal survey mode, the LAT observes the position of every GBM- and BAT-detected burst within 3 hours.  

\section{Sample Definition} \label{sec:SampleDefinition}

We compiled a sample of all GRBs observed by the XRT between the beginning of \Fermi science operations on 2008 August 4 and 2014 March 23.  The majority of bursts in the sample were observed by LAT during its normal survey observations at some time after the BAT trigger and the start of XRT observations.  A small number of bursts were not observed by the LAT owing to pointed observations at the time of the GRB trigger.  For each burst observed by the LAT, we selected good time intervals (GTIs) during which the well-localized afterglow position was within 65$^\circ$ of the LAT z-axis (boresight) beginning after the start of the first XRT observation and ending up to 20 ks post trigger. The sensitivity of the LAT falls as a function of off-axis angle away from the instrument boresight; therefore, intervals during which the burst positions were $>65^\circ$ from the boresight were not considered for this analysis.  Neither XRT nor LAT takes data during SAA passages; therefore we also excluded intervals that occurred during these times.  GRB positions that were at angles larger than 105$^\circ$ with respect to the zenith direction for {\it Fermi}, placing the burst near the Earth's limb, were also excluded.  Observations at such large zenith angles result in emission at the burst location that is dominated by $\gamma$-rays from the Earth's limb produced by interactions of cosmic rays with the Earth's atmosphere.  The resulting sample includes a total of 1156 usable GTIs, for 386 GRBs.  
  

\section{Analysis} \label{sec:Analysis}

\subsection{XRT} \label{sec:AnalysisXRT}

For each burst, we obtained the XRT count-rate light curves from the public XRT team repository hosted at the University of Leicester \citep{Evans07, Evans09} and applied the de-absorbed counts-to-energy-flux conversion factor as determined by the automated late-time spectral fits to the XRT data.  Since the XRT coverage and the LAT GTIs may not always overlap, we fit the XRT light curves with a semi-automated light-curve fitting routine \citep{Racusin09, Racusin2011, Racusin2016} with power laws or broken power laws and gaussian flares (when flaring episodes are present), in order to estimate the X-ray flux during XRT data gaps associated with periods of Earth occultation. We then use the afterglow's time-integrated photon index and associated error to convert the XRT energy flux light curve in the 0.3--10 keV energy range to an extrapolated energy flux light curve in the 0.1--100 GeV energy range.  Note that by selecting only bursts for which there were LAT observations after the start of XRT observations, we avoid the highly uncertain activity of both extrapolating backward in time and to higher energies.  Given the observations of both spectral and temporal variability in early afterglow light curves, including energetic X-ray flares and plateaus followed by sharp drops in flux, this decision avoids making any assumptions about the X-ray behavior prior to the onset of the XRT observations even though it excludes several well-observed LAT bursts for which subsequent XRT observations were made via \Swift target-of-opportunity requests (e$.$g$.$, GRB~080916C and GRB~090926A).


\subsection{LAT} \label{sec:AnalysisLAT}

For each interval in which the GRB was in the LAT FOV, we calculate the 95$\%$ confidence level upper limits, or the observed energy flux with 68$\%$ errors, in the 0.1--100 GeV energy range for LAT nondetections and detections, respectively.  We then compare these values to the expected energy flux  in the 0.1 to 100 GeV energy range from the fit to the XRT data.  The LAT flux estimates are obtained by performing an unbinned likelihood analysis using the standard analysis tools 
(ScienceTools version v10r01p0)\footnote{http://fermi.gsfc.nasa.gov/ssc/}.  For this analysis, we used the `P8R2\_SOURCE\_V6' instrument response functions and selected `Source' class events from a $12^\circ$ radius energy-independent region of interest (ROI) centered on the burst location. 
The size of the ROI is chosen to reflect the 95\% containment radius of the LAT energy-dependent point spread function (PSF) at 100 MeV. The `Source' event class was specifically optimized for the study of point-like sources, with stricter cuts against nonphoton background contamination relative to the `Transient' event class that is typically used to study GRBs on very short timescales \citep{LATPerformancePaper}.

In standard unbinned likelihood fitting of individual sources, the observed distribution of counts for each burst is modeled as a point source using an energy-dependent LAT PSF and a power-law source spectrum with a normalization and photon index that are left as free parameters.  For the purposes of comparing the XRT extrapolation to the LAT data, we fixed the model's photon index to match the value measured by the XRT.  In addition to the point source, Galactic and isotropic background components are also included in the model, as well as all gamma-ray sources in the 3FGL catalog within a source region with a radius of 30$^\circ$ centered on each ROI \citep{Acero2015}.  The Galactic component, \emph{gll\_iem\_v06}, is a spatial and spectral template that accounts for interstellar diffuse gamma-ray emission from the Milky Way.  The normalization of the Galactic component is kept fixed during the fit.  The isotropic component, \emph{iso\_source\_v06}, provides a spectral template to account for all remaining isotropic emission, including contributions from both residual charged-particle backgrounds and the isotropic celestial gamma-ray emission. The normalization of the isotropic component is allowed to vary during the fit.  Both the Galactic and isotropic templates are publicly available\footnote{http://fermi.gsfc.nasa.gov/ssc/data/access/lat/BackgroundModels.html}.

We employ a likelihood ratio test \citep{Neyman1928} to quantify whether there exists a significant excess of counts above the expected background.  We form a test statistic (TS) that is twice the ratio of the likelihood evaluated at the best-fit parameters under a background-only, null hypothesis, i$.$e$.$, a model that does not include a point-source component, to the likelihood evaluated at the best-fit model parameters when including a candidate point source at the center of the ROI \citep{Mattox1996}.  According to Wilks' theorem \citep{Wilks1938}, this ratio is distributed approximately as $\chi^{2}$, so we choose to reject the null hypothesis when the test statistic is greater than TS = 16, roughly equivalent to a 4$\sigma$ rejection criterion for a single degree of freedom.  Using this test statistic as our detection criterion, we estimate the observed LAT flux for bursts with TS $> 16$ and use a profile likelihood method described in more detail in \citet{Ackermann2012} to calculate upper limits for GRBs with TS $<$ 15.

\subsection{Joint XRT/LAT Spectral Fits} \label{sec:AnalysisJointSpectralFits}

For bursts with time intervals during which the high-energy flux extrapolation of the XRT data is equivalent to, or exceeds, the measured LAT flux or upper limit for that period, we also performed joint spectral fits to the XRT and LAT data to investigate the underlying shape of the spectral energy distribution (SED). To simplify the analysis, we only considered intervals with contemporaneous XRT and LAT data.  We refer to this subsample of GTIs as our ``spectroscopic" sample.  

For these fits, the \Swift XRT data, including relevant calibration and response files, were retrieved from the HEASARC archive\footnote{http://heasarc.gsfc.nasa.gov/docs/swift/archive/} and processed with the standard \Swift analysis software (v3.8) included in NASA's HEASOFT software (v6.11).  We use {\it gtbin} to generate the count spectrum of the observed LAT signal and {\it gtbkg} to extract the associated background by computing the predicted counts from all the components of the best-fit likelihood model except the point source associated with the GRB. The LAT instrument response for each interval was computed using {\it gtrspgen}.  

The spectral fits were performed using the XSPEC version 12.7.0 \citep{1996XspecProc}.  Because the number of counts in the LAT energy bins is often in the Poisson regime, we use the PG statistic from XSPEC, since the standard $\chi^2$ statistic is not a reliable estimator of significance for low counts.  For bursts with no detectable emission, the count spectra associated with the modeled signal cannot exceed the background spectra. XSPEC takes this into account by constraining the best-fit model from overpredicting the signal counts in the LAT energy range.  The resulting flux upper limits from these background-only intervals help constrain the hardness of the spectral model. 

For each time interval, we fit two functional forms to the XRT and LAT data; a single power-law (PL) and a broken power-law (BPL) model. Each form is multiplied by models for both fixed Galactic (phabs) and free intrinsic host (zphabs for bursts with known redshift, phabs otherwise) photoelectric absorption and a free cross-calibration constant.  Assuming that any break in the spectrum between the XRT and LAT regimes at late times would be associated with the synchrotron cooling frequency, i.e.\ the frequency at which an electron's cooling time equals the dynamical time of the system, we require the two power-law indices in the BPL model to differ by $\Delta \Gamma = 0.5$ in accordance with the theoretical expectation for electron synchrotron radiation from a forward shock \citep{GranotSari2002}.

We perform a nested model comparison in order to determine if the additional degrees of freedom in the BPL model are warranted over a simpler PL model.  Assuming there are $n_{\rm alt}$ additional free parameters under the alternative model, then the alternative model is statistically preferred at a confidence level according to the difference in the PG-statistic, hereafter referred to as $\Delta$Stat, between the two fits, which is expected to follow a $\chi^2$ distribution for $n_{\rm alt}$ degrees of freedom in the large sample limit.  Requiring that the two power-law indices in the BPL model differ by $\Delta \Gamma = 0.5$ results in a single extra degree of freedom (i$.$e$.$, the break energy) compared to the PL null hypothesis.  Therefore, according to the $\chi^2$ cumulative distribution function, a value of $\Delta$-Stat $> 9$ would represent a $>3\sigma$ improvement in the fit.  We adopt this criterion as the threshold for a statistical preference for a break in the high-energy spectrum.


\section{Results}  \label{sec:Results}

\subsection{XRT Flux Extrapolations}  \label{sec:Results_FluxComparisons}

\begin{figure}[t]
\centering
\includegraphics[width=0.49\textwidth]{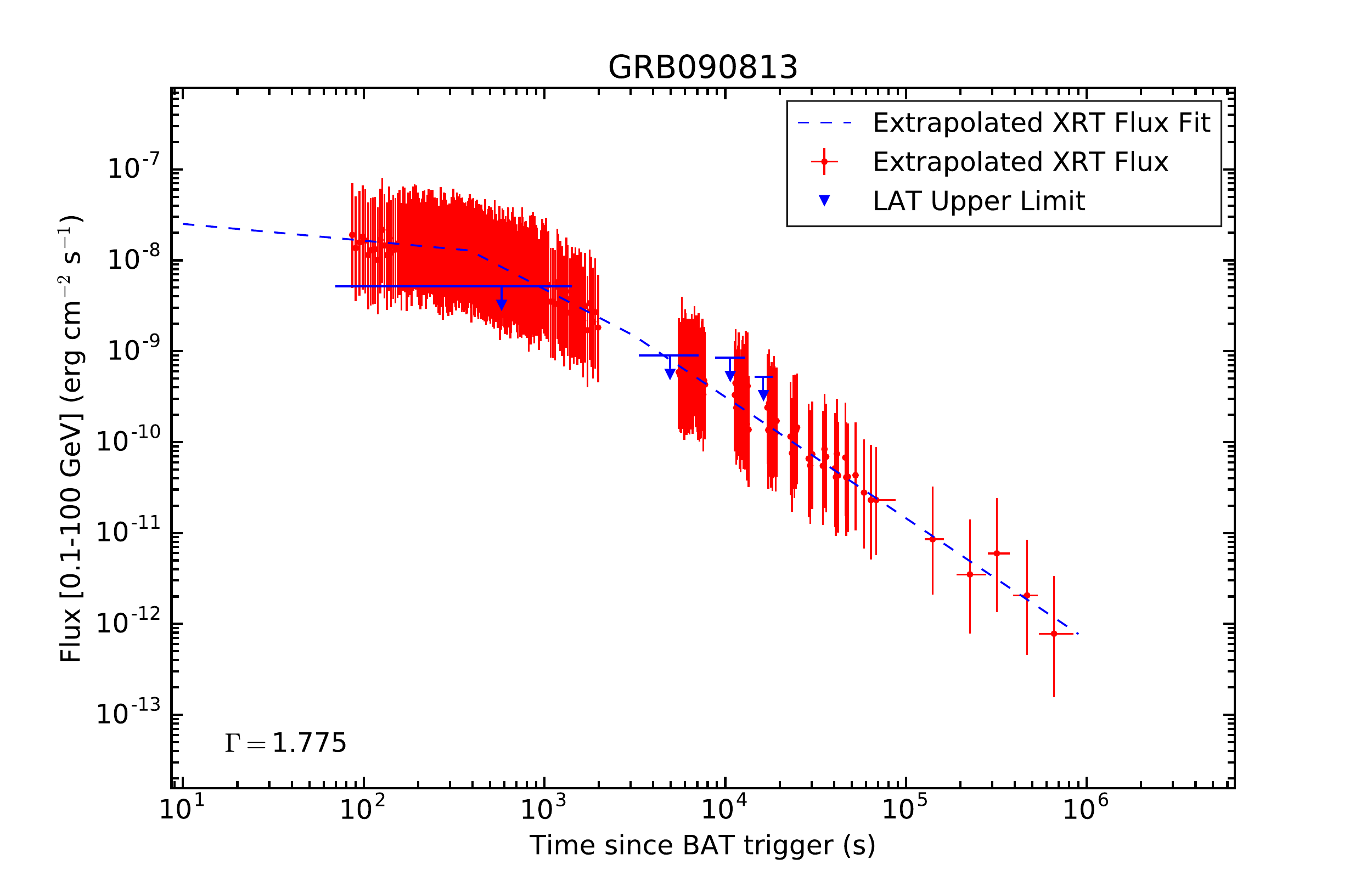}
\includegraphics[width=0.49\textwidth]{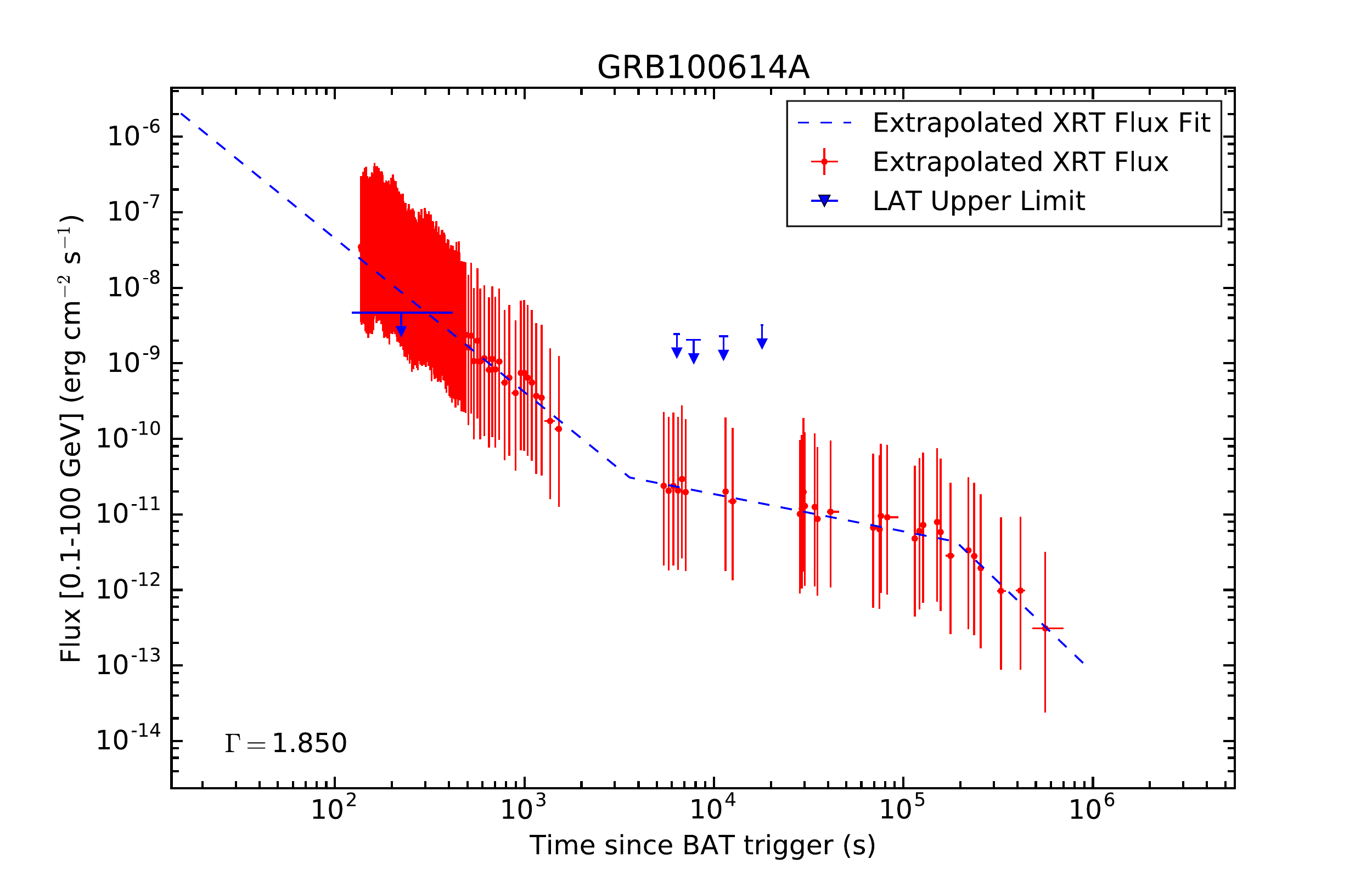}
\caption{Examples of the comparison between the  XRT-extrapolated flux and the LAT observations in the 0.1--100 GeV energy range for GRB~090813 and GRB~100614A.  The $\Gamma$ listed in the lower left corner indicates the time-averaged X-ray photon index used in the extrapolation.  The blue dashed line represents the best-fit power-law segments to the X-ray afterglow flux. Neither burst was detected by the LAT despite both exhibiting bright X-ray afterglows, relatively hard photon indices, and being well observed by the LAT soon after the onset of the afterglow decay.}
\label{Fig:ExampleExtrapolations}
\end{figure}

Examples of comparisons between the XRT fluxes extrapolated into the 0.1--100 GeV energy range and the LAT observations for GRB~090813 and GRB~100614A are shown in Figure \ref{Fig:ExampleExtrapolations}.
The error bars on this XRT-extrapolated LAT-band flux (hereafter referred to as the XRT-extrapolated flux) take into account the propagation of uncertainty of both the X-ray flux and photon index into the LAT energy range.  Both bursts shown in Figure \ref{Fig:ExampleExtrapolations} exhibit bright X-ray afterglows, relatively hard photon indices, and were well observed by the LAT soon after the onset of the afterglow decay.  Neither burst was detected by the LAT, and the estimated upper limits for the energy flux in the 0.1--100 GeV energy range are above or are consistent with the expected flux given the extrapolation of the XRT spectrum.

\begin{figure}
\centering
\includegraphics[width=0.6\textwidth]{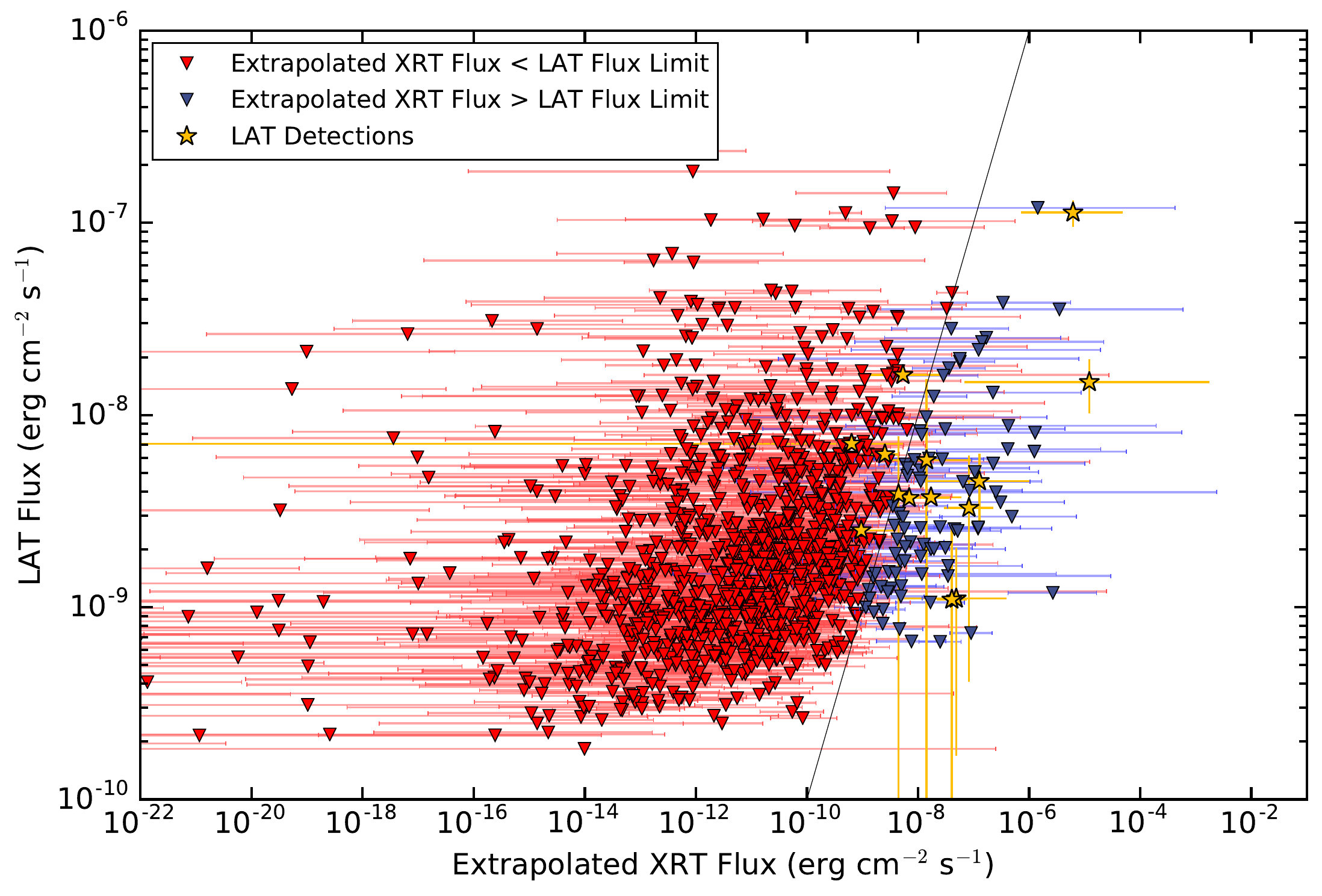}
\caption{The measured LAT flux (yellow stars), or upper limit (downward triangles), versus the XRT-extrapolated flux for a given interval when the burst location was within the LAT FoV.  The black line demarcates the equivalency.  The blue and red colors of the downward triangles represent intervals when the extrapolated flux fell above and below the LAT flux measurements, respectively.  The gold stars represent the LAT detections in our sample.}
\label{Fig:FluxComparison}
\end{figure}

The results of performing the same analysis on all 1156 GTIs in our sample are shown in Figure \ref{Fig:FluxComparison}.  The plot shows the measured LAT flux, or upper limit, versus the XRT-extrapolated flux for a given interval when the burst location was within the LAT FoV. The gold stars represent the LAT detections in our sample, which consist of 14 GTIs for 11 GRBs.  We note that all but one of these detections were announced via the Gamma-ray Coordinates Network (GCN)\footnote{https://gcn.gsfc.nasa.gov}, the two exceptions being GRB~081203A and GRB~120729A, both of which were found through this analysis.  Both these bursts are discussed in greater detail in the 2nd \Fermi LAT GRB catalog (The LAT Collaboration 2018, in prep)

For 91\% of the intervals examined (1055 GTIs), the XRT-extrapolated flux in the LAT energy range fell below the LAT upper limits (i.e.\ to the left of the equivalency line), and therefore were consistent with the LAT nondetections.  The extrapolated fluxes for an additional $\sim$$7\%$ (84 GTIs) were above the LAT upper limits (i.e.\ to the right of the equivalency line). Interestingly, the flux measurements for all of the LAT detections in our sample were either consistent with the XRT extrapolation (4 GTIs) or fell below it (10 GTIs).  None of the LAT detections showed evidence of emission significantly in excess of the flux expected from the extrapolation of the XRT observations. 

\begin{figure}
\centering
\includegraphics[width=0.6\textwidth]{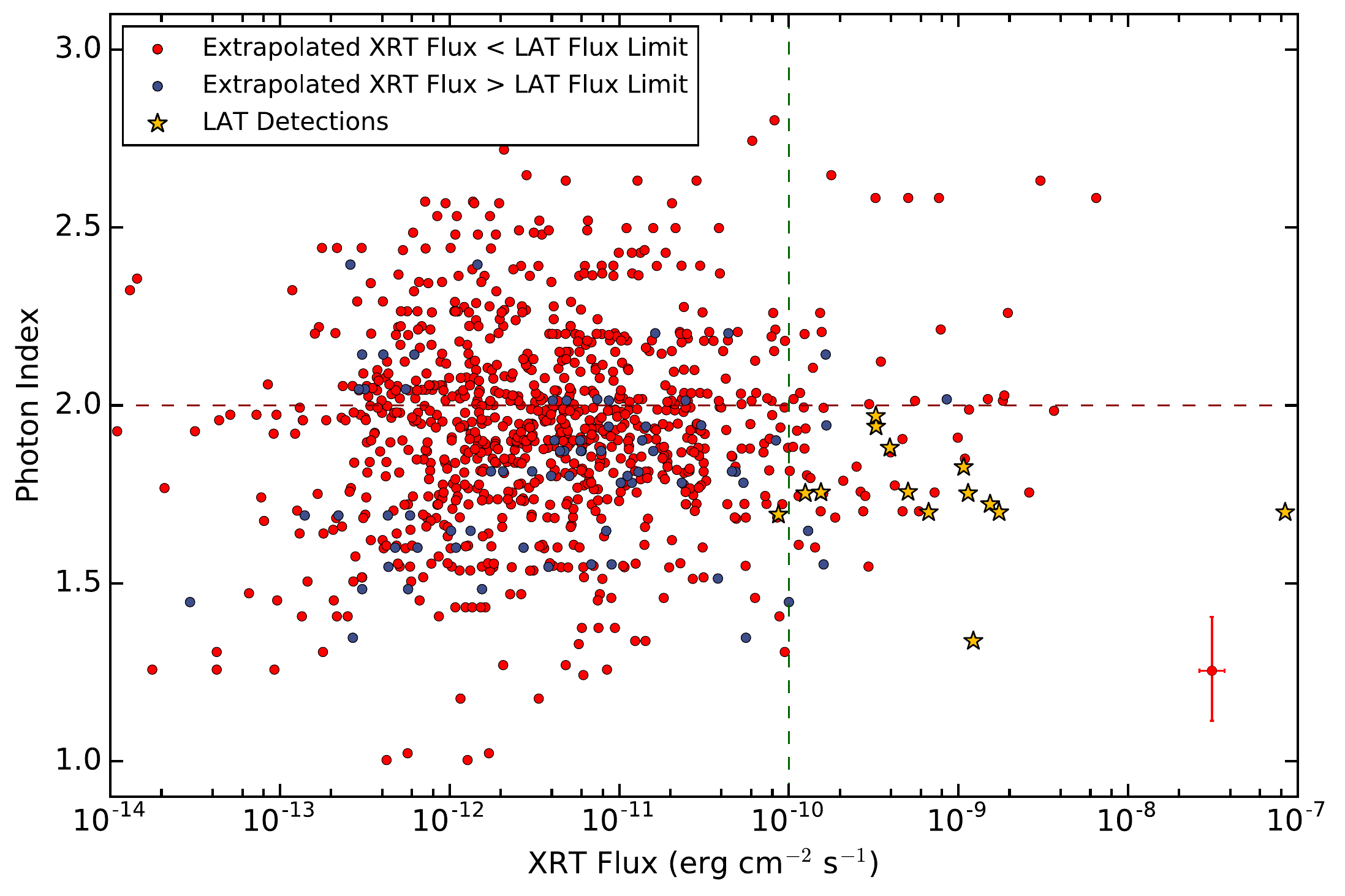}
\caption{The time-average photon index $\Gamma$ vs$.$ the X-ray energy flux as measured by the XRT in the 0.3 to 10 keV energy range. The blue and red symbols represent intervals when the extrapolated flux fell above and below the LAT flux measurements, respectively, and the gold stars represent the LAT detections in our sample. The typical error bar is shown in the bottom right corner, and the vertical and horizontal dashed lines separate the plot into soft/hard and dim/bright quadrants.}
\label{Fig:PhotonIndexVsXRTFlux}
\end{figure}

We examined the X-ray properties of the afterglows during these intervals in Figure \ref{Fig:PhotonIndexVsXRTFlux}, where we plot the X-ray energy flux as measured by the XRT in the 0.3--10 keV energy range versus the associated photon index $\Gamma_{\rm XRT}$.  
The intervals with afterglow emission that would be expected to produce high-energy emission in excess of the LAT sensitivity tend to be spectrally hard, with $\Gamma_{\rm XRT} \lesssim 2$.  They are also drawn from a very wide range of fluxes.  The LAT detections, on the other hand, are drawn exclusively from afterglows that exhibited bright and hard emission, with criteria roughly fulfilling $\Gamma_{\rm XRT} \lesssim 2$ and $F_{\rm XRT} \gtrsim 10^{-10}$ erg cm$^{-2}$ s$^{-1}$ shown as dashed green lines.  The red points that occupy this quadrant of the plot did not have sufficiently deep upper limits for the expected high-energy flux to exceed the LAT sensitivity, so their nondetections are consistent with the LAT observations. The blue points, on the other hand, have deeper LAT upper limits, making their expected high-energy emission inconsistent with the LAT observations.

\begin{figure}
\centering
\includegraphics[width=0.6\textwidth]{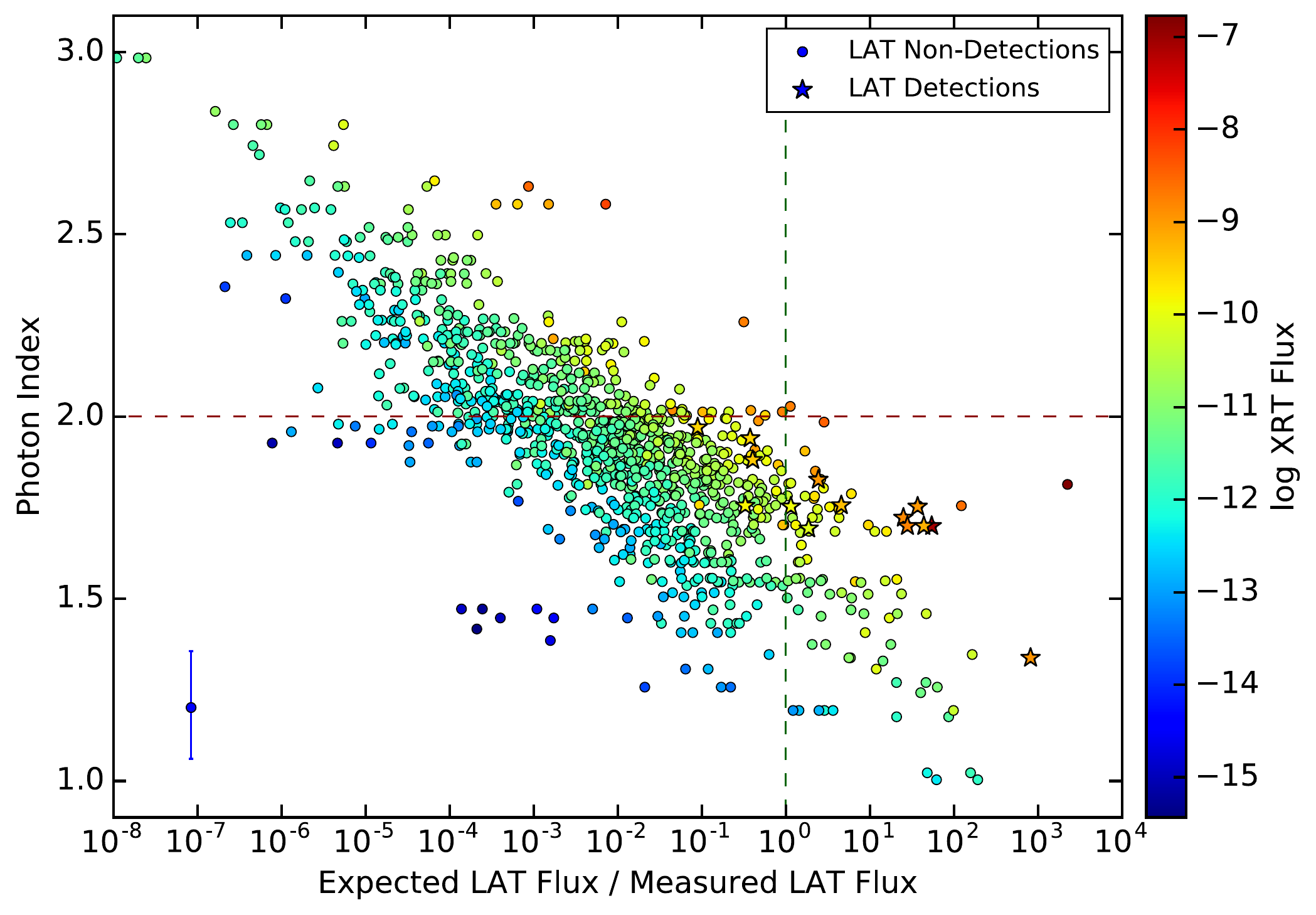}
\caption{The time-averaged afterglow photon index, as measured by XRT, versus the ratio of the XRT-extrapolated flux in the LAT energy range to the LAT upper limit (or measured flux in the case of a detection).  The colors of the symbols shows the XRT energy flux measured during the geometric mean of the afterglow interval, where the geometric mean is defined as the square root of the product of the interval start and end times The green line represents the line of equivalency between the measured LAT flux (or upper limit) and the XRT-extrapolated flux. The typical error bar is shown in the bottom left corner, and the red dashed lines delineates the soft/hard populations and the green dashed line marks the line of equality between the expected and measured LAT flux.}
\label{Fig:PhotonIndexVsXRT2LATRatio}
\end{figure}

We examine the properties of these afterglow intervals after folding in the LAT sensitivity in Figure \ref{Fig:PhotonIndexVsXRT2LATRatio}, where we display the time-averaged photon indices for the afterglows, as measured by XRT, versus the ratio of the XRT-extrapolated fluxes in the LAT energy range to the LAT upper limits (or measured fluxes for detections).  The colors of the symbols now represent the XRT energy fluxes measured during the geometric mean of the afterglow interval.  The geometric mean is defined as the square root of the product of the interval start and end times. The green dashed line represents the line of equivalency between the measured LAT flux (or upper limit) and the XRT-extrapolated flux.  Bursts that fall to the right have X-ray extrapolations that are consistent with the LAT sensitivity, whereas bursts that fall to the left have X-ray extrapolations that exceed the LAT flux measurements.  By construction, all of the blue data points in Figures \ref{Fig:FluxComparison} and \ref{Fig:PhotonIndexVsXRTFlux} lie to the right of the green dashed line.  Again, a general trend is evident wherein the bursts with the hardest afterglow spectra and highest observed XRT fluxes during the intervals in question are the bursts that result in X-ray extrapolations that either exceed the LAT upper limits or result in LAT detections.

\begin{figure}
\centering
\includegraphics[width=0.6\textwidth]{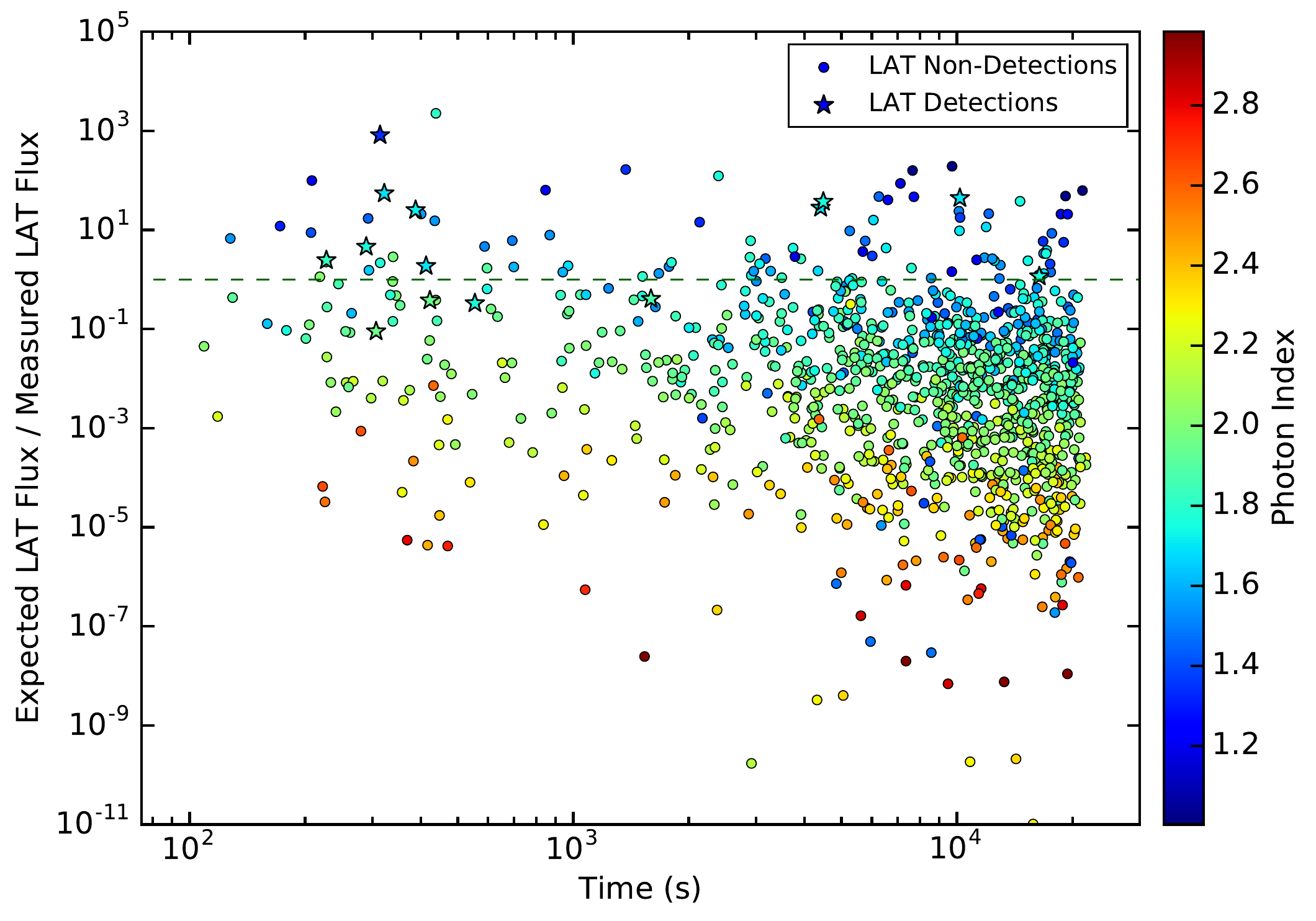}
\caption{The ratio of the XRT-extrapolated flux to the measured LAT flux (or upper limit) vs$.$ the geometric mean of the interval in which the burst position was within the LAT FoV. The colors of the symbols represents the time-average photon index as measured by spectral fits to the late-time XRT data and the stars represent the LAT detections. The vertical green dashed line represents the line of equality between the measured LAT flux (or upper limit) and the XRT-extrapolated flux.}
\label{Fig:XRT2LATRatioVsTime}
\end{figure}

Figure \ref{Fig:XRT2LATRatioVsTime} displays the same results, but now showing the ratio of the XRT-extrapolated flux to the measured LAT flux (or upper limit) versus the geometric mean of the temporal interval in which the burst position was within the LAT FoV. The colors of the symbols represents the time-averaged photon index as measured by spectral fits to the late-time XRT data.  The stars again represent the LAT detections.  Again, we see a general trend of bursts with harder afterglow spectra tending to predict high-energy emission in excess of the LAT sensitivity.  Although X-ray brightness correlates strongly with the time of observation, Figure \ref{Fig:XRT2LATRatioVsTime} demonstrates that many afterglows remain spectrally hard to late times, resulting in afterglow emission that exceeds the LAT sensitivity thousands of seconds after trigger.  Likewise, the LAT detections appear in both early and late-time observations. 

In order to understand what differentiates the afterglow intervals that have expected high-energy emission that is inconsistent with the LAT observations from those with LAT detections, we selected all intervals to the right of the line of equivalency in Figure \ref{Fig:FluxComparison} (i.e.\ the blue data points), as well as all of the LAT-detected bursts (yellow data points), for which simultaneous XRT and LAT data exist. A total of 64 GTIs for 52 bursts fulfill these criteria and form the spectroscopic sample for which we performed additional joint spectral fits, described in the next section.

\subsection{Joint XRT/LAT Spectroscopic Fits}  \label{sec:Results_JointSpectroscopicFits}

\begin{figure}[t]
\centering
\includegraphics[width=0.49\textwidth]{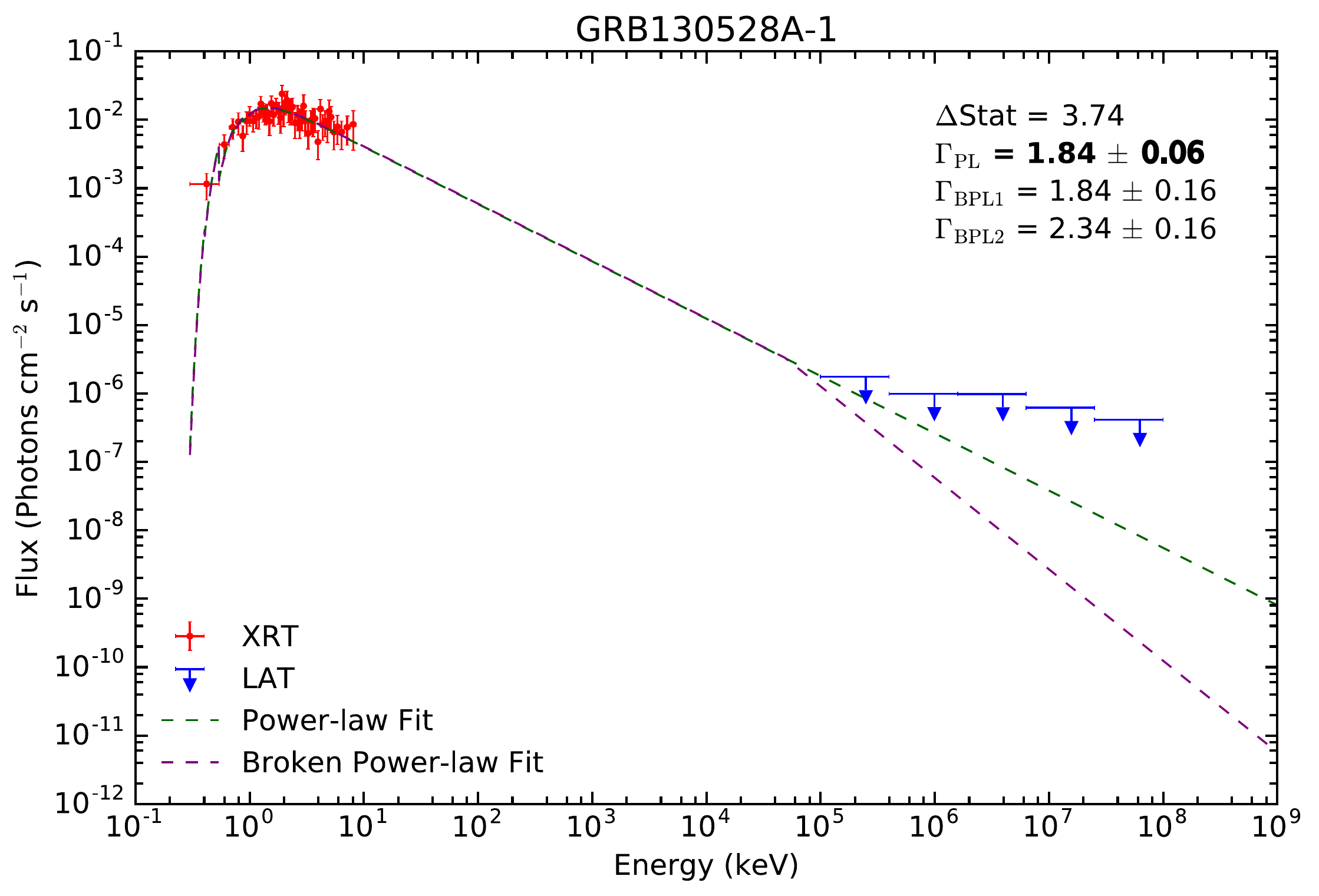}
\includegraphics[width=0.49\textwidth]{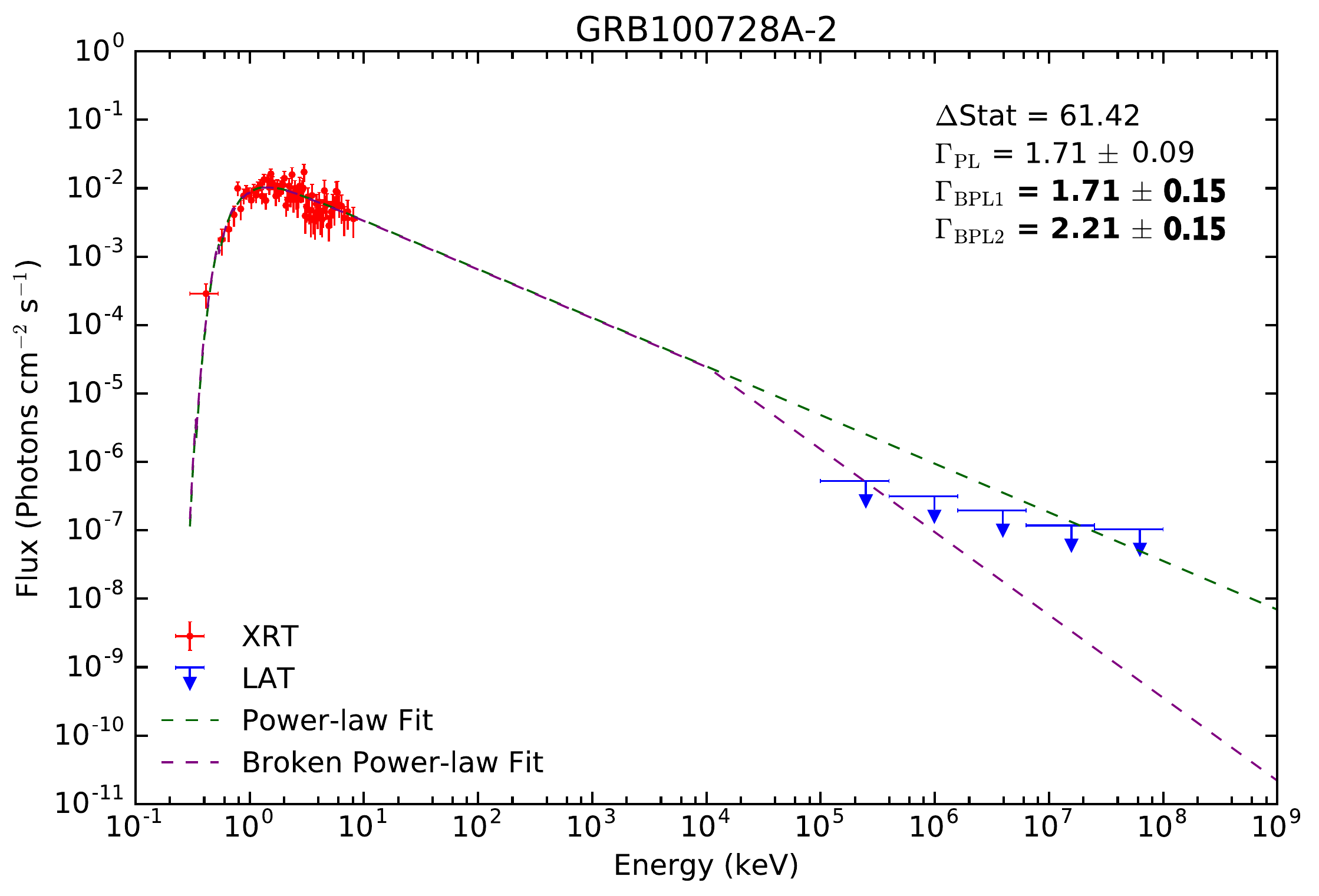}
\caption{Joint spectroscopic fits performed using the contemporaneous XRT and LAT data for GRB~130528A and the second interval of GRB~100728A. The measured XRT spectrum in the 0.3 to 10 keV energy range is shown in red, while the LAT upper limits (95\% confidence level) are shown as blue downward arrows. The green and purple dashed lines represent fits to the data using the single and broken power-law models. The photon indices from the preferred statistically prefered fit is shown in bold.}
\label{Fig:ExampleXSpecFits}
\end{figure}

Two examples of the joint spectroscopic fits performed using the contemporaneous XRT and LAT data for GRB~130528A and GRB~100728A are shown in Figure \ref{Fig:ExampleXSpecFits}.  The measured XRT spectrum in the 0.3 to 10 keV energy range is shown in red, while the LAT upper limits (95$\%$ confidence level) are shown as blue downward arrows.  The green and purple dashed lines represent fits to the data using the single and broken power-law models described in \S\ref{sec:AnalysisJointSpectralFits}.  Neither GRB~130528A nor GRB~100728A were detected by the LAT during the selected intervals (GRB~100728A was detected at an earlier time), so upper limits are shown for emission in the 0.1 to 100 GeV energy range.  Combined with the XRT data, these limits constrain the broadband spectral shape of the afterglow emission from these two bursts.  In the case of GRB~130528A, a single power law covering eight orders of magnitude in energy is consistent with both the XRT and LAT data, whereas a broken power-law is statistically preferred in GRB~100728A, with an $\sim8\sigma$ ($\Delta$Stat $= 64.21$) improvement in the fit over a single power law. 

Of the 64 GTIs in our spectroscopic sample, a total of 52 intervals yielded no LAT-detected emission.  Of these 52 GTIs, 31 (60$\%$) have simultaneous XRT and LAT data that are consistent with being drawn from a spectral distribution that can be represented as a single power law.  An additional 21 GTIs (40$\%$) show a statistical preference, at greater than $3\sigma$ significance, for a spectral break between the XRT and LAT data.  In all but one case, the LAT data can be accommodated by either a power-law or a broken power-law, with a photon index change of $\Delta \Gamma = 0.5$, connecting the contemporaneous XRT and LAT observations.  

A median photon index of $\Gamma_{\rm PL} = 1.98 \pm 0.16$ was measured for the 31 GTIs for which a single power law was adequate to describe both the XRT and LAT data, where we have adopt the standard deviation of the sample as the error on the median.  This is in contrast to the median photon index of $\Gamma_{\rm XRT} = 1.68 \pm 0.21$ for this sample when measured from the XRT data alone.  Therefore, adding the LAT data to the spectral fit softens the estimated spectral shape for these bursts.  For the bursts which show a preference for a break in their broadband afterglow spectra, we find median XRT and LAT photon indices of $\Gamma_{\rm BPL1} = 1.60 \pm 0.13$ and $\Gamma_{\rm BPL2} = 2.10$, where the post break photon index is fixed to $\Gamma_{\rm BPL2}$ = $\Gamma_{\rm BPL1}$ + 0.5.  This is compared to the median photon index of $\Gamma_{\rm XRT} = 1.72 \pm 0.21$ for this sample when estimated from the XRT data alone.  The median spectral fit results are summarized in Table 1.




\subsection{LAT Detections}  \label{sec:Results_JointSpectroscopicFits_LATDetections}

\begin{figure}
\centering
\includegraphics[width=0.48\textwidth]{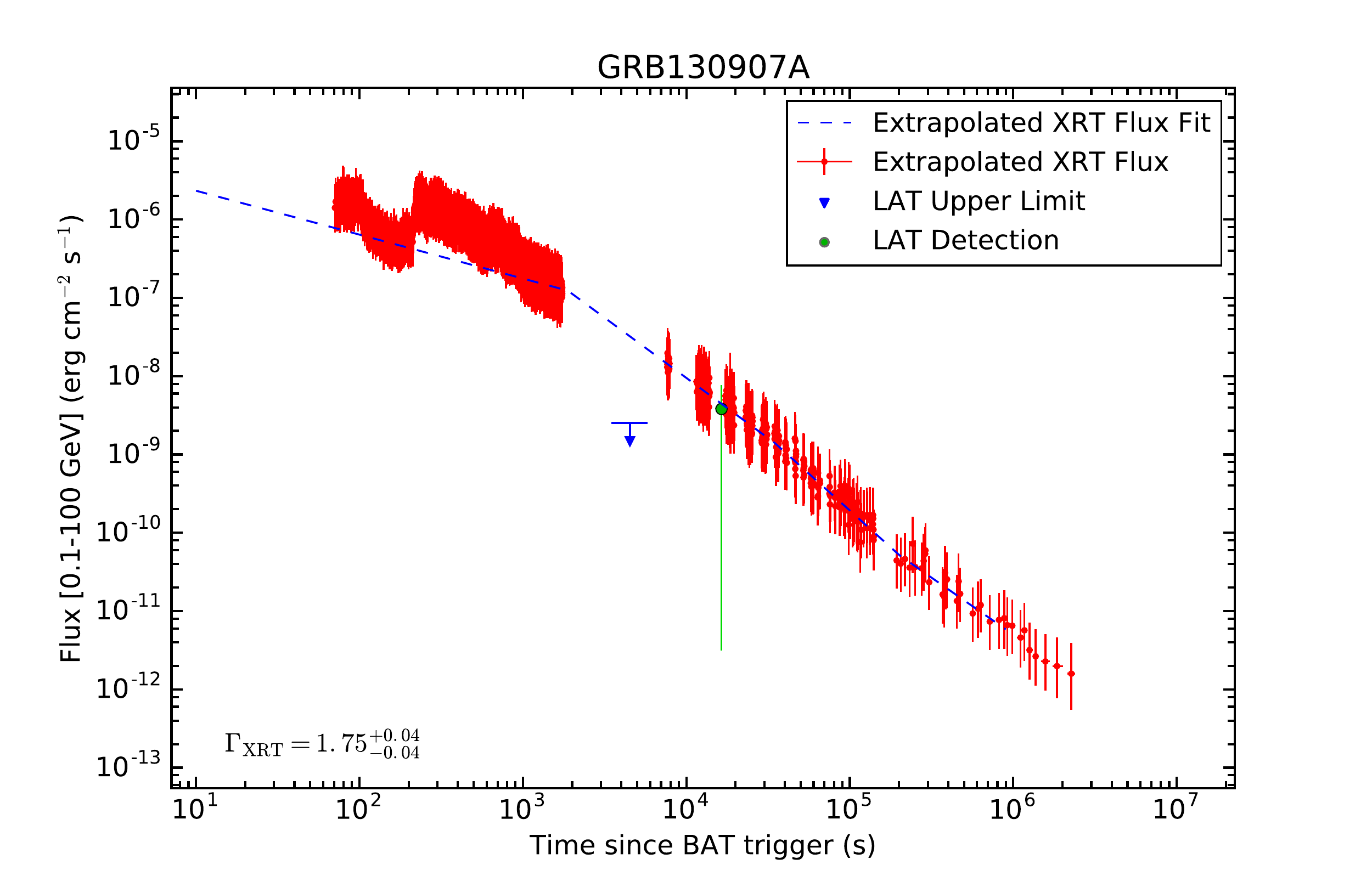}
\includegraphics[width=0.43\textwidth]{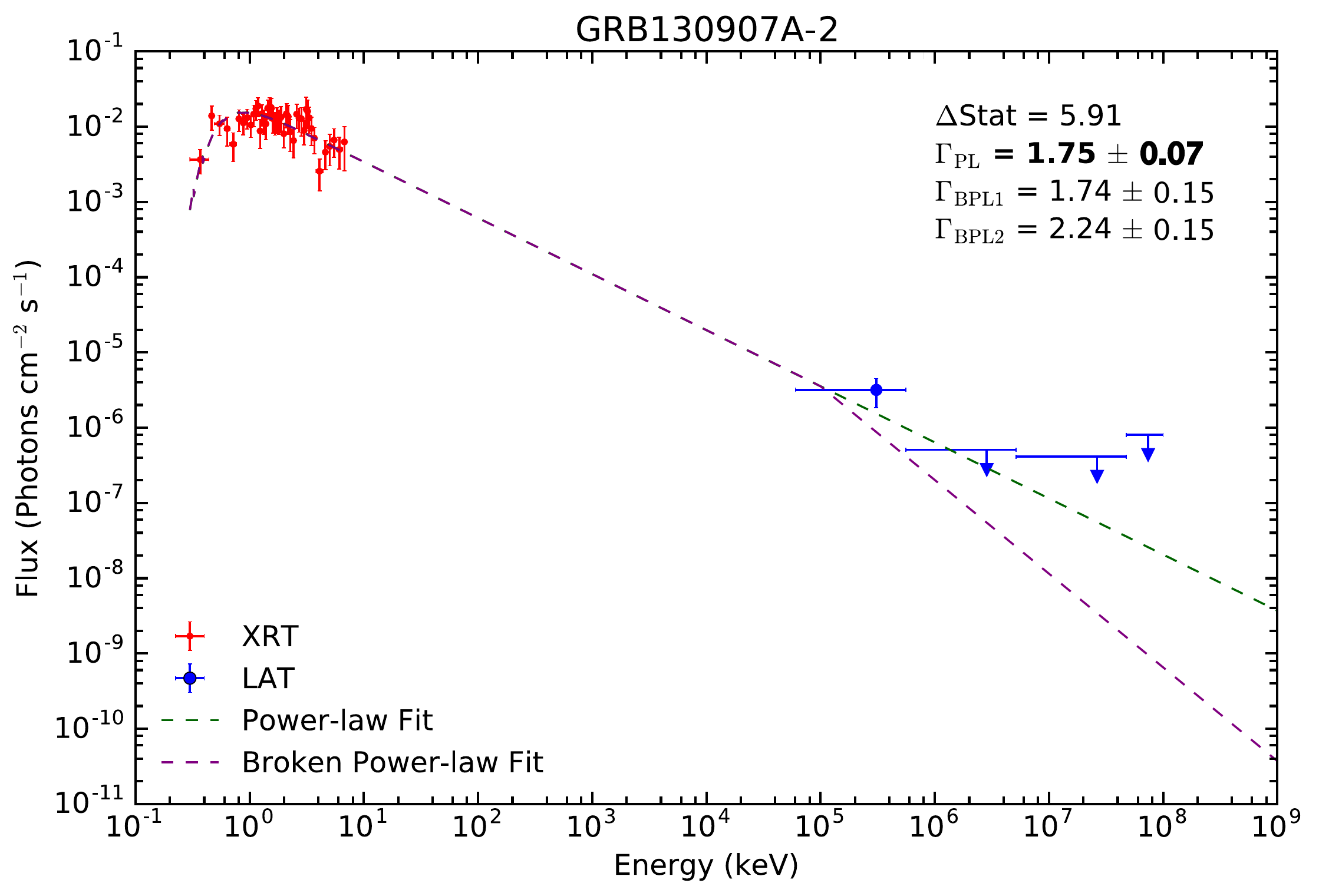}
\includegraphics[width=0.48\textwidth]{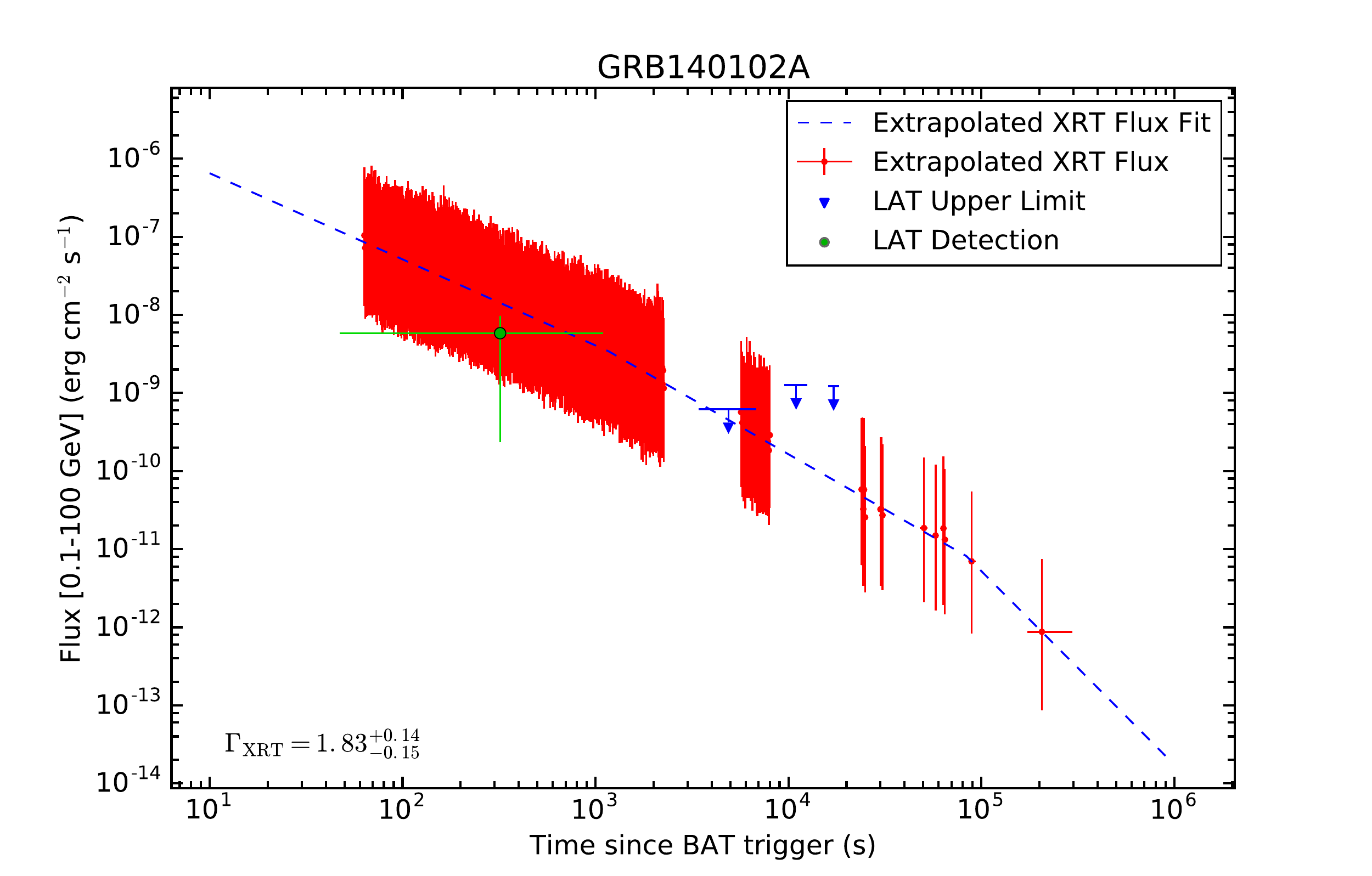}
\includegraphics[width=0.43\textwidth]{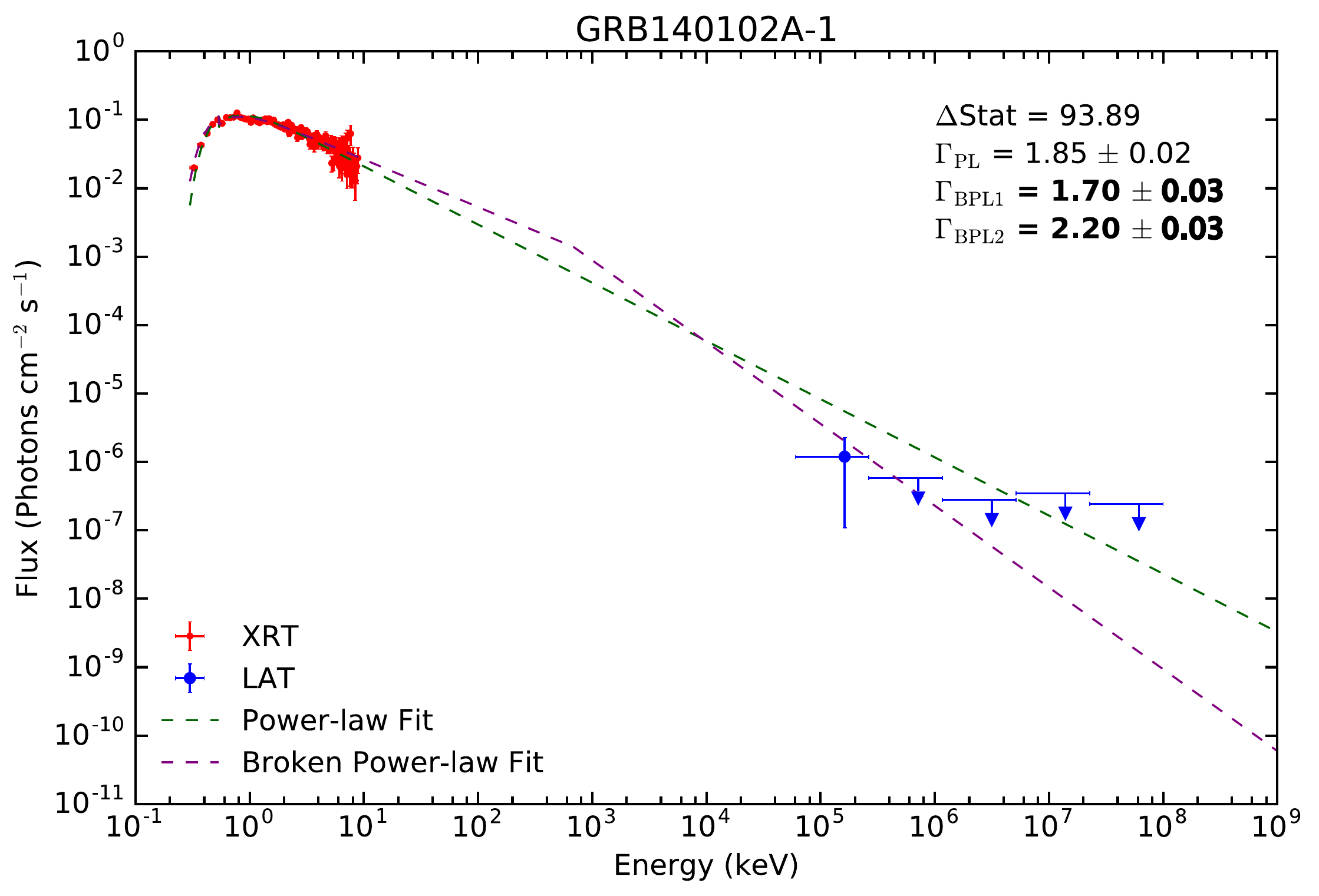}
\includegraphics[width=0.48\textwidth]{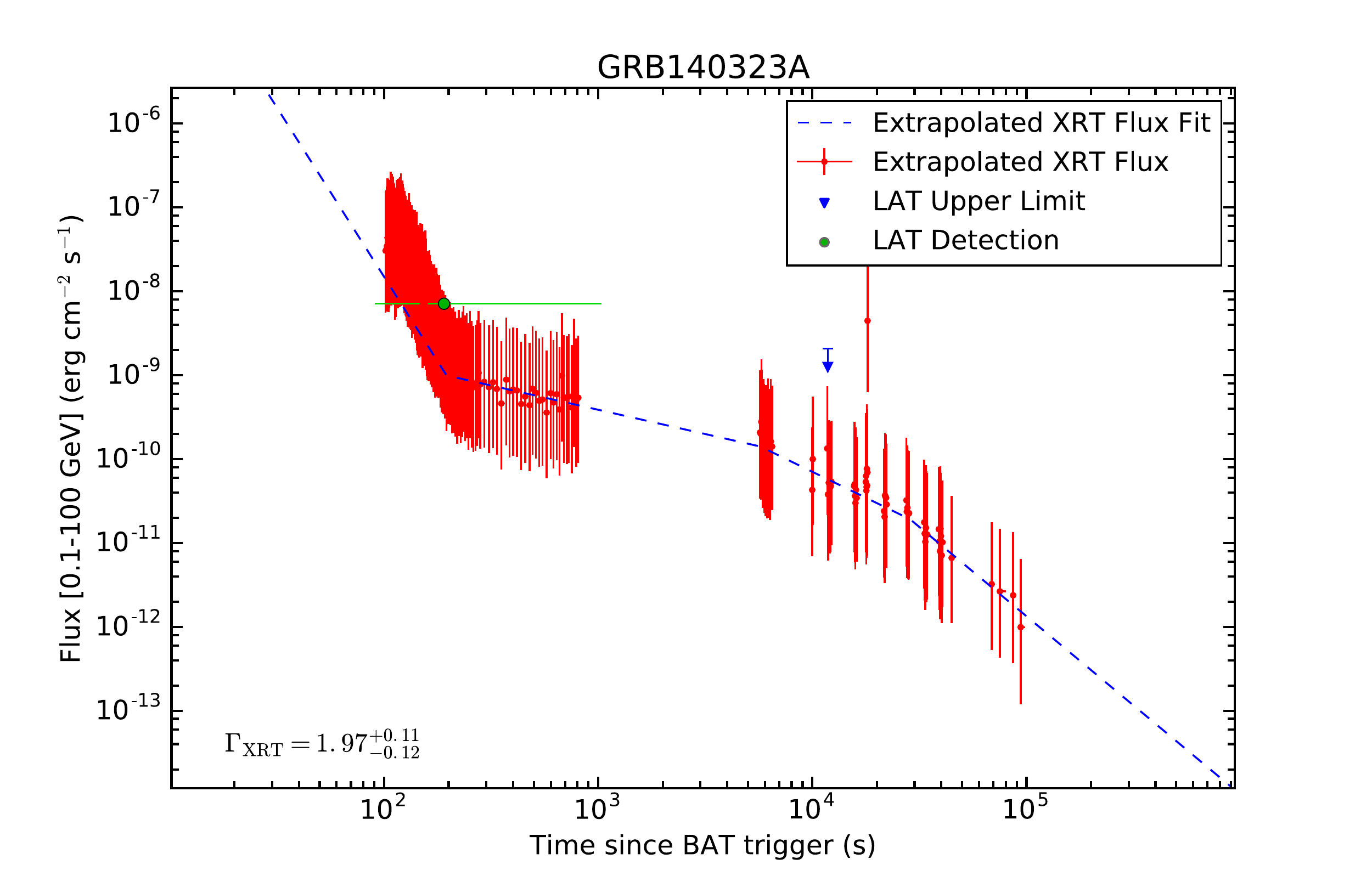}
\includegraphics[width=0.43\textwidth]{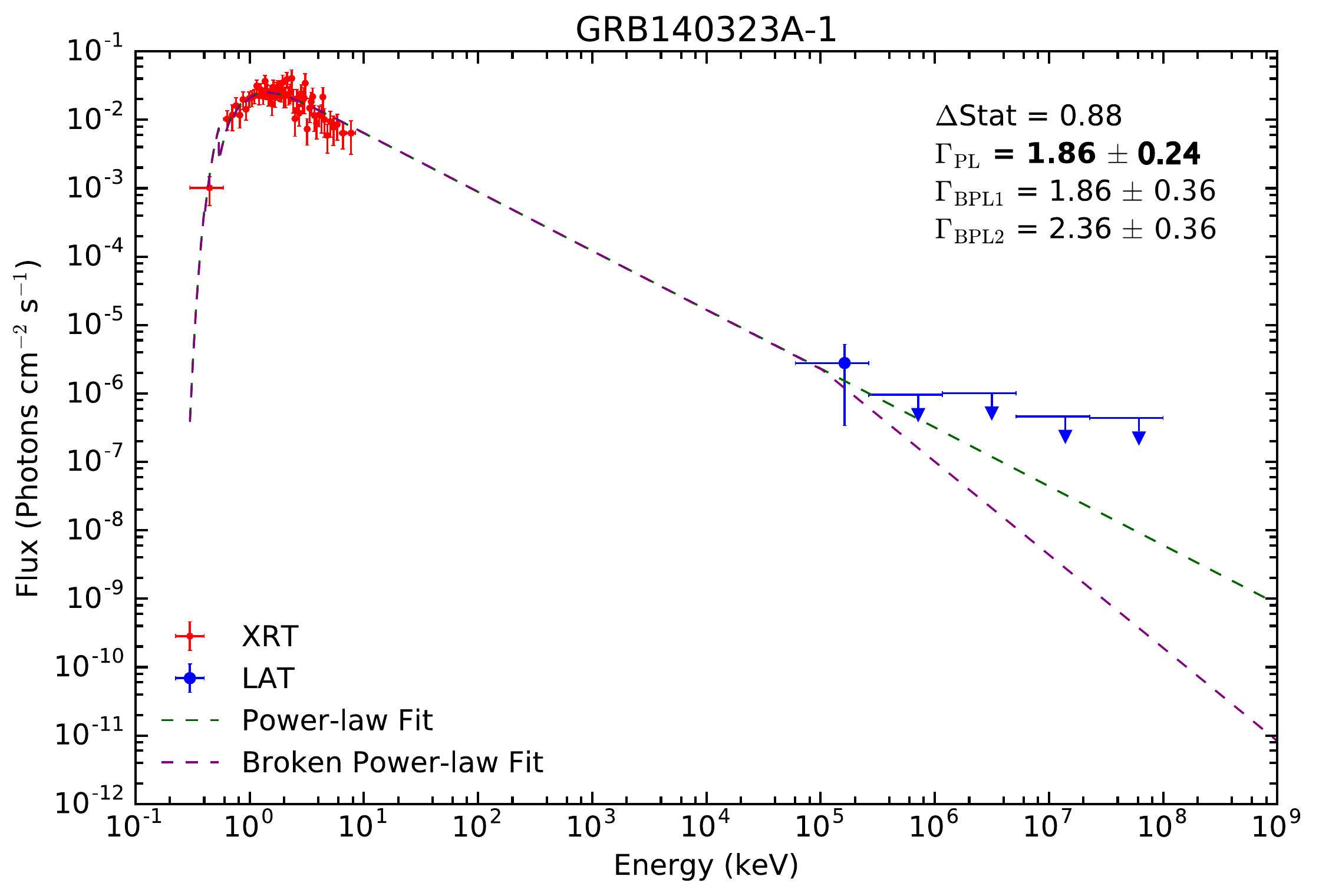}
\caption{The temporal and spectral fits (left and right panels) for the 11 LAT-detected bursts with simultaneous XRT and LAT observations in our sample.  The photon indices $\Gamma_{\rm XRT}$ listed on the temporal plots are derived from fits to only the time-integrated XRT data, whereas the photon indices listed on the spectral fits are obtained through the joint fits of both the XRT and LAT data. The numeric suffix in the title of the spectral plots indicates the temporal interval from which this data was extracted.}
\label{Fig:LATDetections1}
\end{figure}

The temporal and spectral fits for the 11 LAT-detected bursts with contemporaneous XRT and LAT data in our spectroscopic sample are shown in the sub-panels of Figure \ref{Fig:LATDetections1}.  The spectral fits were performed using data extracted from the first detected interval for each burst.  Of the 11 bursts analyzed, 5 show a preference for a break in their broadband spectrum between the XRT and LAT, with the remainder 6 being consistent with a single power law from the X-ray to gamma-ray regimes.  As commented in \S\ref{sec:Results_FluxComparisons}, the flux measurements for all of the LAT detections were either consistent with the XRT extrapolation or fell below it, which is confirmed by the joint spectral fits.  The broadband X-ray and gamma-ray spectral data for the LAT detections are all well fit by either a power-law or a broken power-law model, and show no evidence of high-energy emission significantly in excess of the flux expected from the XRT observations. 

All of the LAT-detected bursts in our sample exhibit bright X-ray afterglows with relatively hard X-ray photon indices (i$.$e$.$, $\Gamma_{\rm XRT} < 2$).  A median photon index of $\Gamma_{\rm PL} = 1.77 \pm 0.04$ was measured for the 6 GTIs for which a single power law was adequate to describe both the XRT and LAT data.  Unlike for the LAT non-detected bursts, this value is consistent with the median photon index of $\Gamma_{\rm XRT} = 1.76 \pm 0.21$ for this sample when estimated from the XRT data alone. For the bursts which show a preference for a break in their broadband afterglow spectrum, we find median XRT and LAT photon indices of $\Gamma_{\rm BPL1} = 1.72 \pm 0.10$ and $\Gamma_{\rm BPL2} = 2.22$. The pre-break photon index is again consistent with the value estimated from the XRT data alone of $\Gamma_{\rm XRT} = 1.70 \pm 0.17$ for this sample. The fit parameters for each individual LAT-detected burst are displayed in Table 2.

Our analysis reveals that a single power law is capable of explaining the broadband emission from GRB~110731A, whereas the emission observed from GRB~130427A and GRB~090510 require a spectral break between the X-ray and gamma-ray regimes.  These results are consistent with those previously reported by \citet{GRB110731A}, \citet{GRB130427A_NuSTAR}, and  \citet{DePasquale2010} respectively.  Conversely, we find that a spectral break is statistically preferred for GRB~100728A, contrary to the findings of \citet{Abdo2011}.  In the latter case, the differing results can likely be attributed to the greater sensitivity of the Pass 8\footnote{https://fermi.gsfc.nasa.gov/ssc/data/analysis/documentation/Pass8\_usage.html} data selection used in this work, compared to the Pass 7 data selection used in previous papers.




\begin{figure}
\centering
\includegraphics[width=0.48\textwidth]{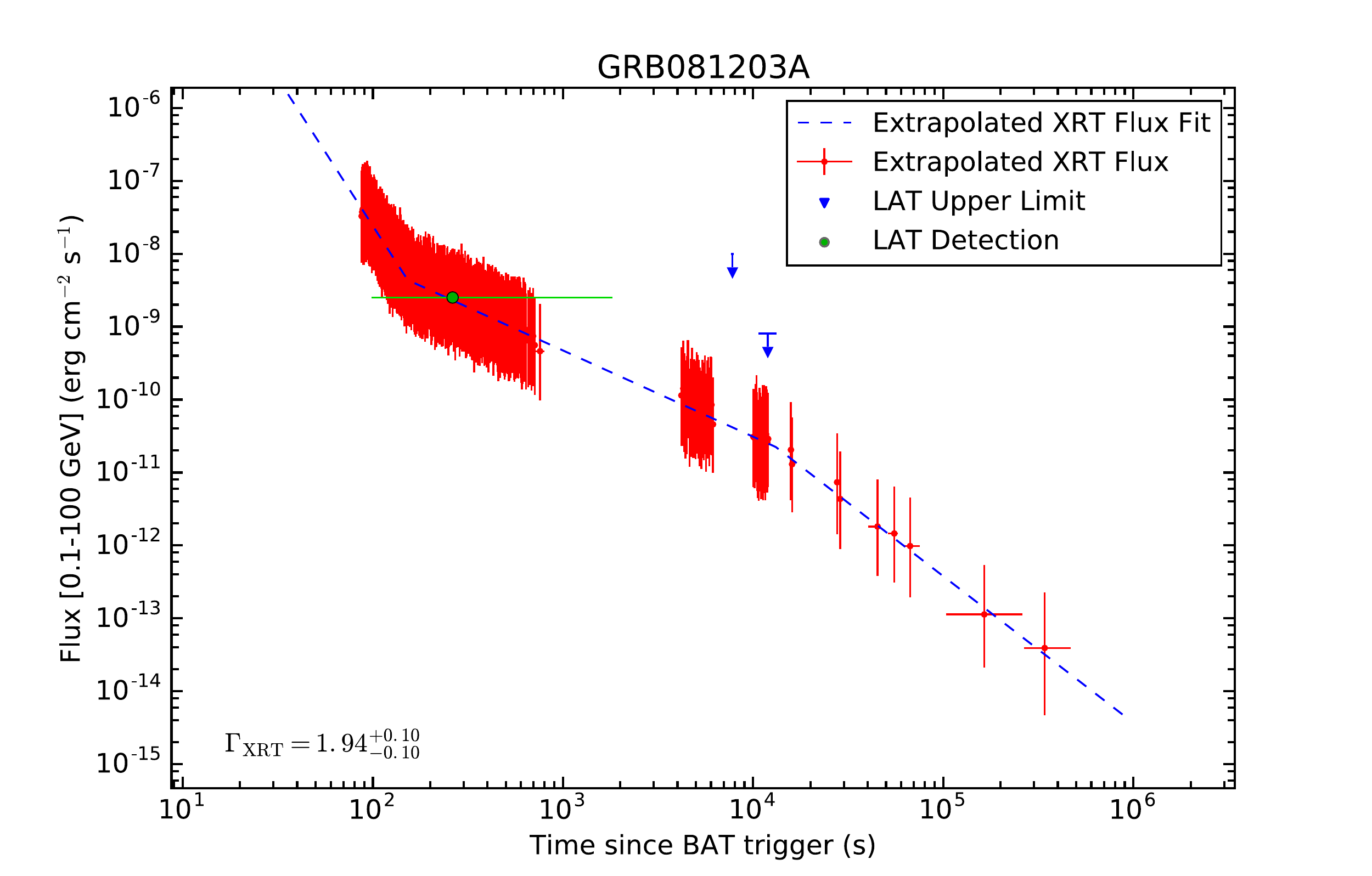}
\includegraphics[width=0.43\textwidth]{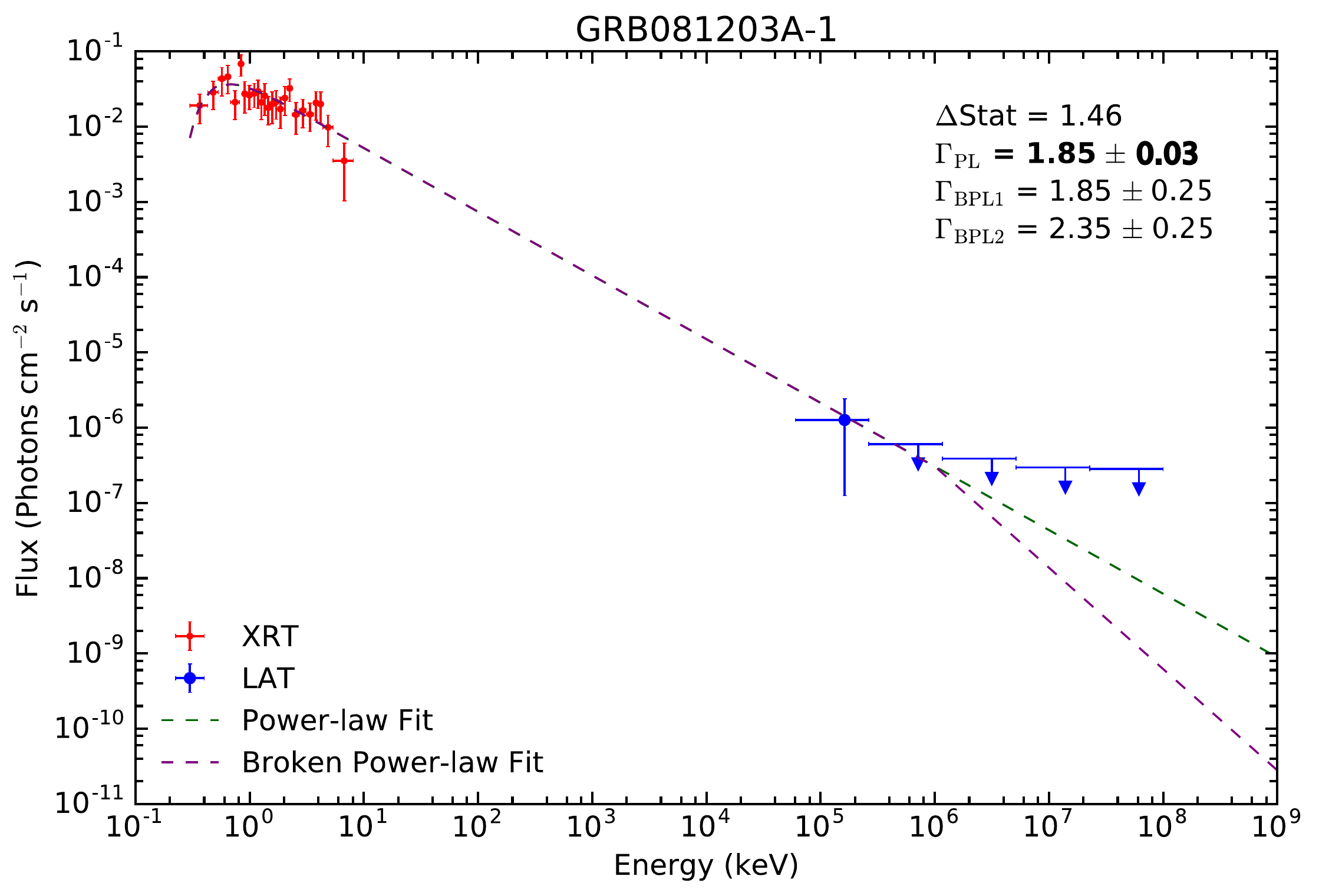}
\includegraphics[width=0.48\textwidth]{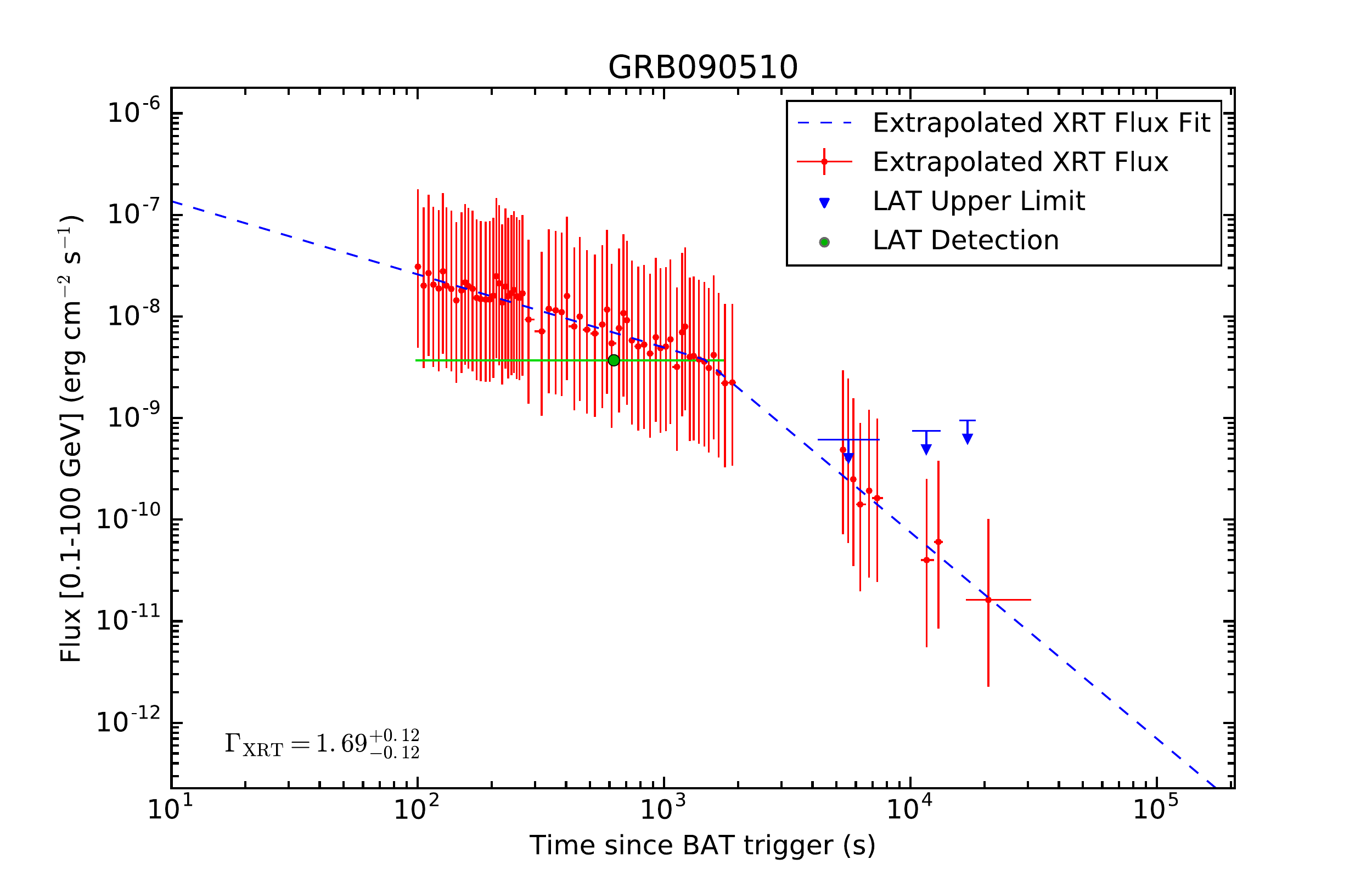}
\includegraphics[width=0.43\textwidth]{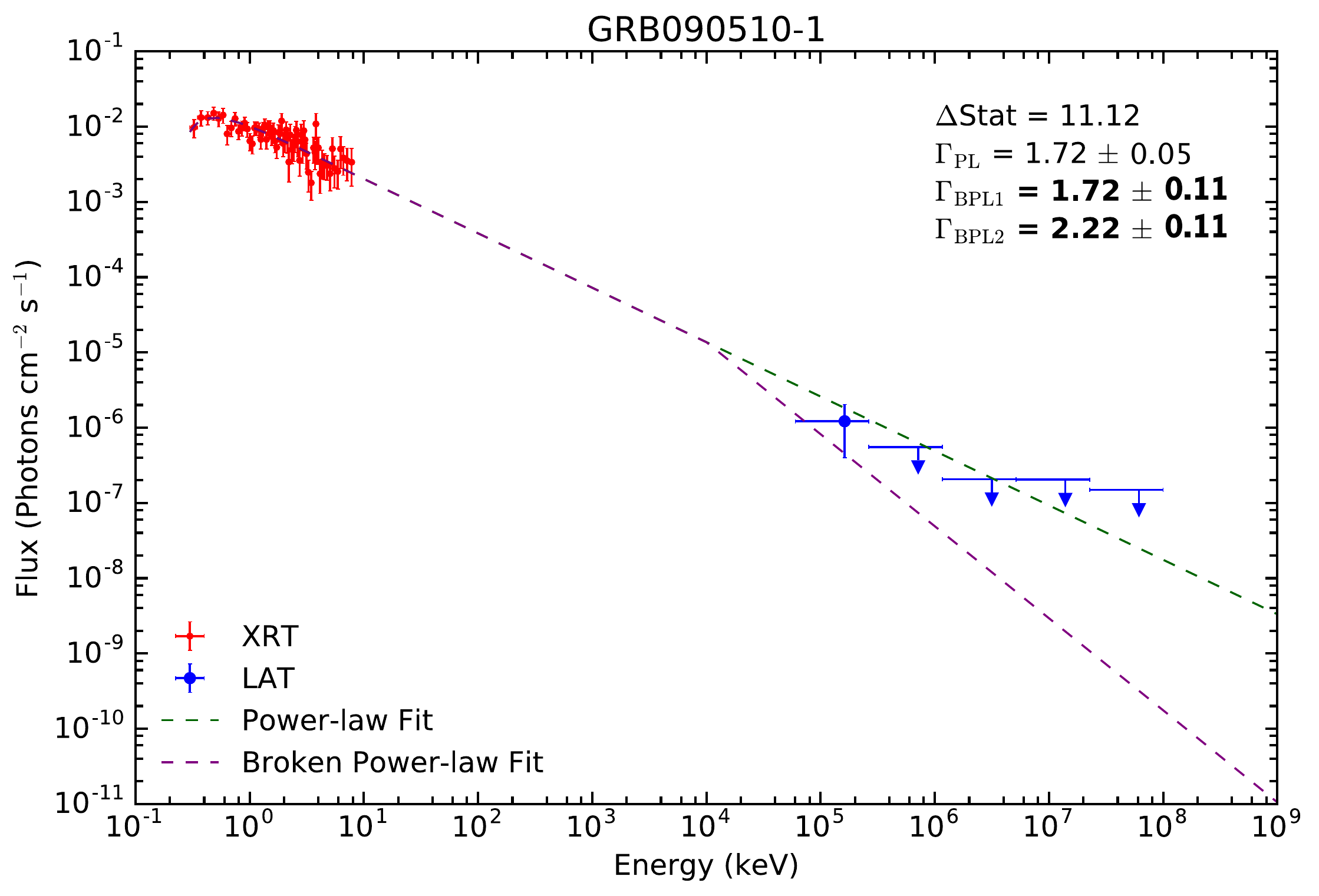}
\includegraphics[width=0.48\textwidth]{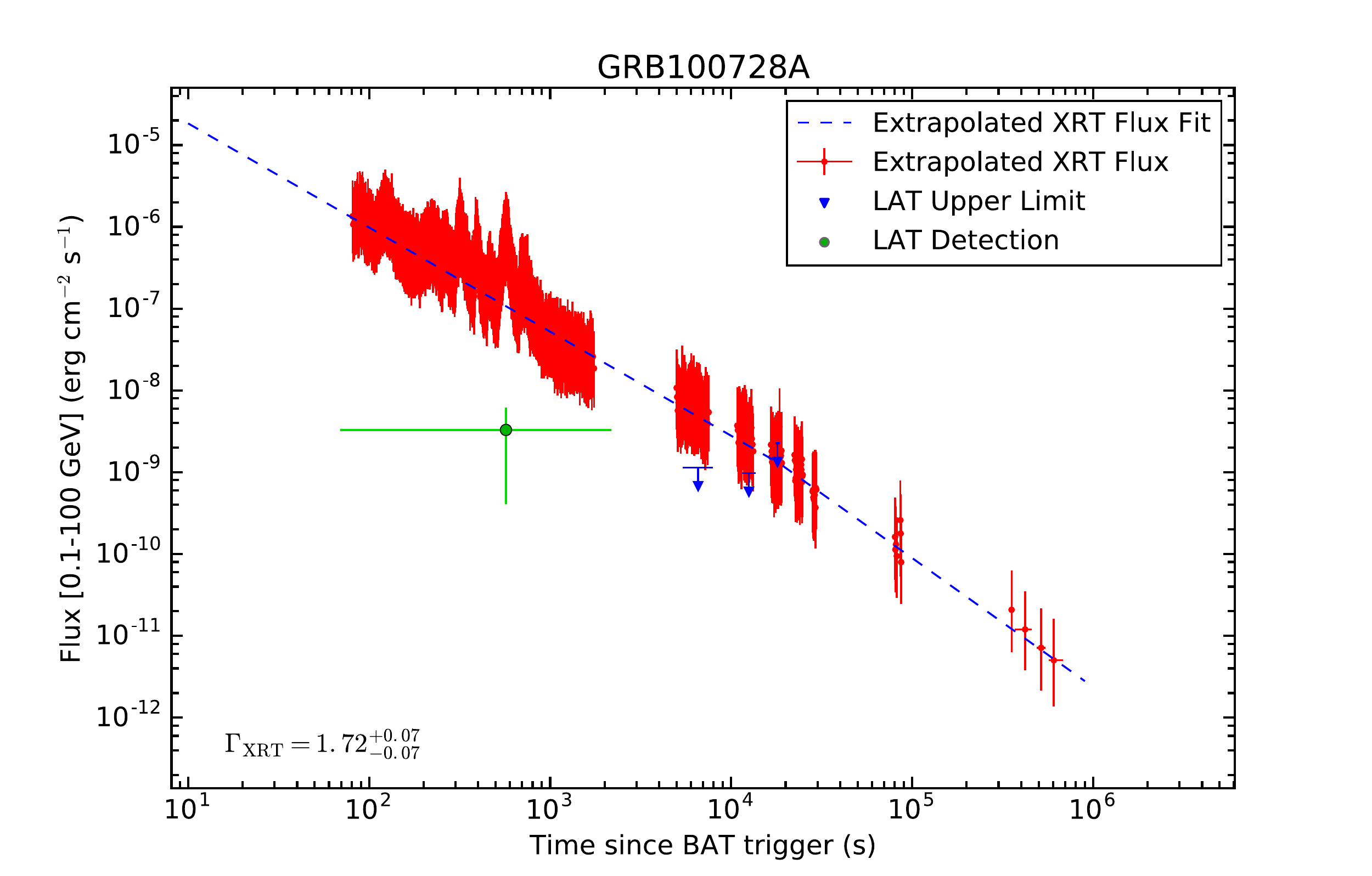}
\includegraphics[width=0.43\textwidth]{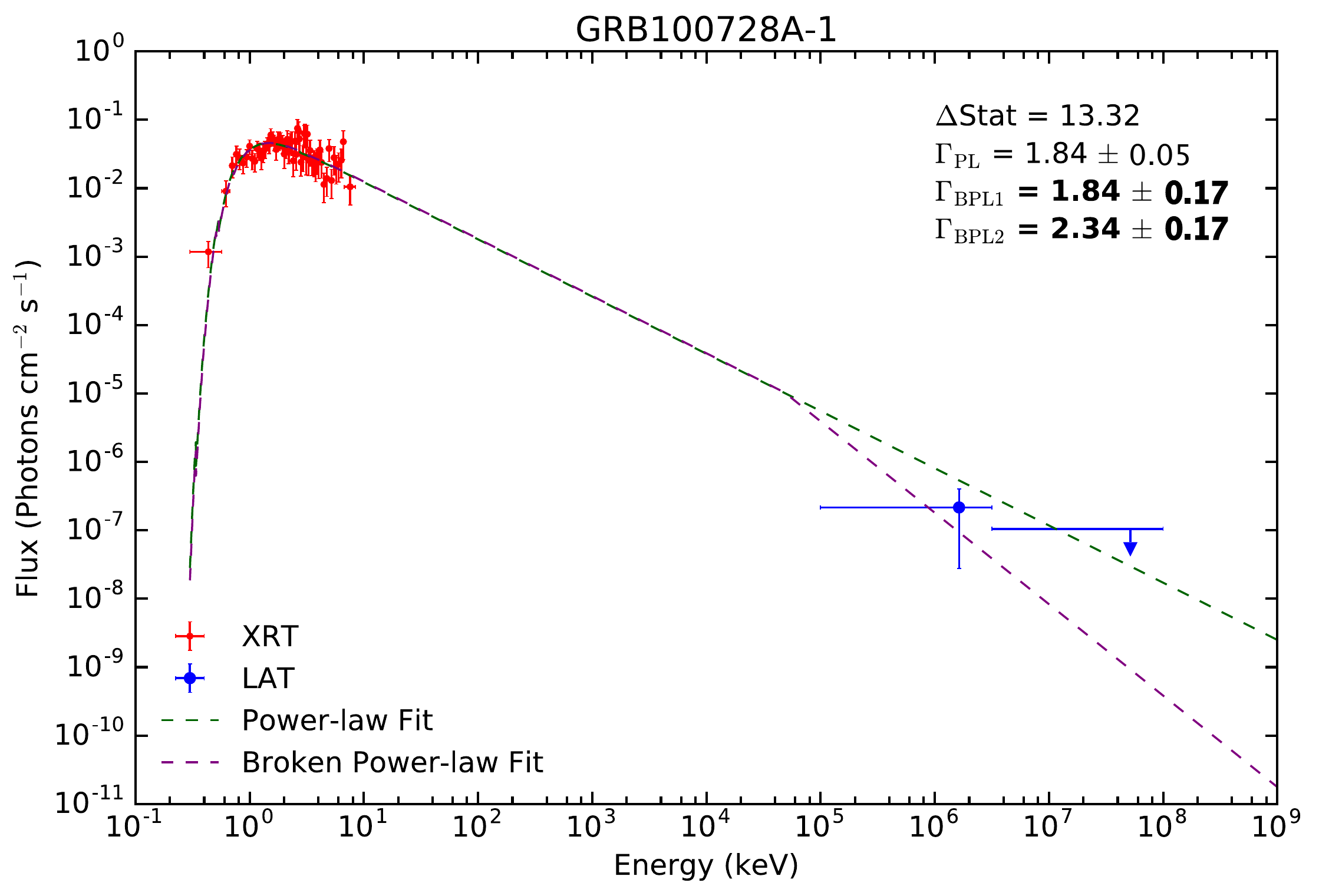}
\includegraphics[width=0.48\textwidth]{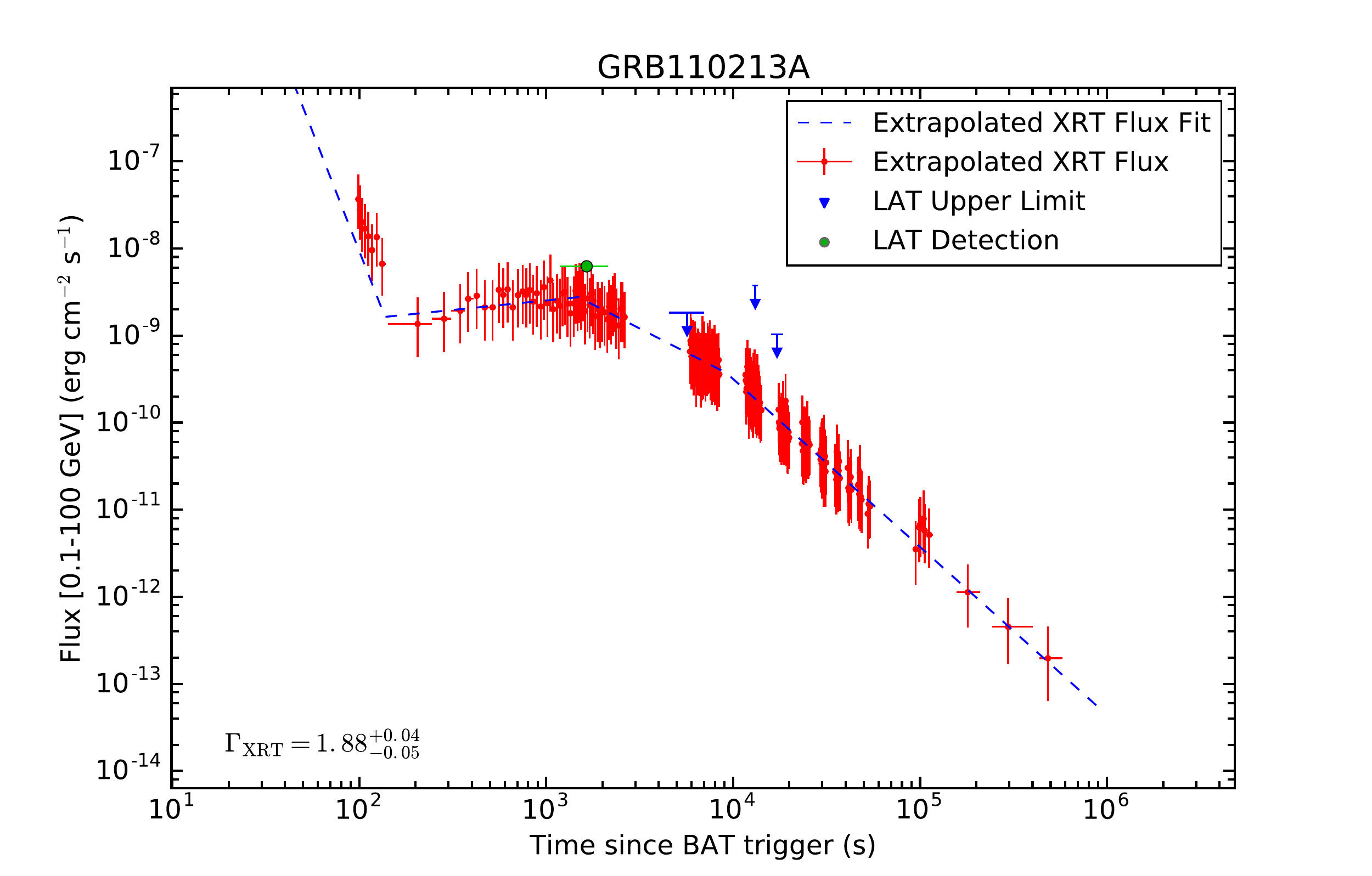}
\includegraphics[width=0.43\textwidth]{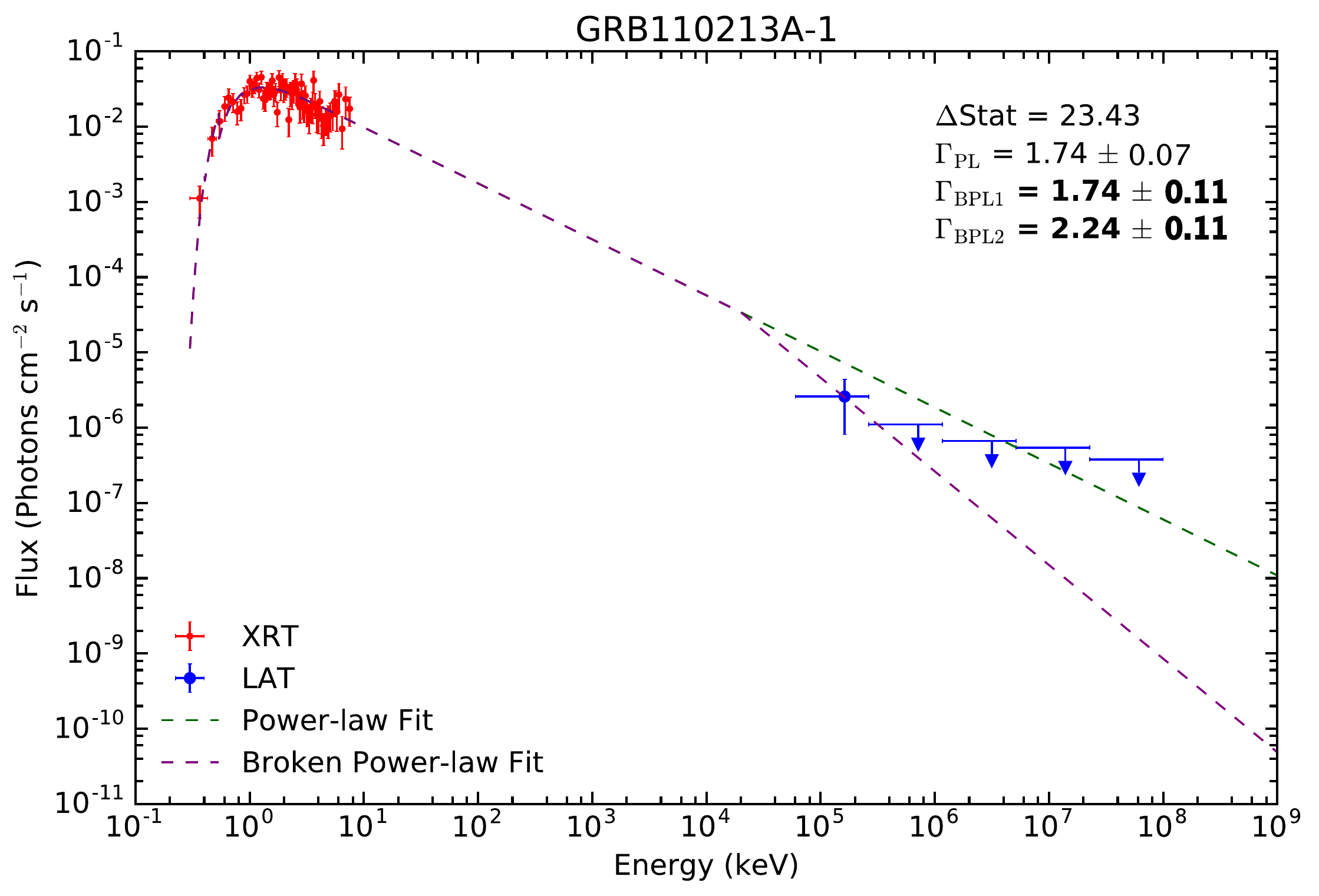}
\end{figure}

\begin{figure}
\centering
\includegraphics[width=0.48\textwidth]{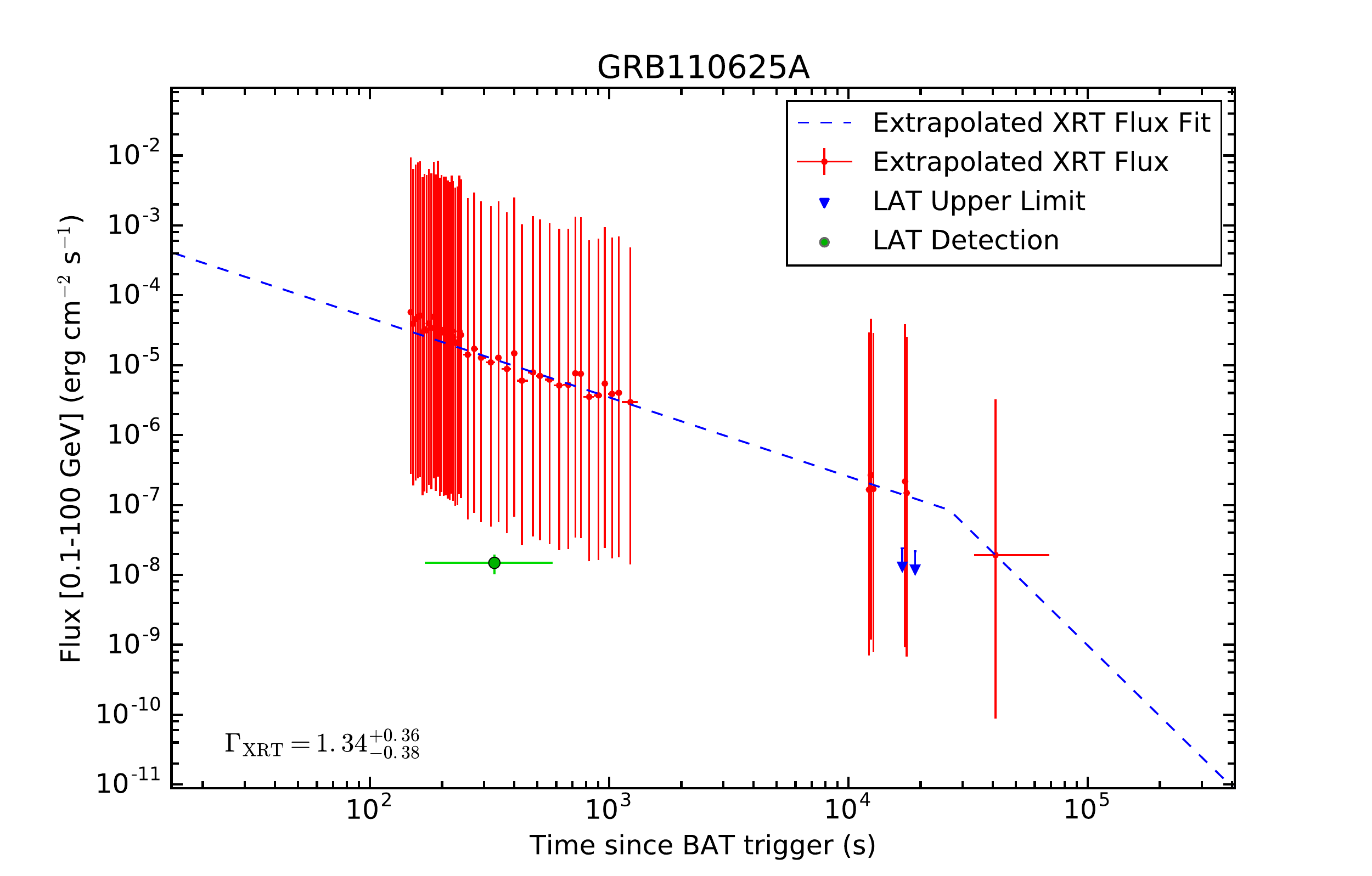}
\includegraphics[width=0.43\textwidth]{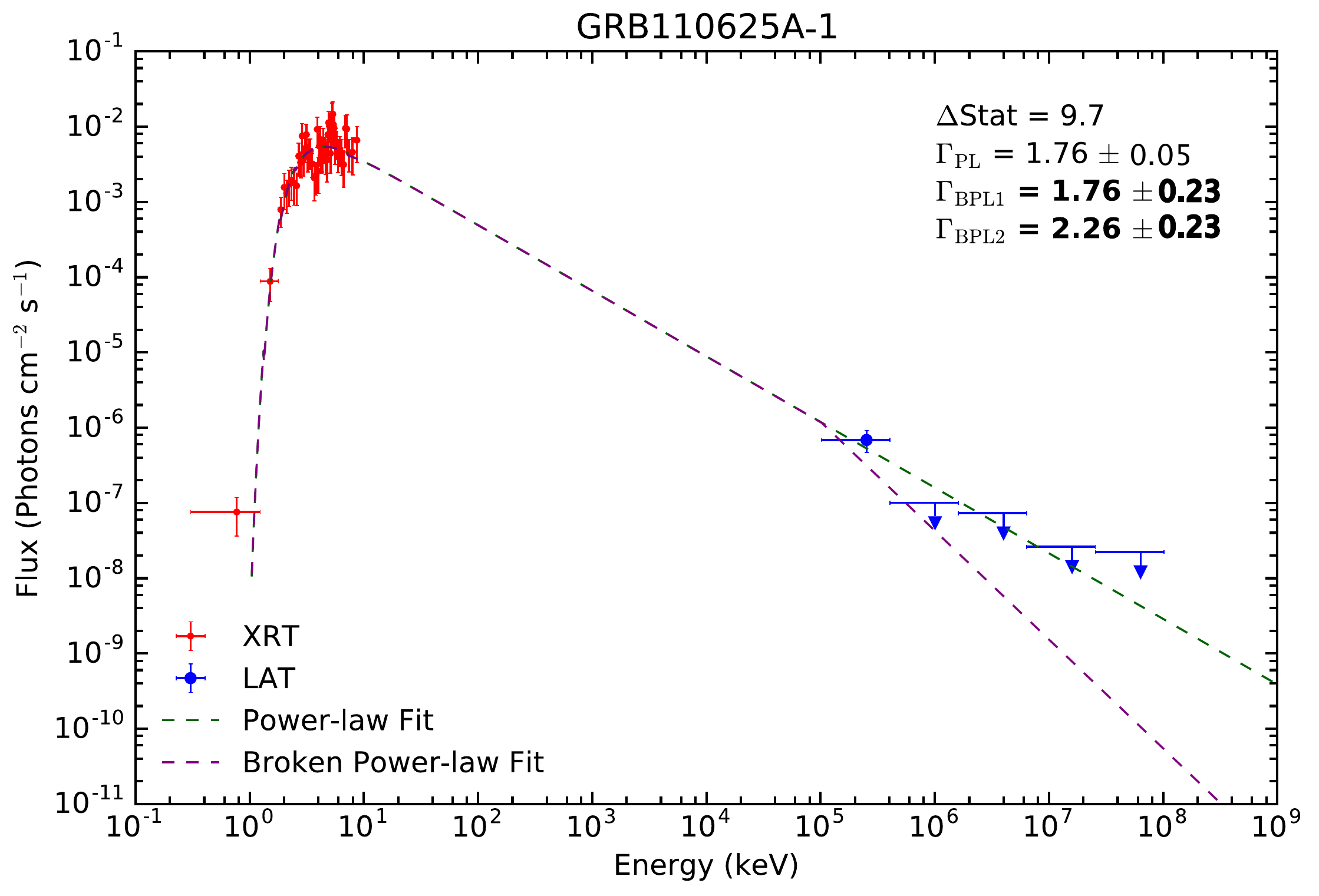}
\includegraphics[width=0.48\textwidth]{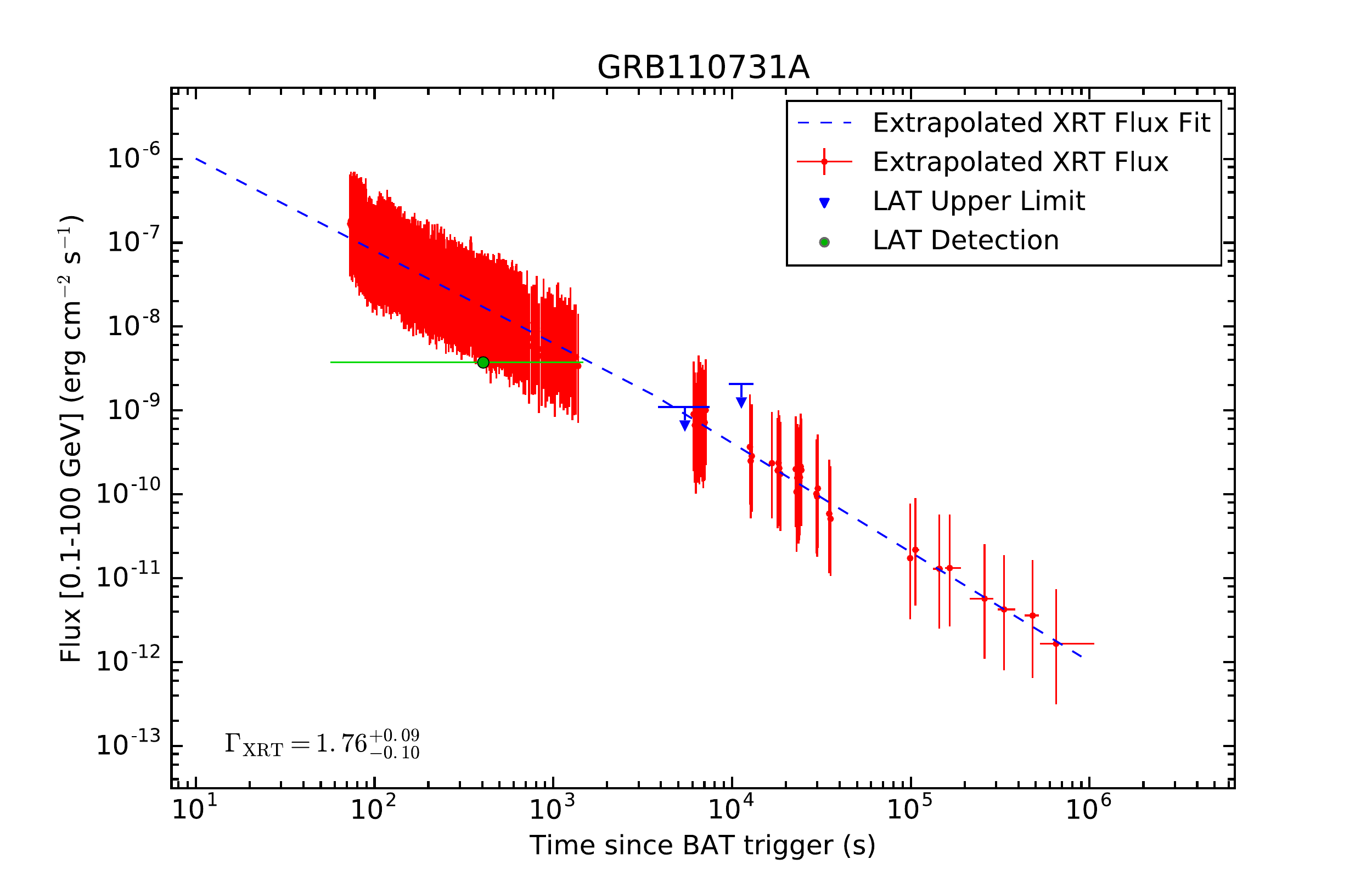}
\includegraphics[width=0.43\textwidth]{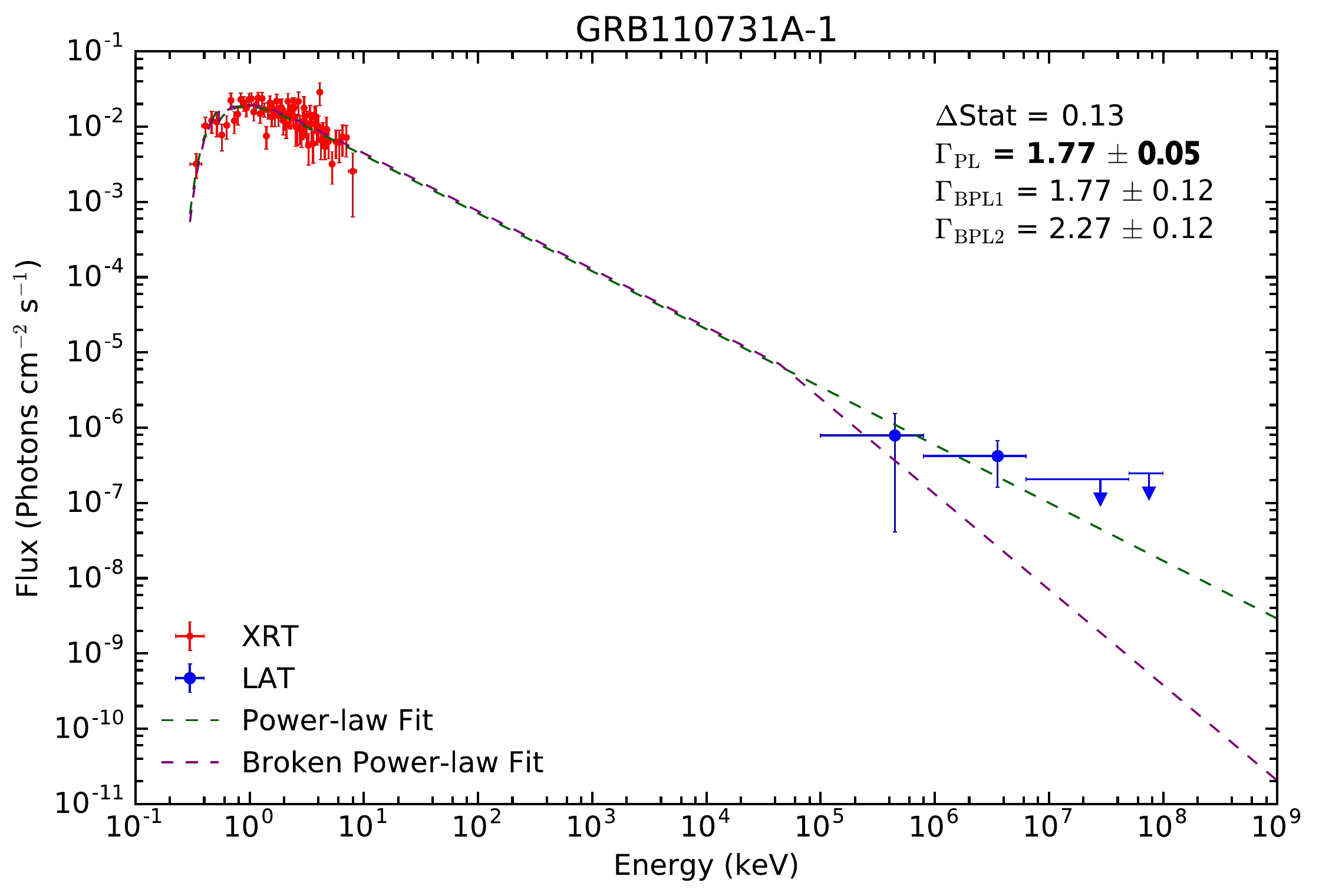}
\includegraphics[width=0.48\textwidth]{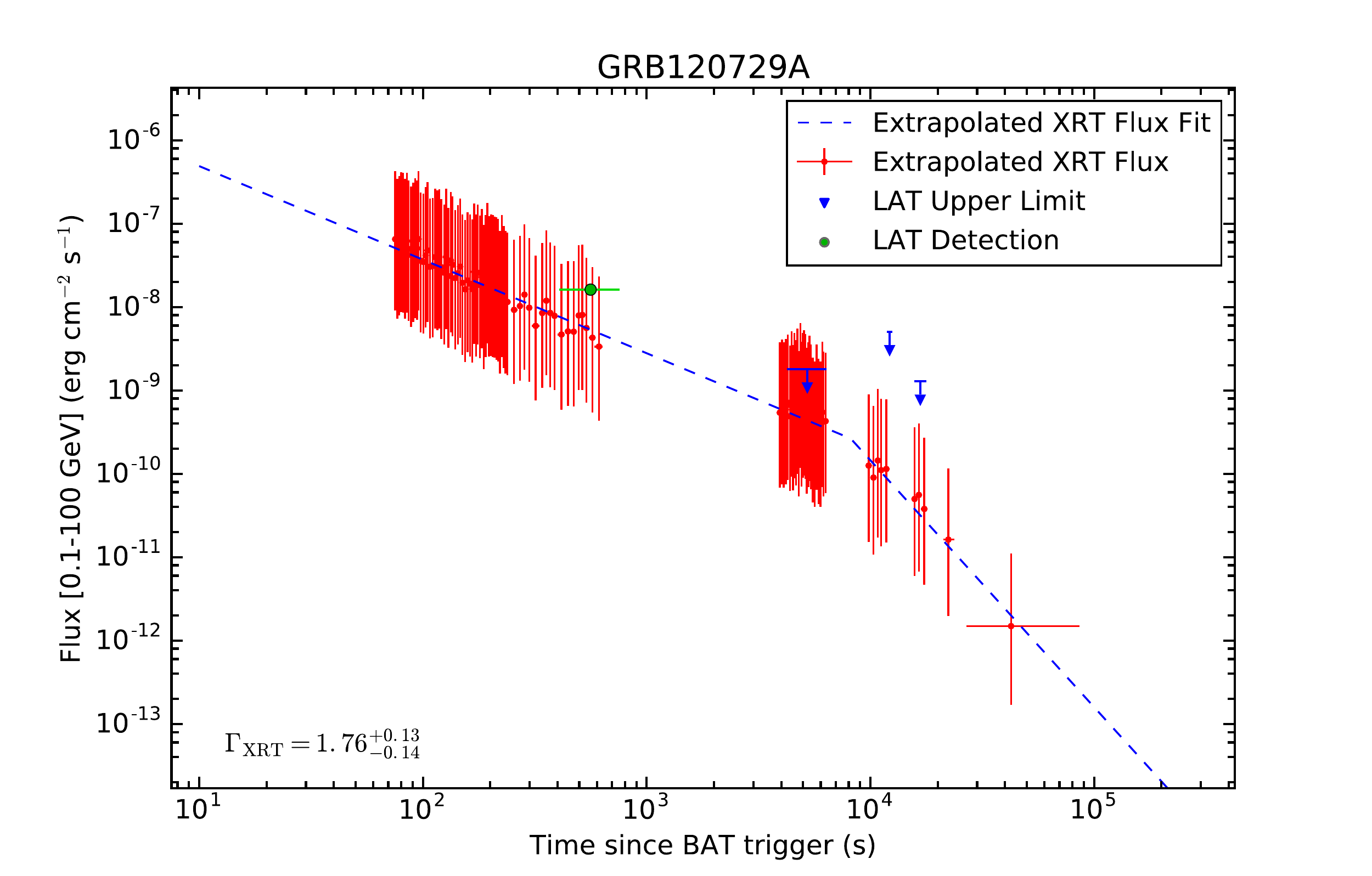}
\includegraphics[width=0.43\textwidth]{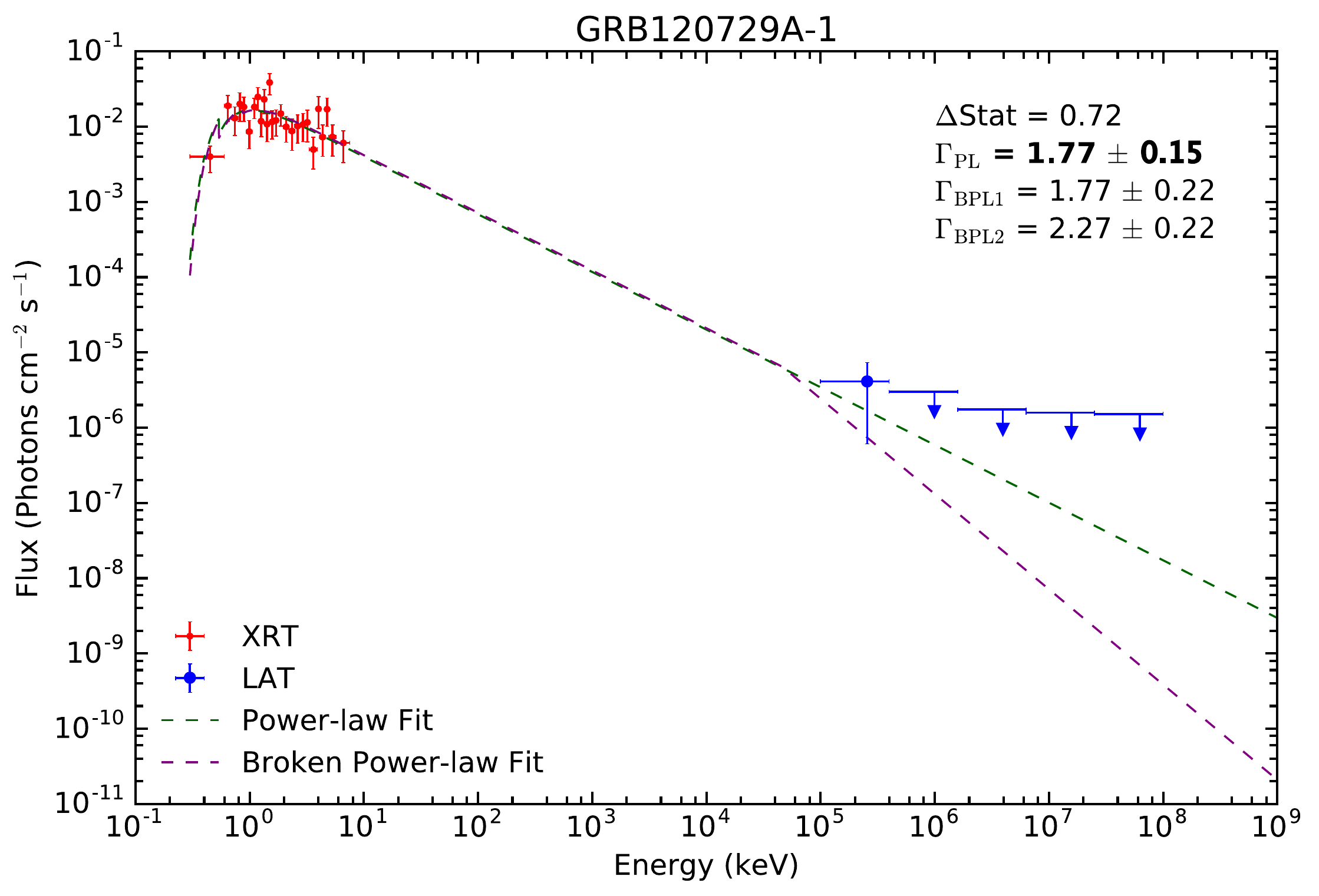}
\includegraphics[width=0.48\textwidth]{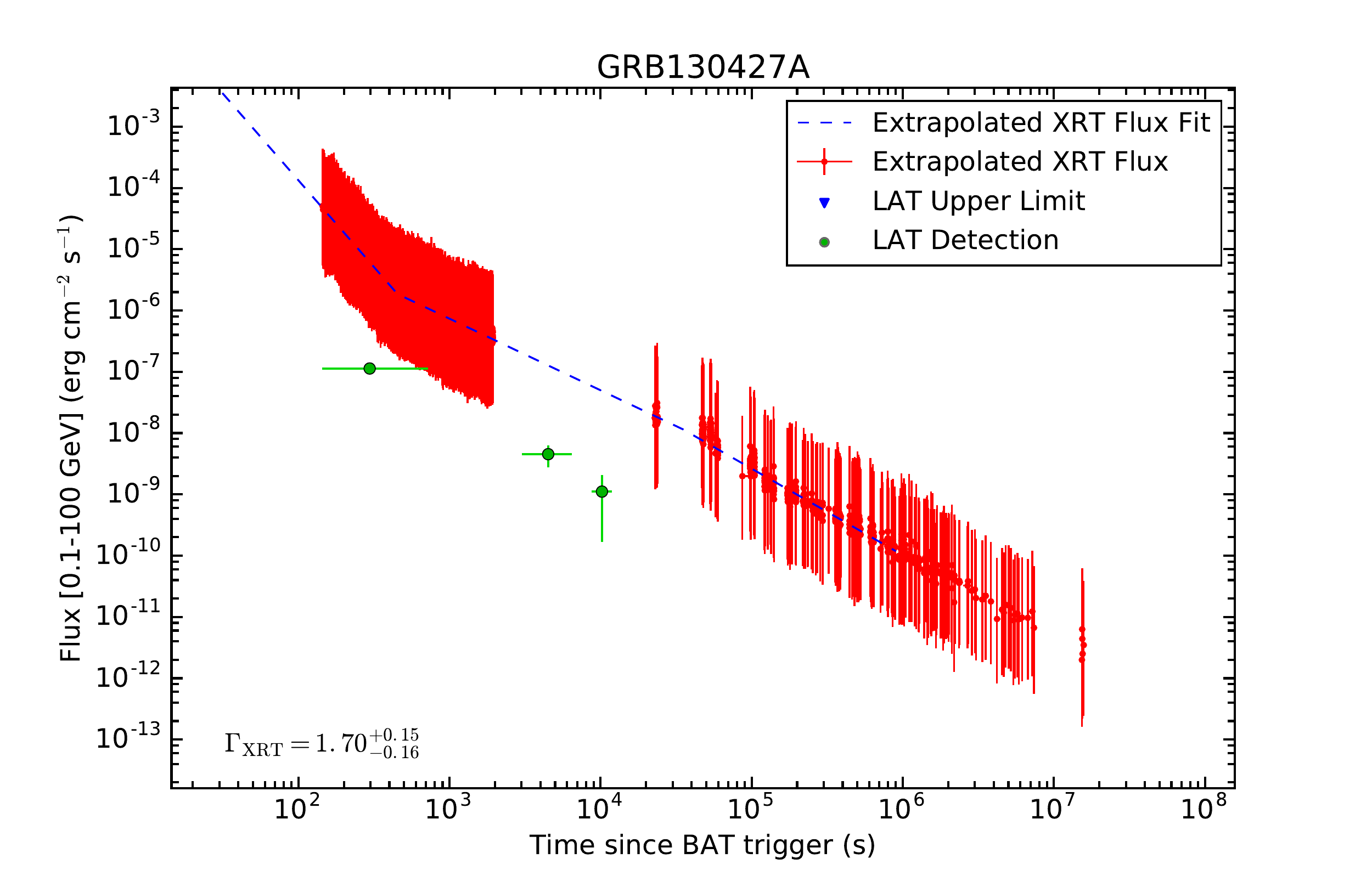}
\includegraphics[width=0.43\textwidth]{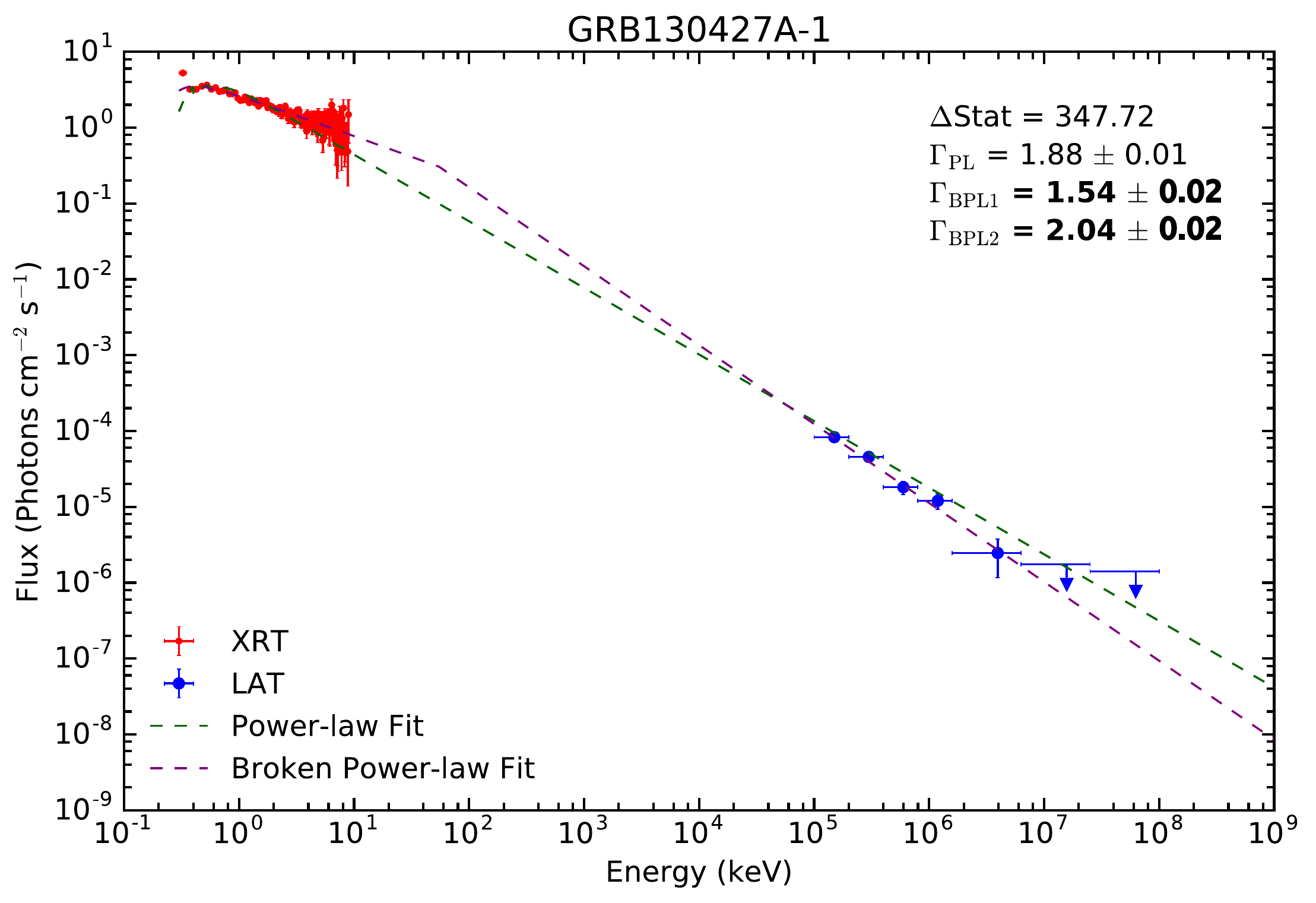}
\end{figure}

\begin{table}[ht]
\centering
\begin{tabular}{l c c c c c c}
\hline\hline
Sample & Best Fit & GTIs & $\Gamma_{\rm XRT}$ &  $\Gamma_{\rm PL}$ & $\Gamma_{\rm BPL1}$ & $\Gamma_{\rm BPL2}$ \\ [0.5ex] 
\hline
LAT nondetections & PL & 31 (58$\%$) & 1.68 $\pm$ 0.21 & 1.98 $\pm$ 0.16 & -- & --  \\
LAT nondetections & BPL & 21 (40$\%$) & 1.72 $\pm$ 0.21 & -- & 1.60 $\pm$ 0.13 & 2.10  \\
LAT Detections & PL & 6 (55$\%$) & 1.76 $\pm$ 0.21 & 1.77 $\pm$ 0.04 & -- & --  \\
LAT Detections &  BPL & 5 (45$\%$) & 1.70 $\pm$ 0.17 & --  & 1.72 $\pm$ 0.10 & 2.22  \\
\hline
\end{tabular}
\label{table:Summary}
\caption{A summary of the median best-fit parameters for the joint XRT/LAT spectral fits outlined in \S\ref{sec:Results_JointSpectroscopicFits} and \S\ref{sec:Results_JointSpectroscopicFits_LATDetections}  }
\end{table}

\begin{table}[ht]
\centering
\begin{tabular}{c | c c | c c c c c c}
\hline\hline
GRB & $\Gamma_{\rm XRT}$ & $\Gamma_{\rm LAT}$ & Best Fit & $\Delta\rm{Stat}$  &  $\Gamma_{\rm PL}$ & $\Gamma_{\rm BPL1}$ & $\Gamma_{\rm BPL2}$ & $E_{\rm b}$ (keV) \\ [0.5ex] 
\hline


081203A 	& 1.94$^{+0.10}_{-0.10}$	& 2.18 $\pm$ 0.36		&  PL 		& 1.5 			& 1.85 $\pm$ 0.03	& 1.85 $\pm$ 0.25	& 2.35 & -- \\
090510A 	& 1.69$^{+0.12}_{-0.12}$	& 2.44 $\pm$ 0.55		&  BPL		& 11.1 		& 1.72 $\pm$ 0.05	& 1.72 $\pm$ 0.11	& 2.22 & 9958 $\pm$  968 \\
100728A 	& 1.72$^{+0.07}_{-0.07}$	& 1.70 $\pm$ 0.22		&  BPL 		& 13.3 		& 1.84 $\pm$ 0.05	& 1.84 $\pm$ 0.17	& 2.34 & 9568  $\pm$ 1045 \\
110213A 	& 1.88$^{+0.04}_{-0.05}$	& 1.60 $\pm$ 0.36		&  BPL 		& 23.4 		& 1.74 $\pm$ 0.07	& 1.74 $\pm$ 0.11	& 2.24 & 10000 $\pm$ 946 \\
110625A 	& 1.34$^{+0.36}_{-0.38}$	& 2.49 $\pm$ 0.22		&  BPL 		& 9.7 			& 1.76 $\pm$ 0.05	& 1.76 $\pm$ 0.23	& 2.26 & 7125 $\pm$ 1060 \\
110731A 	& 1.76$^{+0.09}_{-0.10}$	& 1.69 $\pm$ 0.37		&  PL 		& 0.1 			& 1.77 $\pm$ 0.05	& 1.77 $\pm$ 0.12	& 2.27 & -- \\
120729A 	& 1.76$^{+0.13}_{-0.14}$	& 1.77 $\pm$ 0.35		&  PL 		& 0.7 			& 1.77 $\pm$ 0.15	& 1.77 $\pm$ 0.22	& 2.27 & -- \\
130427A 	& 1.70$^{+0.15}_{-0.16}$	& 2.06 $\pm$ 0.07		&  BPL 		& 347.7 	 	& 1.88 $\pm$ 0.01	& 1.54 $\pm$ 0.02	& 2.04 & 54 $\pm$ 18 \\
130907A 	& 1.75$^{+0.04}_{-0.04}$	& 2.05 $\pm$ 0.35		&  PL 		& 5.9 			& 1.75 $\pm$ 0.07	& 1.74 $\pm$ 0.15	& 2.24 & -- \\
140102A 	& 1.83$^{+0.14}_{-0.15}$	& 1.53 $\pm$ 0.31		&  BPL 		& 93.9 		& 1.85 $\pm$ 0.02	& 1.70 $\pm$ 0.03	& 2.20 & 681 $\pm$ 16 \\
140323A 	& 1.97$^{+0.11}_{-0.12}$	& 1.86 $\pm$ 0.42		&  PL 		& 0.9 			& 1.86 $\pm$ 0.24	& 1.86 $\pm$ 0.36	& 2.36 & -- \\
\hline

\end{tabular}
\label{table:LATDetectedBursts}
\caption{A summary of the best-fit spectral parameters for the LAT-detected population in our sample. $\Gamma_{\rm XRT}$ \& $\Gamma_{\rm LAT}$ are the photon indices obtained from fitting the XRT and LAT GTIs separately, whereas  $\Gamma_{\rm PL}$, $\Gamma_{\rm BPL1}$, and $\Gamma_{\rm BPL2}$ are the photon indices obtained through the joint XRT and LAT fits to power-law (PL) and broken power-law (BPL) models, respectively.  The post-break photon index in the BPL model is fixed to $\Gamma_{\rm BPL2}$ = $\Gamma_{\rm BPL1}$ + 0.5.  A BPL model is statistically preferred at $>3\sigma$ over a simpler PL model when $\Delta\rm{Stat} > 9$. }
\end{table}

\section{Discussion} \label{sec:Discussion}

The results presented in \S\ref{sec:Results_FluxComparisons} reveal that a majority of bursts that are detected by \Swift XRT do not have sufficiently bright afterglows and/or hard spectra to be detected by \Fermi LAT.  Of the 1156 intervals that we analyzed for this study, we found that only a small subset exhibited afterglow emission that could exceed the LAT detection threshold when extrapolated to the 0.1 to 100 GeV energy range.  This finding illustrates that the late-time detection of afterglow emission by the LAT at high energies is relatively uncommon, despite nearly every \emph{Swift}-detected GRB being within the LAT FoV at some point before the end of XRT observations. The bursts that do result in late-time LAT detections exclusively have afterglow intervals with emission brighter than $F_{\rm XRT} \gtrsim 10^{-10}$ erg cm$^{-2}$ s$^{-1}$ and harder than $\Gamma_{\rm XRT} \lesssim 2$.

We performed joint spectral fits of simultaneous XRT and LAT data for 52 GTIs for which no emission was detected by the LAT, but for which their XRT derived afterglow spectra were sufficiently bright and hard that they exceed the LAT upper limits.  These fits reveal that a majority of these cases (58$\%$) can be explained by an afterglow spectrum with a slightly softer photon index when constrained by both the XRT and LAT data, compared to the photon index derived by fits to the XRT data alone.  The remaining LAT nondetections required a break in their afterglow spectra between the XRT and LAT energy ranges, consistent with a cooling break expected in the high-energy regime of electron synchrotron emission from a relativistic blast wave expanding into an external medium.


Of the 11 LAT-detected bursts in our sample, we find that the measured flux in the 0.1--100 GeV energy range is either consistent with, or falls below, the flux expected at these energies from an extrapolation of their afterglow spectra as derived from simultaneous XRT observations.  These results are confirmed by joint spectral fits of XRT and LAT data for these bursts, which show that the broadband X-ray and gamma-ray data are well fit by either a simple power-law, or a broken power-law model that is consistent with a cooling break between the energy ranges of the two instruments. As a result, we find no evidence of high-energy emission significantly in excess of the flux expected from the spectrum predicted by the electron synchrotron model.  

\subsection{On the Nature of the LAT-Detected Population} \label{sec:Discussion:InverseComptonComponents}

An examination of the photon indices derived from the joint spectral fits for the LAT-detected and non-detected bursts suggests a difference between these two populations.  For the LAT non-detected bursts, the median photon index of the spectral component connecting the XRT and LAT data is $\Gamma_{\rm PL} = 1.98 \pm 0.16$.  This value is consistent with the canonical value of $\Gamma \sim 2$ expected from the high-energy component of the electron synchrotron spectrum for both the slow and fast-cooling scenarios, for an assumed power-law electron energy distribution of $p = 2$.  Likewise, the LAT non-detected bursts for which a break between the XRT and LAT was required have median pre- and post-break power-law indices of $\Gamma_{\rm BPL1} = 1.6 \pm 0.13$ and $\Gamma_{\rm BPL2} = 2.1$, again consistent with the expected $\Gamma \sim 2$ post-break value. This indicates that the  cooling break of the synchrotron spectrum lies either below or between the XRT and LAT energy ranges for the LAT nondetections for which we performed joint spectral fits.

By contrast, the LAT-detected bursts with broadband XRT and LAT data that are best fit by a single power-law component yield a harder median photon index of $\Gamma_{\rm PL} = 1.77 \pm 0.04$. The LAT-detected bursts for which a break between the XRT and LAT was required have median values of the pre- and post-break power-law indices $\Gamma_{\rm BPL1} = 1.72 \pm 0.10$ and $\Gamma_{\rm BPL2} = 2.22$.  The cooling break of the synchrotron spectrum for these bursts appears to occur either between or above the XRT and LAT energy ranges for a majority of the LAT-detected bursts.  Not a single LAT-detected burst examined in our analysis has an X-ray photon index that is consistent with the canonical $\Gamma \sim 2$ value expected for the highest-energy component predicted by an electron synchrotron spectrum in either a slow or fast cooling regime.

\begin{figure}[t]
\centering
\includegraphics[width=0.6\textwidth]{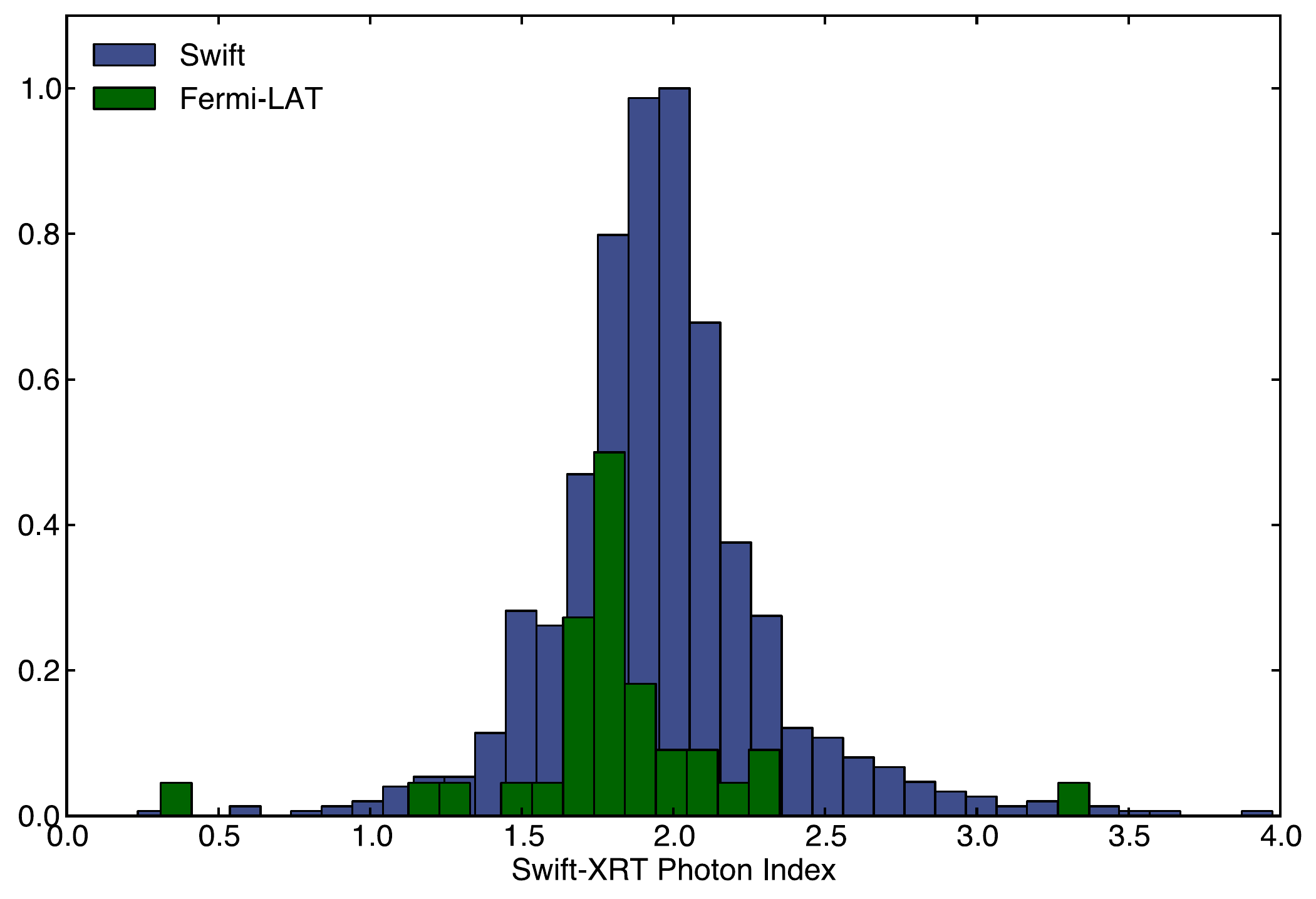}
\caption{A comparison of the X-ray photon index distribution for all \Swift XRT-detected GRBs (blue) and those detected by the LAT (green), for which \Swift XRT observations exist.}
\label{Fig:PhotonIndexDistribution}
\end{figure}

The trend of LAT-detected bursts being spectrally harder in X-ray than their non-detected counterparts can be seen in an examination of the afterglow properties of all LAT-detected bursts observed by the XRT.  Figure \ref{Fig:PhotonIndexDistribution} compares the photon index distributions of all LAT-detected GRBs for which Swift XRT observations exist. A two-sided KS test yields a p-value of 0.0146, rejecting the hypothesis that the two samples are drawn from the same distribution. Here we have dropped the requirement that the LAT detection occurred after the start of the first XRT observations, because we are examining the properties of the afterglows of all LAT-detected bursts and and are not making a joint analysis between the two instruments.  This allows us to include bursts such as GRBs~080916C and 090323A, which were detected by the LAT, but for which XRT observations began after the LAT detections and were therefore excluded from our previous analysis.  The X-ray photon index distribution for all GRB afterglows observed by the XRT peaks at $\Gamma_{\rm XRT} \sim 2$, indicating that the observed emission is consistent with the highest-energy component predicted by an electron synchrotron spectrum in either the slow or fast cooling regimes.  By contrast, the X-ray photon index distribution for LAT-detected bursts peaks at a harder value of $\Gamma_{\rm XRT} \sim 1.8$, again suggesting that the synchrotron spectrum's cooling break lies either between or above the XRT and LAT energy ranges for a majority of the LAT-detected bursts. 

A potentially important effect that we note is that the cooling break frequency ($\nu_{\rm c}$) in the afterglow synchrotron spectrum is expected to be very smooth and possibly extend over $\sim$2--3 decades in photon energy \citep{GranotSari2002}. Therefore, in some cases $\nu_{\rm c}$ might be either (i) near the XRT energy range, in which case $\Gamma_{\rm XRT} >  \Gamma_{1}$ will be inferred, with the spectral index measured by the LAT being $\Gamma_{\rm LAT} <  \Gamma_{2}$, resulting in a measured (or effective) spectral break $\Delta\Gamma_{\rm eff}$ that is less than the theoretical prediction, $\Delta\Gamma_{\rm eff} = \Gamma_{\rm LAT} - \Gamma_{\rm XRT} < \Gamma_{2} - \Gamma_{1} = \Delta\Gamma$, where $\Gamma_{2}$ and $\Gamma_{1}$ are the asymptotic values of the photon index above and below the cooling break, respectively, or (ii) $\nu_{\rm c}$ can be near or within the LAT energy range, in which case $\Gamma_{\rm LAT} < \Gamma_{2}$ can be inferred (while $\Gamma_{\rm XRT} = \Gamma_{1}$) so that again $\Delta\Gamma_{\rm eff}  < \Delta\Gamma$. Therefore, imposing $\Delta\Gamma  = 0.5$ with a broken power-law spectrum may result in inferred $\Gamma_{2}$ and  $\Gamma_{1}$ values that differ from their true values, and thus complicate direct comparison to the theoretical prediction for the asymptotic value of $\Gamma_{2}$, which for p $\sim$ 2--2.5, corresponds to $\Gamma_{2} \sim 2-2.25$.  

We examined the influence that a broad cooling break could have on our results by implementing the smoothly broken power-law (SBPL) spectrum described in \citep{GranotSari2002}, with a fixed sharpness of the break set to s = 0.85.  We fit this model to the XRT and LAT data for GRB~130427A and obtained consistent pre and post break photon indices of $\Gamma_{\rm BPL1} = 1.54 \pm 0.02$ and $\Gamma_{\rm BPL2} = 2.04 \pm 0.02$, whereas the SBPL model returned $\Gamma_{\rm BPL1} = 1.56 \pm 0.07$ and $\Gamma_{\rm BPL2} = 2.06 \pm 0.07$.  We conclude that the large gap in energy between the XRT and LAT data effectively mask the effects of the curvature in the break energy for the SBPL model as long as the spectral break is well within the MeV domain, resulting in asymptotic photon indices in the XRT and LAT energy ranges which are consistent with those obtained using the simpler BPL model.  We present the break energies for the six LAT detected bursts for which a BPL model was preferred over a PL model in Table 2 and show that the break energies are well above the XRT domain or below the LAT domain, with the exception of GRB 130427A, for which we explicitly fit the SBPL model and showed consistency with the simpler BPL model. 

\subsection{Constraining the Circumstellar Environment of LAT-detected GRBs} \label{sec:Discussion:InverseComptonComponents}

The value and time evolution of the cooling frequency, i$.$e$.$ the gyration frequency of an electron whose cooling time equals the dynamical time of the system, in an electron synchrotron spectrum in the slow-cooling regime is heavily dependent on the density profile $\rho_{\rm ext}(r) = A_{*}r^{-k}$ of the circumstellar medium \citep{Chevalier2000, GranotSari2002}.  The cooling frequency is expected to evolve to lower energies with time in a constant density interstellar medium (ISM) ($k = 0$) profile, and evolve to higher energies in a stellar wind ($k = 2$) environment.  

We speculate that the primary difference between the LAT-detected and non-detected populations may be in the type of circumstellar environment in which these bursts occur.  LAT detections may be preferentially selecting GRBs that occur in low wind-like circumburst density profiles for which the synchrotron cooling break begins near the X-ray regime and does not evolve to lower energies; hence the afterglow spectrum above the X-ray regime that remains spectrally hard for longer periods of time.  

The inference that LAT-detected bursts may be preferentially occurring in wind-like environments is consistent with an analysis of the multiwavelength observations of both GRB~110731A \citep{GRB110731A} and GRB~130427A \citep{GRB130427A_NuSTAR}. Using data collected by the XRT, LAT and the Nuclear Spectroscopic Telescope Array (\emph{NuSTAR}), \citet{GRB130427A_NuSTAR} found that a break between the X-ray and gamma-ray regimes best fits the broadband data for GRB~130427A at very late times.  The authors speculate that the cooling break in the afterglow spectra of GRB~130427A may not have evolved with time and remained between the XRT and LAT energy ranges due to a circumstellar density profile that is intermediate between ISM and wind-like circumstellar density profiles.  

Likewise, \citet{GRB110731A} performed broadband modeling of optical, UVOT, BAT, XRT, and LAT data associated with GRB~110731A and found that initially a single power law adequately fit the broadband SED using BAT, GBM and LAT data. At a later time a spectral break was observed between the XRT and LAT data, which was interpreted as a cooling break evolving from low to high frequencies for a GRB blast wave evolving in a wind-like environment. 
Although they concluded that an observed break between the optical and X-ray data can be best explained by the presence of a cooling break between the two regimes, the photon index of $\Gamma = 1.77$ obtained through our joint spectral fits for this burst suggests that this break lies above the LAT energy range. 
Again, the differences between the \citet{GRB110731A} work and this analysis can be likely attributed to the greater sensitivity at low energies of the Pass 8 data used in this work, although we point out that our analysis does not include fits to optical data as were performed by \citet{GRB110731A}.

A preference for LAT-detected GRBs to occur in low-density wind-like circumstellar environments was also found by \citet{Cenko2011}, who modeled the broadband spectral and temporal X-ray, optical, and radio afterglow data of four LAT-detected GRBs: GRB~090323, GRB~090328, GRB~090902B, and GRB~090926A.  The authors found that a wind environment best fit the data for all but GRB~090902B, for which a constant-density ISM environment was preferred.  In this interpretation, the relatively small number of \Swift XRT-detected bursts that have the expected afterglow behavior in a wind-like density profile \citep{Schulze2011} may further explain the relatively small number of LAT detections of bright XRT-detected afterglows. 

\subsection{Constraints on Inverse Compton Emission} \label{sec:Discussion:InverseComptonComponents}

The results summarized in Figure \ref{Fig:FluxComparison} significantly constrain the strength and ubiquity of inverse Compton (IC) emission in the 0.1 to 100 GeV energy range during the XRT and LAT observations that we considered.  Such emission is a natural consequence of nonthermal relativistic blast waves thought to power GRB afterglows, although a definitive detection of IC emission at GeV energies has been elusive in the \Fermi era. IC components can result from upscattering of soft X-ray photons external to the relativistic blast wave, external inverse Compton (EIC) \citep{Fan2006, He2012, Beloborodov2014A}, or synchrotron self-Compton (SSC) in which synchrotron-emitting electrons in the relativistic blast wave upscatter their own synchrotron radiation \citep{Dermer2000, Zhang2001, Sari2001, Wang2013}.  The lack of significant emission in the LAT energy range in excess of the flux expected from the spectra extrapolated from XRT observations requires that any accompanying IC components must be subdominant to the high-energy tail of the synchrotron spectrum, or peak above the LAT energy range we considered for this analysis.

We can examine these constraints more closely if we consider that the ratio of the peak flux of the synchrotron and SSC components, or Compton $Y$ parameter, in the slow-cooling regime, scales as $\propto (\epsilon_{e} / \epsilon_{B})^{1/2} (\gamma_{m} / \gamma_{c})^{p-2}$.  Here $\epsilon_{e}$ and $\epsilon_{B}$ are the fractional-energy densities of the relativistic electrons and magnetic field, and $\gamma_{m}$ and $\gamma_{c}$ represent the minimum injection energy and the typical electron Lorentz factor above which the relativistic electrons radiate a significant fraction of their energy on the dynamical timescale,  respectively \citep{Sari2001}.  A relativistic blast wave with a large fraction of its total energy stored in energetic electrons (large $\epsilon_{e}$) and/or low magnetic field density (extremely small $\epsilon_{B}$), is expected to generate prominent SSC emission, which is in disagreement with our observations.  This could point to a blast wave in the synchrotron-dominated regime in which a larger fraction of its total energy is stored in the magnetic field density (large $\epsilon_{B}$) \citep{Zhang2001}.  Alternatively, the blast wave could be in the Klein-Nishina dominated regime in which $Y < 1$, even though $\epsilon_{e}$/$\epsilon_{B} \gg 1$ because of the Klein-Nishina reduction to the electron-photon scattering cross-section. Both scenarios could suppress the SSC component, making it undetectable in the LAT energy range.

On the other hand, the peak frequency of the SSC component scales roughly as $E^{\rm SSC}_{pk} =  \gamma_{c}^2 E^{\rm syn}_{pk}$, with $E^{\rm syn}_{pk} = E_{c}$ in the slow-cooling regime, where $E_{c}$ is the energy of the cooling break.  Therefore, a non-detection of strong SSC emission could also imply that $E^{\rm SSC}_{pk}$ is beyond the LAT energy range we considered.  Assuming that $E^{\rm syn}_{pk}$ lies between or above the XRT and LAT energy range during our observations, this could be accommodated with a moderate value of $\gamma_{c}$ of 100--1000.  We note, though, that since the SSC component is expected to span several orders of magnitude in energy around $E^{\rm SSC}_{pk}$ \citep{Sari2001}, requiring the spectral upturn due to the SSC component to be above the LAT energy range is far more demanding.  Likewise, the non-detection of the SSC component at late times, when the cooling break has potentially evolved into the X-ray regime, places even further constraints on this scenario.

The widely discussed detection of high-energy photons with energies $>10$ GeV hours after the onset of GRB~130427A has been attributed to SSC emission by \citet{Tam2013} and \citet{Wang2013}.  \citet{GRB130427A_LAT} and \citet{GRB130427A_NuSTAR}, on the other hand, both argue that the high-energy light curve and spectra are consistent with a single electron synchrotron spectrum throughout the evolution of the extended emission.  Here we draw similar conclusions from the three intervals for which we compared the XRT and LAT data for GRB~130427A.  The extension of the XRT spectra over-predicts the emission expected in the 0.1 to 100 GeV energy range and suggests that a break exists between the two energy ranges.  Our joint spectral fit to the first of these three intervals ($t_0\sim300$ sec post trigger) shows that the broadband SED can be well described by a single electron synchrotron spectrum with a cooling break between the X-ray and gamma-ray regimes, matching the conclusions of \citet{GRB130427A_NuSTAR} at much later times.

The non-detection of IC emission is also notable in GRB~100728A and GRB~110213A, both of which were detected by the LAT and which showed energetic X-ray flares and a significant X-ray plateau lasting roughly $\sim2000$ sec, respectively.  These light curve features have been proposed to be the result of late-time energy injection due to continued activity of the central engine \citep{Burrows2005, Fan2005, Zhang2006, Panaitescu2008} and SSC emission at GeV energies could be expected in such a scenario.  For both bursts, our analysis finds that the contemporaneous XRT and LAT observations are consistent with a single spectral component.  In the case of GRB 100728A we find weak evidence of a break in the broadband spectrum, consistent with a cooling break in an electron synchrotron spectrum.  These results point to synchrotron-dominated emission during the flare and plateau afterglow components, and the non-detection of IC emission again suggests a shocked external medium with a strong magnetic field, an extremely high $\gamma_{c}$ value so as to have avoided the production of a dominant SSC component at GeV energies, or a blast wave in the Klein-Nishina dominated regime so as to suppress electron-photon scattering.

\section{Conclusions} \label{sec:Conclusions}

We have used joint observations by the \Swift XRT and the \Fermi LAT of GRB afterglows to investigate the nature of long-lived, high-energy emission observed by \Fermi LAT.  By extrapolating the XRT derived spectra of \emph{Swift}-detected GRBs, we compared the expected flux in the 0.1 to 100 GeV energy range to the LAT upper limits for the periods in which the burst position was within the LAT FoV. We found that only a small subset of bursts exhibit afterglow emission that could exceed the LAT detection threshold when extrapolated to the 0.1 to 100 GeV energy range.  Bursts that do result in late-time LAT detections are almost exclusively drawn from afterglows that exhibit emission brighter than $F_{\rm XRT} \gtrsim 10^{-10}$ erg cm$^{-2}$ s$^{-1}$ and harder than $\Gamma_{\rm XRT} \lesssim 2$.

Joint broadband spectral fits of XRT and LAT data reveal that a majority of LAT nondetections of relatively bright X-ray afterglows can be explained by an afterglow spectrum with a slightly softer photon index when constrained by both the XRT and LAT data, compared to the photon index derived by fits to the XRT data alone.  The remaining LAT nondetections are consistent with a cooling break in the predicted electron synchrotron spectrum between the XRT and LAT energy ranges.  Such a break is sufficient to suppress the high-energy emission below the LAT detection threshold.  On the other hand, the broadband spectra of LAT-detected bursts are best modeled by spectral components that indicate that the cooling break in the synchrotron spectrum lies either between or above the XRT and LAT energy ranges.  

Since the value and time evolution of the cooling frequency in an electron synchrotron spectrum is strongly dependent on the density profile of the circumstellar medium, we speculate that the primary difference between bursts with afterglow detections by the LAT and the non-detected population may be the type of circumstellar environment.  Late-time LAT detections may be preferentially selecting GRBs that occur in low-density wind-like circumburst environments for which the synchrotron cooling break begins near the X-ray regime and does not evolve to lower energies, resulting in an afterglow spectrum above the X-ray regime that remains spectrally hard for longer periods of time, enhancing the detectability of the afterglow in the LAT energy range. 

We find no evidence of high-energy emission significantly in excess of the flux expected from the spectrum predicted by the electron synchrotron model.  In addition, joint spectral fits of contemporaneous XRT and LAT observations of an episode of energetic X-ray flaring in GRB~100728A and a significant X-ray plateau in GRB~110213A find that the XRT and LAT data are consistent with a single spectral component.  The lack of excess emission at high energies points to two possibilities: 1) a shocked external medium in which the energy density in the magnetic field is elevated or comparable to that of the relativistic electrons behind the shock, precluding the production of a dominant SSC component in the LAT energy range at late times, or 2) the peak of the SSC emission is beyond the 0.1 to 100 GeV energy range we considered.

\medskip
\noindent The \textit{Fermi} LAT Collaboration acknowledges generous ongoing support
from a number of agencies and institutes that have supported both the
development and the operation of the LAT as well as scientific data analysis.
These include the National Aeronautics and Space Administration and the
Department of Energy in the United States, the Commissariat \`a l'Energie Atomique
and the Centre National de la Recherche Scientifique / Institut National de Physique
Nucl\'eaire et de Physique des Particules in France, the Agenzia Spaziale Italiana
and the Istituto Nazionale di Fisica Nucleare in Italy, the Ministry of Education,
Culture, Sports, Science and Technology (MEXT), High Energy Accelerator Research
Organization (KEK) and Japan Aerospace Exploration Agency (JAXA) in Japan, and
the K.~A.~Wallenberg Foundation, the Swedish Research Council and the
Swedish National Space Board in Sweden.
 
Additional support for science analysis during the operations phase is gratefully
acknowledged from the Istituto Nazionale di Astrofisica in Italy and the Centre
National d'\'Etudes Spatiales in France. This work performed in part under DOE
Contract DE-AC02-76SF00515.

\clearpage


\bibliography{ms}

\begin{thebibliography}{}
\expandafter\ifx\csname natexlab\endcsname\relax\def\natexlab#1{#1}\fi
\providecommand{\url}[1]{\href{#1}{#1}}
\providecommand{\dodoi}[1]{doi:~\href{http://doi.org/#1}{\nolinkurl{#1}}}
\providecommand{\doeprint}[1]{\href{http://ascl.net/#1}{\nolinkurl{http://ascl.net/#1}}}
\providecommand{\doarXiv}[1]{\href{https://arxiv.org/abs/#1}{\nolinkurl{https://arxiv.org/abs/#1}}}

\bibitem[{{Abdo} {et~al.}(2009{\natexlab{a}}){Abdo}, {Ackermann}, {Arimoto},
  {Asano}, {Atwood}, {Axelsson}, {Baldini}, {Ballet}, {Band}, {Barbiellini}, \&
  et~al.}]{Abdo2009a}
{Abdo}, A.~A., {Ackermann}, M., {Arimoto}, M., {et~al.} 2009{\natexlab{a}},
  Science, 323, 1688, \dodoi{10.1126/science.1169101}

\bibitem[{{Abdo} {et~al.}(2009{\natexlab{b}}){Abdo}, {Ackermann}, {Asano},
  {Atwood}, {Axelsson}, {Baldini}, {Ballet}, {Band}, {Barbiellini}, {Bastieri},
  {Bechtol}, {Bellazzini}, {Berenji}, {Bhat}, {Bissaldi}, {Bloom}, {Bonamente},
  {Borgland}, {Bouvier}, {Bregeon}, {Brez}, {Briggs}, {Brigida}, {Bruel},
  {Burnett}, {Caliandro}, {Cameron}, {Caraveo}, {Casandjian}, {Cecchi},
  {Chaplin}, {Chekhtman}, {Cheung}, {Chiang}, {Ciprini}, {Claus},
  {Cohen-Tanugi}, {Cominsky}, {Connaughton}, {Conrad}, {Cutini}, {Dermer}, {de
  Angelis}, {de Palma}, {Digel}, {Silva}, {Drell}, {Dubois}, {Dumora},
  {Farnier}, {Favuzzi}, {Focke}, {Frailis}, {Fukazawa}, {Fusco}, {Gargano},
  {Gasparrini}, {Gehrels}, {Germani}, {Gibby}, {Giebels}, {Giglietto},
  {Giordano}, {Glanzman}, {Godfrey}, {Goldstein}, {Granot}, {Grenier},
  {Grondin}, {Grove}, {Guillemot}, {Guiriec}, {Hanabata}, {Harding},
  {Hayashida}, {Hays}, {Hughes}, {J{\'o}hannesson}, {Johnson}, {Johnson},
  {Kamae}, {Katagiri}, {Kataoka}, {Kawai}, {Kerr}, {Kn{\"o}dlseder},
  {Kocevski}, {Komin}, {Kouveliotou}, {Kuehn}, {Kuss}, {Latronico}, {Longo},
  {Loparco}, {Lott}, {Lovellette}, {Lubrano}, {Makeev}, {Mazziotta}, {McBreen},
  {McEnery}, {McGlynn}, {Meegan}, {Meurer}, {Michelson}, {Mitthumsiri},
  {Mizuno}, {Monte}, {Monzani}, {Moretti}, {Morselli}, {Moskalenko}, {Murgia},
  {Nakamori}, {Nolan}, {Norris}, {Nuss}, {Ohno}, {Ohsugi}, {Omodei}, {Orlando},
  {Ormes}, {Ozaki}, {Paciesas}, {Paneque}, {Panetta}, {Parent}, {Pelassa},
  {Pepe}, {Pesce-Rollins}, {Piron}, {Porter}, {Preece}, {Rain{\`o}}, {Rando},
  {Razzano}, {Razzaque}, {Reimer}, {Reposeur}, {Ritz}, {Rochester},
  {Rodriguez}, {Roth}, {Ryde}, {Sadrozinski}, {Sanchez}, {Sander}, {Saz
  Parkinson}, {Scargle}, {Sgr{\`o}}, {Siskind}, {Smith}, {Smith}, {Spandre},
  {Spinelli}, {Stamatikos}, {Strickman}, {Suson}, {Tajima}, {Takahashi},
  {Tanaka}, {Thayer}, {Thayer}, {Tibaldo}, {Torres}, {Tosti}, {Tramacere},
  {Uchiyama}, {Usher}, {van der Horst}, {Vasileiou}, {Vilchez}, {Vitale}, {von
  Kienlin}, {Waite}, {Wang}, {Wilson-Hodge}, {Winer}, {Wood}, {Ylinen}, \&
  {Ziegler}}]{Abdo2009b}
{Abdo}, A.~A., {Ackermann}, M., {Asano}, K., {et~al.} 2009{\natexlab{b}}, \apj,
  707, 580, \dodoi{10.1088/0004-637X/707/1/580}

\bibitem[{{Abdo} {et~al.}(2011){Abdo}, {Ackermann}, {Ajello}, {Baldini},
  {Ballet}, {Barbiellini}, {Baring}, {Bastieri}, {Bechtol}, {Bellazzini},
  {Berenji}, {Bhat}, {Bissaldi}, {Blandford}, {Bonamente}, {Bonnell},
  {Borgland}, {Bouvier}, {Bregeon}, {Brigida}, {Bruel}, {Buehler}, {Buson},
  {Caliandro}, {Cameron}, {Caraveo}, {Casandjian}, {Cecchi}, {Charles},
  {Chekhtman}, {Chiang}, {Ciprini}, {Claus}, {Connaughton}, {Conrad}, {Cutini},
  {de Angelis}, {de Palma}, {Dermer}, {Silva}, {Drell}, {Dubois}, {Favuzzi},
  {Fukazawa}, {Fusco}, {Gargano}, {Gehrels}, {Germani}, {Giglietto}, {Giommi},
  {Giordano}, {Giroletti}, {Glanzman}, {Godfrey}, {Granot}, {Grenier},
  {Guiriec}, {Hadasch}, {Hanabata}, {Hughes}, {J{\'o}hannesson}, {Johnson},
  {Kamae}, {Katagiri}, {Kataoka}, {Kerr}, {Kn{\"o}dlseder}, {Kuss}, {Lande},
  {Latronico}, {Lee}, {Longo}, {Loparco}, {Lott}, {Lubrano}, {Mazziotta},
  {McEnery}, {M{\'e}sz{\'a}ros}, {Michelson}, {Mizuno}, {Moiseev}, {Monzani},
  {Morselli}, {Moskalenko}, {Murgia}, {Nakamori}, {Naumann-Godo}, {Nolan},
  {Norris}, {Nuss}, {Ohsugi}, {Okumura}, {Omodei}, {Orlando}, {Paciesas},
  {Pelassa}, {Pesce-Rollins}, {Pierbattista}, {Piron}, {Porter}, {Racusin},
  {Rain{\`o}}, {Razzano}, {Razzaque}, {Reimer}, {Reimer}, {Reyes}, {Roth},
  {Sadrozinski}, {Sgr{\`o}}, {Siskind}, {Smith}, {Sonbas}, {Spandre},
  {Spinelli}, {Stamatikos}, {Strickman}, {Takahashi}, {Tanaka}, {Tanaka},
  {Thayer}, {Thayer}, {Torres}, {Tosti}, {Troja}, {Uehara}, {Usher},
  {Vandenbroucke}, {Vasileiou}, {Vianello}, {Vilchez}, {Vitale}, {von Kienlin},
  {Waite}, {Wang}, {Winer}, {Wood}, {Yamazaki}, {Yang}, {Ziegler}, {Piro}, \&
  {Fermi Collaboration}}]{Abdo2011}
{Abdo}, A.~A., {Ackermann}, M., {Ajello}, M., {et~al.} 2011, \apjl, 734, L27,
  \dodoi{10.1088/2041-8205/734/2/L27}

\bibitem[{{Acero} {et~al.}(2015){Acero}, {Ackermann}, {Ajello}, {Albert},
  {Atwood}, {Axelsson}, {Baldini}, {Ballet}, {Barbiellini}, {Bastieri},
  {Belfiore}, {Bellazzini}, {Bissaldi}, {Blandford}, {Bloom}, {Bogart},
  {Bonino}, {Bottacini}, {Bregeon}, {Britto}, {Bruel}, {Buehler}, {Burnett},
  {Buson}, {Caliandro}, {Cameron}, {Caputo}, {Caragiulo}, {Caraveo},
  {Casandjian}, {Cavazzuti}, {Charles}, {Chaves}, {Chekhtman}, {Cheung},
  {Chiang}, {Chiaro}, {Ciprini}, {Claus}, {Cohen-Tanugi}, {Cominsky}, {Conrad},
  {Cutini}, {D'Ammando}, {de Angelis}, {DeKlotz}, {de Palma}, {Desiante},
  {Digel}, {Di Venere}, {Drell}, {Dubois}, {Dumora}, {Favuzzi}, {Fegan},
  {Ferrara}, {Finke}, {Franckowiak}, {Fukazawa}, {Funk}, {Fusco}, {Gargano},
  {Gasparrini}, {Giebels}, {Giglietto}, {Giommi}, {Giordano}, {Giroletti},
  {Glanzman}, {Godfrey}, {Grenier}, {Grondin}, {Grove}, {Guillemot}, {Guiriec},
  {Hadasch}, {Harding}, {Hays}, {Hewitt}, {Hill}, {Horan}, {Iafrate}, {Jogler},
  {J{\'o}hannesson}, {Johnson}, {Johnson}, {Johnson}, {Johnson}, {Kamae},
  {Kataoka}, {Katsuta}, {Kuss}, {La Mura}, {Landriu}, {Larsson}, {Latronico},
  {Lemoine-Goumard}, {Li}, {Li}, {Longo}, {Loparco}, {Lott}, {Lovellette},
  {Lubrano}, {Madejski}, {Massaro}, {Mayer}, {Mazziotta}, {McEnery},
  {Michelson}, {Mirabal}, {Mizuno}, {Moiseev}, {Mongelli}, {Monzani},
  {Morselli}, {Moskalenko}, {Murgia}, {Nuss}, {Ohno}, {Ohsugi}, {Omodei},
  {Orienti}, {Orlando}, {Ormes}, {Paneque}, {Panetta}, {Perkins},
  {Pesce-Rollins}, {Piron}, {Pivato}, {Porter}, {Racusin}, {Rando}, {Razzano},
  {Razzaque}, {Reimer}, {Reimer}, {Reposeur}, {Rochester}, {Romani},
  {Salvetti}, {S{\'a}nchez-Conde}, {Saz Parkinson}, {Schulz}, {Siskind},
  {Smith}, {Spada}, {Spandre}, {Spinelli}, {Stephens}, {Strong}, {Suson},
  {Takahashi}, {Takahashi}, {Tanaka}, {Thayer}, {Thayer}, {Thompson},
  {Tibaldo}, {Tibolla}, {Torres}, {Torresi}, {Tosti}, {Troja}, {Van Klaveren},
  {Vianello}, {Winer}, {Wood}, {Wood}, {Zimmer}, \& {Fermi-LAT
  Collaboration}}]{Acero2015}
{Acero}, F., {Ackermann}, M., {Ajello}, M., {et~al.} 2015, \apjs, 218, 23,
  \dodoi{10.1088/0067-0049/218/2/23}

\bibitem[{{Ackermann} {et~al.}(2010){Ackermann}, {Ajello}, {Baldini}, {Ballet},
  {Barbiellini}, {Baring}, {Bastieri}, {Bechtol}, {Bellazzini}, {Berenji},
  {Bhat}, {Bissaldi}, {Blandford}, {Bonamente}, {Borgland}, {Bouvier},
  {Bregeon}, {Brez}, {Briggs}, {Brigida}, {Bruel}, {Buehler}, {Buson},
  {Caliandro}, {Cameron}, {Caraveo}, {Carrigan}, {Casandjian}, {Cecchi}, {{\c
  C}elik}, {Charles}, {Chekhtman}, {Chiang}, {Ciprini}, {Claus},
  {Cohen-Tanugi}, {Connaughton}, {Conrad}, {Cutini}, {Dermer}, {de Angelis},
  {de Palma}, {Digel}, {Silva}, {Drell}, {Dubois}, {Favuzzi}, {Fegan},
  {Ferrara}, {Frailis}, {Fukazawa}, {Fusco}, {Gargano}, {Gasparrini},
  {Gehrels}, {Germani}, {Giglietto}, {Giommi}, {Giordano}, {Giroletti},
  {Glanzman}, {Godfrey}, {Granot}, {Grenier}, {Grove}, {Guillemot}, {Guiriec},
  {Hadasch}, {Hays}, {Horan}, {Hughes}, {J{\'o}hannesson}, {Johnson},
  {Johnson}, {Kamae}, {Katagiri}, {Kippen}, {Kn{\"o}dlseder}, {Kocevski},
  {Kuss}, {Lande}, {Latronico}, {Lee}, {Llena Garde}, {Longo}, {Loparco},
  {Lovellette}, {Lubrano}, {Makeev}, {Mazziotta}, {McBreen}, {McEnery},
  {McGlynn}, {Meegan}, {Mehault}, {M{\'e}sz{\'a}ros}, {Michelson}, {Mizuno},
  {Moiseev}, {Monte}, {Monzani}, {Moretti}, {Morselli}, {Moskalenko}, {Murgia},
  {Nakajima}, {Nakamori}, {Naumann-Godo}, {Nolan}, {Norris}, {Nuss}, {Ohno},
  {Ohsugi}, {Okumura}, {Omodei}, {Orlando}, {Ormes}, {Ozaki}, {Paciesas},
  {Paneque}, {Panetta}, {Parent}, {Pelassa}, {Pepe}, {Pesce-Rollins},
  {Petrosian}, {Piron}, {Porter}, {Preece}, {Racusin}, {Rain{\`o}}, {Rando},
  {Rau}, {Razzano}, {Razzaque}, {Reimer}, {Reimer}, {Ripken}, {Roth}, {Ryde},
  {Sadrozinski}, {Sander}, {Scargle}, {Schalk}, {Sgr{\`o}}, {Siskind}, {Smith},
  {Spandre}, {Spinelli}, {Stamatikos}, {Strickman}, {Suson}, {Tajima},
  {Takahashi}, {Tanaka}, {Thayer}, {Thayer}, {Tibaldo}, {Torres}, {Tosti},
  {Tramacere}, {Uehara}, {Usher}, {Vandenbroucke}, {van der Horst},
  {Vasileiou}, {Vilchez}, {Vitale}, {von Kienlin}, {Waite}, {Wang},
  {Wilson-Hodge}, {Winer}, {Wood}, {Wu}, {Yamazaki}, {Yang}, {Ylinen},
  {Ziegler}, {Fermi LAT Collaboration}, \& {Fermi GBM
  Collaboration}}]{Ackermann2010}
{Ackermann}, M., {Ajello}, M., {Baldini}, L., {et~al.} 2010, \apjl, 717, L127,
  \dodoi{10.1088/2041-8205/717/2/L127}

\bibitem[{{Ackermann} {et~al.}(2012{\natexlab{a}}){Ackermann}, {Ajello},
  {Albert}, {Allafort}, {Atwood}, {Axelsson}, {Baldini}, {Ballet},
  {Barbiellini}, {Bastieri}, {Bechtol}, {Bellazzini}, {Bissaldi}, {Blandford},
  {Bloom}, {Bogart}, {Bonamente}, {Borgland}, {Bottacini}, {Bouvier}, {Brandt},
  {Bregeon}, {Brigida}, {Bruel}, {Buehler}, {Burnett}, {Buson}, {Caliandro},
  {Cameron}, {Caraveo}, {Casandjian}, {Cavazzuti}, {Cecchi}, {{\c C}elik},
  {Charles}, {Chaves}, {Chekhtman}, {Cheung}, {Chiang}, {Ciprini}, {Claus},
  {Cohen-Tanugi}, {Conrad}, {Corbet}, {Cutini}, {D'Ammando}, {Davis}, {de
  Angelis}, {DeKlotz}, {de Palma}, {Dermer}, {Digel}, {Silva}, {Drell},
  {Drlica-Wagner}, {Dubois}, {Favuzzi}, {Fegan}, {Ferrara}, {Focke}, {Fortin},
  {Fukazawa}, {Funk}, {Fusco}, {Gargano}, {Gasparrini}, {Gehrels}, {Giebels},
  {Giglietto}, {Giordano}, {Giroletti}, {Glanzman}, {Godfrey}, {Grenier},
  {Grove}, {Guiriec}, {Hadasch}, {Hayashida}, {Hays}, {Horan}, {Hou}, {Hughes},
  {Jackson}, {Jogler}, {J{\'o}hannesson}, {Johnson}, {Johnson}, {Johnson},
  {Kamae}, {Katagiri}, {Kataoka}, {Kerr}, {Kn{\"o}dlseder}, {Kuss}, {Lande},
  {Larsson}, {Latronico}, {Lavalley}, {Lemoine-Goumard}, {Longo}, {Loparco},
  {Lott}, {Lovellette}, {Lubrano}, {Mazziotta}, {McConville}, {McEnery},
  {Mehault}, {Michelson}, {Mitthumsiri}, {Mizuno}, {Moiseev}, {Monte},
  {Monzani}, {Morselli}, {Moskalenko}, {Murgia}, {Naumann-Godo}, {Nemmen},
  {Nishino}, {Norris}, {Nuss}, {Ohno}, {Ohsugi}, {Okumura}, {Omodei},
  {Orienti}, {Orlando}, {Ormes}, {Paneque}, {Panetta}, {Perkins},
  {Pesce-Rollins}, {Pierbattista}, {Piron}, {Pivato}, {Porter}, {Racusin},
  {Rain{\`o}}, {Rando}, {Razzano}, {Razzaque}, {Reimer}, {Reimer}, {Reposeur},
  {Reyes}, {Ritz}, {Rochester}, {Romoli}, {Roth}, {Sadrozinski}, {Sanchez},
  {Saz Parkinson}, {Sbarra}, {Scargle}, {Sgr{\`o}}, {Siegal-Gaskins},
  {Siskind}, {Spandre}, {Spinelli}, {Stephens}, {Suson}, {Tajima}, {Takahashi},
  {Tanaka}, {Thayer}, {Thayer}, {Thompson}, {Tibaldo}, {Tinivella}, {Tosti},
  {Troja}, {Usher}, {Vandenbroucke}, {Van Klaveren}, {Vasileiou}, {Vianello},
  {Vitale}, {Waite}, {Wallace}, {Winer}, {Wood}, {Wood}, {Wood}, {Yang}, \&
  {Zimmer}}]{LATPerformancePaper}
{Ackermann}, M., {Ajello}, M., {Albert}, A., {et~al.} 2012{\natexlab{a}},
  \apjs, 203, 4, \dodoi{10.1088/0067-0049/203/1/4}

\bibitem[{{Ackermann} {et~al.}(2012{\natexlab{b}}){Ackermann}, {Ajello},
  {Baldini}, {Barbiellini}, {Baring}, {Bechtol}, {Bellazzini}, {Blandford},
  {Bloom}, {Bonamente}, {Borgland}, {Bottacini}, {Bouvier}, {Brigida},
  {Buehler}, {Buson}, {Caliandro}, {Cameron}, {Cecchi}, {Charles}, {Chekhtman},
  {Chiang}, {Ciprini}, {Claus}, {Cohen-Tanugi}, {Cutini}, {D'Ammando}, {de
  Palma}, {Dermer}, {Silva}, {Drell}, {Drlica-Wagner}, {Favuzzi}, {Fukazawa},
  {Fusco}, {Gargano}, {Gasparrini}, {Gehrels}, {Germani}, {Giglietto},
  {Giordano}, {Giroletti}, {Glanzman}, {Granot}, {Grenier}, {Grove}, {Hadasch},
  {Hanabata}, {Harding}, {Hays}, {Horan}, {J{\'o}hannesson}, {Kataoka},
  {Kn{\"o}dlseder}, {Kocevski}, {Kuss}, {Lande}, {Longo}, {Loparco},
  {Lovellette}, {Lubrano}, {Mazziotta}, {McEnery}, {McGlynn}, {Michelson},
  v~{Mitthumsiri}, {Monzani}, {Moretti}, {Morselli}, {Moskalenko}, {Murgia},
  {Naumann-Godo}, {Norris}, {Nuss}, {Nymark}, {Ohsugi}, {Okumura}, {Omodei},
  {Orlando}, {Panetta}, {Parent}, {Pelassa}, {Pesce-Rollins}, {Piron},
  {Pivato}, {Racusin}, {Rain{\`o}}, {Rando}, {Razzaque}, {Reimer}, {Reimer},
  {Ritz}, {Ryde}, {Sgr{\`o}}, {Siskind}, {Sonbas}, {Spandre}, {Spinelli},
  {Stamatikos}, {Stawarz}, {Suson}, {Takahashi}, {Tanaka}, {Thayer}, {Thayer},
  {Tibaldo}, {Tinivella}, {Tosti}, {Uehara}, {Vandenbroucke}, {Vasileiou},
  {Vianello}, {Vitale}, {Waite}, {Fermi Gamma-ray Burst Monitor Team},
  {Connaughton}, {Briggs}, {Guirec}, {Goldstein}, {Burgess}, {Bhat},
  {Bissaldi}, {Camero-Arranz}, {Fishman}, {Fitzpatrick}, {Foley}, {Gruber},
  {Jenke}, {Kippen}, {Kouveliotou}, {McBreen}, {Meegan}, {Paciesas}, {Preece},
  {Rau}, {Tierney}, {van der Horst}, {von Kienlin}, {Wilson-Hodge}, \&
  {Xiong}}]{Ackermann2012}
{Ackermann}, M., {Ajello}, M., {Baldini}, L., {et~al.} 2012{\natexlab{b}},
  \apj, 754, 121, \dodoi{10.1088/0004-637X/754/2/121}

\bibitem[{{Ackermann} {et~al.}(2013{\natexlab{a}}){Ackermann}, {Ajello},
  {Asano}, {Baldini}, {Barbiellini}, {Baring}, {Bastieri}, {Bellazzini},
  {Blandford}, {Bonamente}, {Borgland}, {Bottacini}, {Bregeon}, {Brigida},
  {Bruel}, {Buehler}, {Buson}, {Caliandro}, {Cameron}, {Caraveo}, {Cecchi},
  {Charles}, {Chaves}, {Chekhtman}, {Chiang}, {Ciprini}, {Claus},
  {Cohen-Tanugi}, {Conrad}, {Cutini}, {D'Ammando}, {de Angelis}, {de Palma},
  {Dermer}, {Silva}, {Drell}, {Drlica-Wagner}, {Favuzzi}, {Fegan}, {Focke},
  {Franckowiak}, {Fukazawa}, {Fusco}, {Gargano}, {Gasparrini}, {Gehrels},
  {Giglietto}, {Giordano}, {Giroletti}, {Glanzman}, {Godfrey}, {Granot},
  {Greiner}, {Grenier}, {Grove}, {Guiriec}, {Hadasch}, {Hanabata}, {Hayashida},
  {Hays}, {Hughes}, {Jackson}, {Jogler}, {J{\'o}hannesson}, {Johnson},
  {Kn{\"o}dlseder}, {Kocevski}, {Kuss}, {Lande}, {Larsson}, {Latronico},
  {Longo}, {Loparco}, {Lovellette}, {Lubrano}, {Mazziotta}, {McEnery},
  {Mehault}, {M{\'e}sz{\'a}ros}, {Michelson}, {Mitthumsiri}, {Mizuno}, {Monte},
  {Monzani}, {Moretti}, {Morselli}, {Moskalenko}, {Murgia}, {Naumann-Godo},
  {Norris}, {Nuss}, {Nymark}, {Ohno}, {Ohsugi}, {Omodei}, {Orienti}, {Orlando},
  {Paneque}, {Perkins}, {Pesce-Rollins}, {Piron}, {Pivato}, {Racusin},
  {Rain{\`o}}, {Rando}, {Razzano}, {Razzaque}, {Reimer}, {Reimer}, {Romoli},
  {Roth}, {Ryde}, {Sanchez}, {Sgr{\`o}}, {Siskind}, {Sonbas}, {Spinelli},
  {Stamatikos}, {Takahashi}, {Tanaka}, {Thayer}, {Thayer}, {Tibaldo},
  {Tinivella}, {Tosti}, {Troja}, {Usher}, {Vandenbroucke}, {Vasileiou},
  {Vianello}, {Vitale}, {Waite}, {Winer}, {Wood}, {Yang}, {Gruber}, {Bhat},
  {Bissaldi}, {Briggs}, {Burgess}, {Connaughton}, {Foley}, {Kippen},
  {Kouveliotou}, {McBreen}, {McGlynn}, {Paciesas}, {Pelassa}, {Preece}, {Rau},
  {van der Horst}, {von Kienlin}, {Kann}, {Filgas}, {Klose}, {Kr{\"u}hler},
  {Fukui}, {Sako}, {Tristram}, {Oates}, {Ukwatta}, \&
  {Littlejohns}}]{GRB110731A}
{Ackermann}, M., {Ajello}, M., {Asano}, K., {et~al.} 2013{\natexlab{a}}, \apj,
  763, 71, \dodoi{10.1088/0004-637X/763/2/71}

\bibitem[{{Ackermann} {et~al.}(2013{\natexlab{b}}){Ackermann}, {Ajello},
  {Asano}, {Axelsson}, {Baldini}, {Ballet}, {Barbiellini}, {Bastieri},
  {Bechtol}, {Bellazzini}, {Bhat}, {Bissaldi}, {Bloom}, {Bonamente}, {Bonnell},
  {Bouvier}, {Brandt}, {Bregeon}, {Brigida}, {Bruel}, {Buehler}, {Burgess},
  {Buson}, {Byrne}, {Caliandro}, {Cameron}, {Caraveo}, {Cecchi}, {Charles},
  {Chaves}, {Chekhtman}, {Chiang}, {Chiaro}, {Ciprini}, {Claus},
  {Cohen-Tanugi}, {Connaughton}, {Conrad}, {Cutini}, {D'Ammando}, {de Angelis},
  {de Palma}, {Dermer}, {Desiante}, {Digel}, {Dingus}, {Di Venere}, {Drell},
  {Drlica-Wagner}, {Dubois}, {Favuzzi}, {Ferrara}, {Fitzpatrick}, {Foley},
  {Franckowiak}, {Fukazawa}, {Fusco}, {Gargano}, {Gasparrini}, {Gehrels},
  {Germani}, {Giglietto}, {Giommi}, {Giordano}, {Giroletti}, {Glanzman},
  {Godfrey}, {Goldstein}, {Granot}, {Grenier}, {Grove}, {Gruber}, {Guiriec},
  {Hadasch}, {Hanabata}, {Hayashida}, {Horan}, {Hou}, {Hughes}, {Inoue},
  {Jackson}, {Jogler}, {J{\'o}hannesson}, {Johnson}, {Johnson}, {Kamae},
  {Kataoka}, {Kawano}, {Kippen}, {Kn{\"o}dlseder}, {Kocevski}, {Kouveliotou},
  {Kuss}, {Lande}, {Larsson}, {Latronico}, {Lee}, {Longo}, {Loparco},
  {Lovellette}, {Lubrano}, {Massaro}, {Mayer}, {Mazziotta}, {McBreen},
  {McEnery}, {McGlynn}, {Michelson}, {Mizuno}, {Moiseev}, {Monte}, {Monzani},
  {Moretti}, {Morselli}, {Murgia}, {Nemmen}, {Nuss}, {Nymark}, {Ohno},
  {Ohsugi}, {Omodei}, {Orienti}, {Orlando}, {Paciesas}, {Paneque}, {Panetta},
  {Pelassa}, {Perkins}, {Pesce-Rollins}, {Piron}, {Pivato}, {Porter}, {Preece},
  {Racusin}, {Rain{\`o}}, {Rando}, {Rau}, {Razzano}, {Razzaque}, {Reimer},
  {Reimer}, {Reposeur}, {Ritz}, {Romoli}, {Roth}, {Ryde}, {Saz Parkinson},
  {Schalk}, {Sgr{\`o}}, {Siskind}, {Sonbas}, {Spandre}, {Spinelli}, {Suson},
  {Tajima}, {Takahashi}, {Takeuchi}, {Tanaka}, {Thayer}, {Thayer}, {Thompson},
  {Tibaldo}, {Tierney}, {Tinivella}, {Torres}, {Tosti}, {Troja}, {Tronconi},
  {Usher}, {Vandenbroucke}, {van der Horst}, {Vasileiou}, {Vianello}, {Vitale},
  {von Kienlin}, {Winer}, {Wood}, {Wood}, {Xiong}, \& {Yang}}]{Ackermann2013}
---. 2013{\natexlab{b}}, \apjs, 209, 11, \dodoi{10.1088/0067-0049/209/1/11}

\bibitem[{{Ackermann} {et~al.}(2014){Ackermann}, {Ajello}, {Asano}, {Atwood},
  {Axelsson}, {Baldini}, {Ballet}, {Barbiellini}, {Baring}, {Bastieri},
  {Bechtol}, {Bellazzini}, {Bissaldi}, {Bonamente}, {Bregeon}, {Brigida},
  {Bruel}, {Buehler}, {Burgess}, {Buson}, {Caliandro}, {Cameron}, {Caraveo},
  {Cecchi}, {Chaplin}, {Charles}, {Chekhtman}, {Cheung}, {Chiang}, {Chiaro},
  {Ciprini}, {Claus}, {Cleveland}, {Cohen-Tanugi}, {Collazzi}, {Cominsky},
  {Connaughton}, {Conrad}, {Cutini}, {D'Ammando}, {de Angelis}, {DeKlotz}, {de
  Palma}, {Dermer}, {Desiante}, {Diekmann}, {Di Venere}, {Drell},
  {Drlica-Wagner}, {Favuzzi}, {Fegan}, {Ferrara}, {Finke}, {Fitzpatrick},
  {Focke}, {Franckowiak}, {Fukazawa}, {Funk}, {Fusco}, {Gargano}, {Gehrels},
  {Germani}, {Gibby}, {Giglietto}, {Giles}, {Giordano}, {Giroletti}, {Godfrey},
  {Granot}, {Grenier}, {Grove}, {Gruber}, {Guiriec}, {Hadasch}, {Hanabata},
  {Harding}, {Hayashida}, {Hays}, {Horan}, {Hughes}, {Inoue}, {Jogler},
  {J{\'o}hannesson}, {Johnson}, {Kawano}, {Kn{\"o}dlseder}, {Kocevski}, {Kuss},
  {Lande}, {Larsson}, {Latronico}, {Longo}, {Loparco}, {Lovellette}, {Lubrano},
  {Mayer}, {Mazziotta}, {McEnery}, {Michelson}, {Mizuno}, {Moiseev}, {Monzani},
  {Moretti}, {Morselli}, {Moskalenko}, {Murgia}, {Nemmen}, {Nuss}, {Ohno},
  {Ohsugi}, {Okumura}, {Omodei}, {Orienti}, {Paneque}, {Pelassa}, {Perkins},
  {Pesce-Rollins}, {Petrosian}, {Piron}, {Pivato}, {Porter}, {Racusin},
  {Rain{\`o}}, {Rando}, {Razzano}, {Razzaque}, {Reimer}, {Reimer}, {Ritz},
  {Roth}, {Ryde}, {Sartori}, {Parkinson}, {Scargle}, {Schulz}, {Sgr{\`o}},
  {Siskind}, {Sonbas}, {Spandre}, {Spinelli}, {Tajima}, {Takahashi}, {Thayer},
  {Thayer}, {Thompson}, {Tibaldo}, {Tinivella}, {Torres}, {Tosti}, {Troja},
  {Usher}, {Vandenbroucke}, {Vasileiou}, {Vianello}, {Vitale}, {Winer}, {Wood},
  {Yamazaki}, {Younes}, {Yu}, {Zhu}, {Bhat}, {Briggs}, {Byrne}, {Foley},
  {Goldstein}, {Jenke}, {Kippen}, {Kouveliotou}, {McBreen}, {Meegan},
  {Paciesas}, {Preece}, {Rau}, {Tierney}, {van der Horst}, {von Kienlin},
  {Wilson-Hodge}, {Xiong}, {Cusumano}, {La Parola}, \&
  {Cummings}}]{GRB130427A_LAT}
---. 2014, Science, 343, 42, \dodoi{10.1126/science.1242353}

\bibitem[{{Ackermann} {et~al.}(2016){Ackermann}, {Ajello}, {Anderson},
  {Atwood}, {Axelsson}, {Baldini}, {Barbiellini}, {Bastieri}, {Bellazzini},
  {Bhat}, {Bissaldi}, {Bonino}, {Bottacini}, {Brandt}, {Bregeon}, {Bruel},
  {Buehler}, {Buson}, {Caliandro}, {Cameron}, {Caragiulo}, {Caraveo}, {Cecchi},
  {Charles}, {Chekhtman}, {Chiang}, {Chiaro}, {Ciprini}, {Claus},
  {Cohen-Tanugi}, {Conrad}, {Cutini}, {D'Ammando}, {de Angelis}, {de Palma},
  {Desiante}, {Di Venere}, {Drell}, {Favuzzi}, {Focke}, {Franckowiak}, {Funk},
  {Fusco}, {Gargano}, {Gasparrini}, {Gehrels}, {Giglietto}, {Giordano},
  {Giroletti}, {Godfrey}, {Grenier}, {Grove}, {Guiriec}, {Hewitt}, {Hill},
  {Horan}, {J{\'o}hannesson}, {Kocevski}, {Kouveliotou}, {Kuss}, {Larsson},
  {Li}, {Li}, {Longo}, {Loparco}, {Lovellette}, {Lubrano}, {Mayer},
  {Mazziotta}, {McEnery}, {Michelson}, {Mizuno}, {Monzani}, {Morselli},
  {Murgia}, {Nemmen}, {Nuss}, {Ohno}, {Ohsugi}, {Omodei}, {Orienti}, {Orlando},
  {Paneque}, {Perkins}, {Pesce-Rollins}, {Piron}, {Pivato}, {Porter},
  {Racusin}, {Rain{\`o}}, {Rando}, {Razzano}, {Reimer}, {Reimer}, {Schaal},
  {Schulz}, {Sgr{\`o}}, {Siskind}, {Spada}, {Spandre}, {Spinelli}, {Takahashi},
  {Thayer}, {Tibaldo}, {Tinivella}, {Torres}, {Tosti}, {Troja}, {Vianello},
  {von Kienlin}, {Werner}, \& {Wood}}]{LATStackingAnalysis}
{Ackermann}, M., {Ajello}, M., {Anderson}, B., {et~al.} 2016, \apj, 822, 68,
  \dodoi{10.3847/0004-637X/822/2/68}

\bibitem[{{Arnaud}(1996)}]{1996XspecProc}
{Arnaud}, K.~A. 1996, in Astronomical Society of the Pacific Conference Series,
  Vol. 101, Astronomical Data Analysis Software and Systems V, ed. G.~H.
  {Jacoby} \& J.~{Barnes}, 17

\bibitem[{{Atwood} {et~al.}(2009){Atwood}, {Abdo}, {Ackermann}, {Althouse},
  {Anderson}, {Axelsson}, {Baldini}, {Ballet}, {Band}, {Barbiellini}, \&
  et~al.}]{Atwood:09}
{Atwood}, W.~B., {Abdo}, A.~A., {Ackermann}, M., {et~al.} 2009, \apj, 697,
  1071, \dodoi{10.1088/0004-637X/697/2/1071}

\bibitem[{{Barthelmy} {et~al.}(2005){Barthelmy}, {Barbier}, {Cummings},
  {Fenimore}, {Gehrels}, {Hullinger}, {Krimm}, {Markwardt}, {Palmer},
  {Parsons}, {Sato}, {Suzuki}, {Takahashi}, {Tashiro}, \&
  {Tueller}}]{Barthelmy05}
{Barthelmy}, S.~D., {Barbier}, L.~M., {Cummings}, J.~R., {et~al.} 2005, \ssr,
  120, 143, \dodoi{10.1007/s11214-005-5096-3}

\bibitem[{{Beloborodov} {et~al.}(2014{\natexlab{a}}){Beloborodov},
  {Hasco{\"e}t}, \& {Vurm}}]{Beloborodov2014}
{Beloborodov}, A.~M., {Hasco{\"e}t}, R., \& {Vurm}, I. 2014{\natexlab{a}},
  \apj, 788, 36, \dodoi{10.1088/0004-637X/788/1/36}

\bibitem[{{Beloborodov} {et~al.}(2014{\natexlab{b}}){Beloborodov},
  {Hasco{\"e}t}, \& {Vurm}}]{Beloborodov2014A}
---. 2014{\natexlab{b}}, \apj, 788, 36, \dodoi{10.1088/0004-637X/788/1/36}

\bibitem[{{Beniamini} {et~al.}(2011){Beniamini}, {Guetta}, {Nakar}, \&
  {Piran}}]{Beniamini2011}
{Beniamini}, P., {Guetta}, D., {Nakar}, E., \& {Piran}, T. 2011, \mnras, 416,
  3089, \dodoi{10.1111/j.1365-2966.2011.19259.x}

\bibitem[{{Burrows} {et~al.}(2005{\natexlab{a}}){Burrows}, {Hill}, {Nousek},
  {Kennea}, {Wells}, {Osborne}, {Abbey}, {Beardmore}, {Mukerjee}, {Short},
  {Chincarini}, {Campana}, {Citterio}, {Moretti}, {Pagani}, {Tagliaferri},
  {Giommi}, {Capalbi}, {Tamburelli}, {Angelini}, {Cusumano}, {Br{\"a}uninger},
  {Burkert}, \& {Hartner}}]{Burrows05}
{Burrows}, D.~N., {Hill}, J.~E., {Nousek}, J.~A., {et~al.} 2005{\natexlab{a}},
  \ssr, 120, 165, \dodoi{10.1007/s11214-005-5097-2}

\bibitem[{{Burrows} {et~al.}(2005{\natexlab{b}}){Burrows}, {Romano}, {Falcone},
  {Kobayashi}, {Zhang}, {Moretti}, {O'Brien}, {Goad}, {Campana}, {Page},
  {Angelini}, {Barthelmy}, {Beardmore}, {Capalbi}, {Chincarini}, {Cummings},
  {Cusumano}, {Fox}, {Giommi}, {Hill}, {Kennea}, {Krimm}, {Mangano},
  {Marshall}, {M{\'e}sz{\'a}ros}, {Morris}, {Nousek}, {Osborne}, {Pagani},
  {Perri}, {Tagliaferri}, {Wells}, {Woosley}, \& {Gehrels}}]{Burrows2005}
{Burrows}, D.~N., {Romano}, P., {Falcone}, A., {et~al.} 2005{\natexlab{b}},
  Science, 309, 1833, \dodoi{10.1126/science.1116168}

\bibitem[{{Cenko} {et~al.}(2011){Cenko}, {Frail}, {Harrison}, {Haislip},
  {Reichart}, {Butler}, {Cobb}, {Cucchiara}, {Berger}, {Bloom}, {Chandra},
  {Fox}, {Perley}, {Prochaska}, {Filippenko}, {Glazebrook}, {Ivarsen},
  {Kasliwal}, {Kulkarni}, {LaCluyze}, {Lopez}, {Morgan}, {Pettini}, \&
  {Rana}}]{Cenko2011}
{Cenko}, S.~B., {Frail}, D.~A., {Harrison}, F.~A., {et~al.} 2011, \apj, 732,
  29, \dodoi{10.1088/0004-637X/732/1/29}

\bibitem[{{Chevalier} \& {Li}(2000)}]{Chevalier2000}
{Chevalier}, R.~A., \& {Li}, Z.-Y. 2000, \apj, 536, 195, \dodoi{10.1086/308914}

\bibitem[{{De Pasquale} {et~al.}(2010){De Pasquale}, {Schady}, {Kuin}, {Page},
  {Curran}, {Zane}, {Oates}, {Holland}, {Breeveld}, {Hoversten}, {Chincarini},
  {Grupe}, {Abdo}, {Ackermann}, {Ajello}, {Axelsson}, {Baldini}, {Ballet},
  {Barbiellini}, {Baring}, {Bastieri}, {Bechtol}, {Bellazzini}, {Berenji},
  {Bissaldi}, {Blandford}, {Bloom}, {Bonamente}, {Borgland}, {Bouvier},
  {Bregeon}, {Brez}, {Briggs}, {Brigida}, {Bruel}, {Burnett}, {Buson},
  {Caliandro}, {Cameron}, {Caraveo}, {Carrigan}, {Casandjian}, {Cecchi}, {{\c
  C}elik}, {Chekhtman}, {Chiang}, {Ciprini}, {Claus}, {Cohen-Tanugi},
  {Connaughton}, {Conrad}, {Dermer}, {de Angelis}, {de Palma}, {Dingus},
  {Silva}, {Drell}, {Dubois}, {Dumora}, {Farnier}, {Favuzzi}, {Fegan},
  {Fishman}, {Focke}, {Frailis}, {Fukazawa}, {Funk}, {Fusco}, {Gargano},
  {Gasparrini}, {Gehrels}, {Germani}, {Giglietto}, {Giordano}, {Glanzman},
  {Godfrey}, {Granot}, {Greiner}, {Grenier}, {Grove}, {Guillemot}, {Guiriec},
  {Harding}, {Hayashida}, {Hays}, {Horan}, {Hughes}, {Jackson},
  {J{\'o}hannesson}, {Johnson}, {Johnson}, {Kamae}, {Katagiri}, {Kataoka},
  {Kawai}, {Kerr}, {Kippen}, {Kn{\"o}dlseder}, {Kocevski}, {Kuss}, {Lande},
  {Latronico}, {Lemoine-Goumard}, {Longo}, {Loparco}, {Lott}, {Lovellette},
  {Lubrano}, {Makeev}, {Mazziotta}, {McEnery}, {McGlynn}, {Meegan},
  {M{\'e}sz{\'a}ros}, {Meurer}, {Michelson}, {Mitthumsiri}, {Mizuno}, {Monte},
  {Monzani}, {Moretti}, {Morselli}, {Moskalenko}, {Murgia}, {Nolan}, {Norris},
  {Nuss}, {Ohno}, {Ohsugi}, {Omodei}, {Orlando}, {Ormes}, {Paciesas},
  {Paneque}, {Panetta}, {Parent}, {Pelassa}, {Pepe}, {Pesce-Rollins}, {Piron},
  {Porter}, {Preece}, {Rain{\`o}}, {Rando}, {Razzano}, {Reimer}, {Reimer},
  {Reposeur}, {Ritz}, {Rochester}, {Rodriguez}, {Roth}, {Ryde}, {Sadrozinski},
  {Sander}, {Saz Parkinson}, {Scargle}, {Schalk}, {Sgr{\`o}}, {Siskind},
  {Smith}, {Spandre}, {Spinelli}, {Stamatikos}, {Starck}, {Stecker},
  {Strickman}, {Suson}, {Tajima}, {Takahashi}, {Tanaka}, {Thayer}, {Thayer},
  {Thompson}, {Tibaldo}, {Toma}, {Torres}, {Tosti}, {Tramacere}, {Uchiyama},
  {Uehara}, {Usher}, {van der Horst}, {Vasileiou}, {Vilchez}, {Vitale}, {von
  Kienlin}, {Waite}, {Wang}, {Winer}, {Wood}, {Wu}, {Yamazaki}, {Ylinen}, \&
  {Ziegler}}]{DePasquale2010}
{De Pasquale}, M., {Schady}, P., {Kuin}, N.~P.~M., {et~al.} 2010, \apjl, 709,
  L146, \dodoi{10.1088/2041-8205/709/2/L146}

\bibitem[{{Dermer} {et~al.}(2000){Dermer}, {Chiang}, \& {Mitman}}]{Dermer2000}
{Dermer}, C.~D., {Chiang}, J., \& {Mitman}, K.~E. 2000, \apj, 537, 785,
  \dodoi{10.1086/309061}

\bibitem[{{Evans} {et~al.}(2007){Evans}, {Beardmore}, {Page}, {Tyler},
  {Osborne}, {Goad}, {O'Brien}, {Vetere}, {Racusin}, {Morris}, {Burrows},
  {Capalbi}, {Perri}, {Gehrels}, \& {Romano}}]{Evans07}
{Evans}, P.~A., {Beardmore}, A.~P., {Page}, K.~L., {et~al.} 2007, \aap, 469,
  379, \dodoi{10.1051/0004-6361:20077530}

\bibitem[{{Evans} {et~al.}(2009){Evans}, {Beardmore}, {Page}, {Osborne},
  {O'Brien}, {Willingale}, {Starling}, {Burrows}, {Godet}, {Vetere}, {Racusin},
  {Goad}, {Wiersema}, {Angelini}, {Capalbi}, {Chincarini}, {Gehrels}, {Kennea},
  {Margutti}, {Morris}, {Mountford}, {Pagani}, {Perri}, {Romano}, \&
  {Tanvir}}]{Evans09}
---. 2009, \mnras, 397, 1177, \dodoi{10.1111/j.1365-2966.2009.14913.x}

\bibitem[{{Fan} \& {Piran}(2006)}]{Fan2006}
{Fan}, Y., \& {Piran}, T. 2006, \mnras, 370, L24,
  \dodoi{10.1111/j.1745-3933.2006.00181.x}

\bibitem[{{Fan} \& {Wei}(2005)}]{Fan2005}
{Fan}, Y.~Z., \& {Wei}, D.~M. 2005, \mnras, 364, L42,
  \dodoi{10.1111/j.1745-3933.2005.00102.x}

\bibitem[{{Ghisellini} {et~al.}(2010){Ghisellini}, {Ghirlanda}, {Nava}, \&
  {Celotti}}]{Ghisellini2010}
{Ghisellini}, G., {Ghirlanda}, G., {Nava}, L., \& {Celotti}, A. 2010, \mnras,
  403, 926, \dodoi{10.1111/j.1365-2966.2009.16171.x}

\bibitem[{{Granot} \& {Sari}(2002)}]{GranotSari2002}
{Granot}, J., \& {Sari}, R. 2002, \apj, 568, 820, \dodoi{10.1086/338966}

\bibitem[{{He} {et~al.}(2012){He}, {Zhang}, {Wang}, {Li}, \&
  {M{\'e}sz{\'a}ros}}]{He2012}
{He}, H.-N., {Zhang}, B.-B., {Wang}, X.-Y., {Li}, Z., \& {M{\'e}sz{\'a}ros}, P.
  2012, \apj, 753, 178, \dodoi{10.1088/0004-637X/753/2/178}

\bibitem[{{Kouveliotou} {et~al.}(2013){Kouveliotou}, {Granot}, {Racusin},
  {Bellm}, {Vianello}, {Oates}, {Fryer}, {Boggs}, {Christensen}, {Craig},
  {Dermer}, {Gehrels}, {Hailey}, {Harrison}, {Melandri}, {McEnery}, {Mundell},
  {Stern}, {Tagliaferri}, \& {Zhang}}]{GRB130427A_NuSTAR}
{Kouveliotou}, C., {Granot}, J., {Racusin}, J.~L., {et~al.} 2013, \apjl, 779,
  L1, \dodoi{10.1088/2041-8205/779/1/L1}

\bibitem[{{Kumar} \& {Barniol Duran}(2009)}]{KumarBarniolDuran2009}
{Kumar}, P., \& {Barniol Duran}, R. 2009, \mnras, 400, L75,
  \dodoi{10.1111/j.1745-3933.2009.00766.x}

\bibitem[{{Mattox} {et~al.}(1996){Mattox}, {Bertsch}, {Chiang}, {Dingus},
  {Digel}, {Esposito}, {Fierro}, {Hartman}, {Hunter}, {Kanbach}, {Kniffen},
  {Lin}, {Macomb}, {Mayer-Hasselwander}, {Michelson}, {von Montigny},
  {Mukherjee}, {Nolan}, {Ramanamurthy}, {Schneid}, {Sreekumar}, {Thompson}, \&
  {Willis}}]{Mattox1996}
{Mattox}, J.~R., {Bertsch}, D.~L., {Chiang}, J., {et~al.} 1996, \apj, 461, 396,
  \dodoi{10.1086/177068}

\bibitem[{{Maxham} {et~al.}(2011){Maxham}, {Zhang}, \& {Zhang}}]{Maxham2011}
{Maxham}, A., {Zhang}, B.-B., \& {Zhang}, B. 2011, \mnras, 415, 77,
  \dodoi{10.1111/j.1365-2966.2011.18648.x}

\bibitem[{{Neyman} \& {Pearson}(1928)}]{Neyman1928}
{Neyman}, J., \& {Pearson}, E.~S. 1928, Biometrika, 20, 175,
  \dodoi{10.1007/s11214-005-5097-2}

\bibitem[{{Panaitescu}(2008)}]{Panaitescu2008}
{Panaitescu}, A. 2008, \mnras, 383, 1143,
  \dodoi{10.1111/j.1365-2966.2007.12607.x}

\bibitem[{{Racusin} {et~al.}(2016){Racusin}, {Oates}, {de Pasquale}, \&
  {Kocevski}}]{Racusin2016}
{Racusin}, J.~L., {Oates}, S.~R., {de Pasquale}, M., \& {Kocevski}, D. 2016,
  \apj, 826, 45, \dodoi{10.3847/0004-637X/826/1/45}

\bibitem[{{Racusin} {et~al.}(2009){Racusin}, {Liang}, {Burrows}, {Falcone},
  {Sakamoto}, {Zhang}, {Zhang}, {Evans}, \& {Osborne}}]{Racusin09}
{Racusin}, J.~L., {Liang}, E.~W., {Burrows}, D.~N., {et~al.} 2009, \apj, 698,
  43, \dodoi{10.1088/0004-637X/698/1/43}

\bibitem[{{Racusin} {et~al.}(2011){Racusin}, {Oates}, {Schady}, {Burrows}, {de
  Pasquale}, {Donato}, {Gehrels}, {Koch}, {McEnery}, {Piran}, {Roming},
  {Sakamoto}, {Swenson}, {Troja}, {Vasileiou}, {Virgili}, {Wanderman}, \&
  {Zhang}}]{Racusin2011}
{Racusin}, J.~L., {Oates}, S.~R., {Schady}, P., {et~al.} 2011, \apj, 738, 138,
  \dodoi{10.1088/0004-637X/738/2/138}

\bibitem[{{Razzaque} {et~al.}(2010){Razzaque}, {Dermer}, \&
  {Finke}}]{Razzaque2010a}
{Razzaque}, S., {Dermer}, C.~D., \& {Finke}, J.~D. 2010, The Open Astronomy
  Journal, 3, 150, \dodoi{10.2174/1874381101003010150}

\bibitem[{{Roming} {et~al.}(2005){Roming}, {Kennedy}, {Mason}, {Nousek}, {Ahr},
  {Bingham}, {Broos}, {Carter}, {Hancock}, {Huckle}, {Hunsberger}, {Kawakami},
  {Killough}, {Koch}, {McLelland}, {Smith}, {Smith}, {Soto}, {Boyd},
  {Breeveld}, {Holland}, {Ivanushkina}, {Pryzby}, {Still}, \&
  {Stock}}]{Roming05}
{Roming}, P.~W.~A., {Kennedy}, T.~E., {Mason}, K.~O., {et~al.} 2005, \ssr, 120,
  95, \dodoi{10.1007/s11214-005-5095-4}

\bibitem[{{Sari} \& {Esin}(2001)}]{Sari2001}
{Sari}, R., \& {Esin}, A.~A. 2001, \apj, 548, 787, \dodoi{10.1086/319003}

\bibitem[{{Schulze} {et~al.}(2011){Schulze}, {Klose}, {Bj{\"o}rnsson},
  {Jakobsson}, {Kann}, {Rossi}, {Kr{\"u}hler}, {Greiner}, \&
  {Ferrero}}]{Schulze2011}
{Schulze}, S., {Klose}, S., {Bj{\"o}rnsson}, G., {et~al.} 2011, \aap, 526, A23,
  \dodoi{10.1051/0004-6361/201015581}

\bibitem[{{Tam} {et~al.}(2013){Tam}, {Tang}, {Hou}, {Liu}, \& {Wang}}]{Tam2013}
{Tam}, P.-H.~T., {Tang}, Q.-W., {Hou}, S.-J., {Liu}, R.-Y., \& {Wang}, X.-Y.
  2013, \apjl, 771, L13, \dodoi{10.1088/2041-8205/771/1/L13}

\bibitem[{{Vianello} {et~al.}(2015){Vianello}, {Omodei}, \& {Fermi/LAT
  collaboration}}]{Vianello2015}
{Vianello}, G., {Omodei}, N., \& {Fermi/LAT collaboration}. 2015, ArXiv
  e-prints.
\newblock \doarXiv{1502.03122}

\bibitem[{{Wang} {et~al.}(2013){Wang}, {Liu}, \& {Lemoine}}]{Wang2013}
{Wang}, X.-Y., {Liu}, R.-Y., \& {Lemoine}, M. 2013, \apjl, 771, L33,
  \dodoi{10.1088/2041-8205/771/2/L33}

\bibitem[{Wilks(1938)}]{Wilks1938}
Wilks, S.~S. 1938, Ann. Math. Statist., 9, 60, \dodoi{10.1214/aoms/1177732360}

\bibitem[{{Yassine} {et~al.}(2017){Yassine}, {Piron}, {Mochkovitch}, \&
  {Daigne}}]{Yassine2017}
{Yassine}, M., {Piron}, F., {Mochkovitch}, R., \& {Daigne}, F. 2017, \aap, 606,
  A93, \dodoi{10.1051/0004-6361/201630353}

\bibitem[{{Zhang} {et~al.}(2006){Zhang}, {Fan}, {Dyks}, {Kobayashi},
  {M{\'e}sz{\'a}ros}, {Burrows}, {Nousek}, \& {Gehrels}}]{Zhang2006}
{Zhang}, B., {Fan}, Y.~Z., {Dyks}, J., {et~al.} 2006, \apj, 642, 354,
  \dodoi{10.1086/500723}

\bibitem[{{Zhang} \& {M{\'e}sz{\'a}ros}(2001)}]{Zhang2001}
{Zhang}, B., \& {M{\'e}sz{\'a}ros}, P. 2001, \apj, 559, 110,
  \dodoi{10.1086/322400}

\bibitem[{{Zhang} {et~al.}(2011){Zhang}, {Zhang}, {Liang}, {Fan}, {Wu},
  {Pe'er}, {Maxham}, {Gao}, \& {Dong}}]{Zhang2011}
{Zhang}, B.-B., {Zhang}, B., {Liang}, E.-W., {et~al.} 2011, \apj, 730, 141,
  \dodoi{10.1088/0004-637X/730/2/141}

\end{thebibliography}

\end{document}